\documentclass[12pt]{article}
\pdfoutput=1

\usepackage{multirow}
\usepackage[pdftex]{graphicx}
\usepackage{amssymb}
\usepackage{amsmath}
\usepackage{cite}

\usepackage{longtable}
\usepackage{booktabs}
\usepackage{array}
\usepackage{ulem}

\usepackage{xcolor}
\definecolor{tit}{rgb}{0.1,0.2,0.4}

\renewcommand{\arraystretch}{1.2} 

\usepackage{subfig}

\newcommand{\eq}[1]{\begin{equation} #1 \end{equation}}

\newcommand{\eqa}[1]{\begin{eqnarray} #1 \end{eqnarray}}

\newcommand{\av}[1]{\langle #1 \rangle}

\newcommand{\GeV}{\,{\rm GeV}}
\newcommand{\MeV}{\,{\rm MeV}}

\newcommand{\op}{\mathcal{O}}

\newcommand{\Eq}[1]{Eq.~(\ref{#1})}

\newcommand{\Sec}[1]{Section~\ref{#1}}
\newcommand{\App}[1]{Appendix~\ref{#1}}
\newcommand{\Reff}[1]{Ref.~\cite{#1}}
\newcommand{\Tab}[1]{Table~\ref{#1}}
\newcommand{\Fig}[1]{Figure~\ref{#1}}
\newcommand{\im}{{\rm Im}}
\newcommand{\re}{{\rm Re}}
\newcommand{\A}{{\cal A}}
\newcommand{\F}{{\cal F}}
\newcommand{\G}{{\cal G}}
\newcommand{\D}{{\cal D}}
\newcommand{\N}{{\cal N}}
\newcommand{\B}{{\cal B}}

\renewcommand{\H}{{\cal H}}
\renewcommand{\S}{{\cal S}}

\begin{document}

\allowdisplaybreaks

\begin{flushright}
SI-HEP-2023-02, P3H-23-004,
Nikhef-2023-004\\
\end{flushright}

$\ $
\vspace{-2mm}
\begin{center}
\fontsize{16}{20}\selectfont
\bf 
Light-Cone Sum Rules for $S$-wave $B\to K\pi$ Form Factors
\end{center}

\vspace{2mm}

\begin{center}
{\rm S\'ebastien Descotes-Genon{$^{\, a}$}, Alexander Khodjamirian{$^{\, b}$},
Javier Virto{$^{\, c,d}$}} and K. Keri Vos{$^{\, e,f}$}\\[5mm]
{\it\small
{$^{\, a}$} 
Universit\'e Paris-Saclay, CNRS/IN2P3, IJCLab, 
91405 Orsay, France\\[2mm]
{$^{\, b}$} 
Center for Particle Physics Siegen, \\
Theoretische Physik 1, Universit\"at Siegen, 57068 Siegen, Germany\\[2mm]
{$^{\, c}$}
Departament de Física Quàntica i Astrofísica, Universitat de Barcelona,\\
Martí Franquès 1, E08028 Barcelona, Catalunya\\[2mm]
{$^{\, d}$}
Institut de Ciències del Cosmos (ICCUB), Universitat de Barcelona,\\
Martí Franquès 1, E08028 Barcelona, Catalunya\\[2mm]
{$^{\, e}$} Gravitational 
Waves and Fundamental Physics (GWFP),\\ 
Maastricht University, Duboisdomein 30,
NL-6229 GT Maastricht, the
Netherlands\\[0.3cm]
{$^{\, f}$} Nikhef, Science Park 105,
NL-1098 XG Amsterdam, the Netherlands}
\end{center}

\vspace{1mm}
\begin{abstract}\noindent
\vspace{-5mm}

We derive a set of light-cone sum rules relating the $S$-wave $B\to K \pi$ hadronic form factors to the $B$-meson light-cone distribution amplitudes (LCDAs), taking into account the complete set of LCDAs up to and including twist four. 
These results complement the sum rules for the $P$-wave $B\to K \pi$ form factors obtained earlier.
We then use the new sum rules to estimate the $S$-wave contributions to $B\to K \pi\ell\ell$ decays as a function of the $K\pi$ invariant  mass. 
We pay particular attention to the fact that the $S$-wave $K\pi$ spectrum cannot be modelled by a sum of Breit-Wigner resonances, and employ a more consistent dispersive coupled-channel approach. We compare our predictions for branching ratios and angular observables with LHCb measurements in two different kinematic regions, around $K^*(892)$ and $K^*_0(1430)$. We observe an overall compatibility and discuss possible improvements of our model to obtain a better description of the $B\to K\pi$ form factors over a large kinematic range.

\end{abstract}

\newpage

\setcounter{tocdepth}{2}
\tableofcontents

\newpage

%%%%%%%%%%%%%%%%%%%%%%%%%%%%%%%%%%%%%%%%%%%%%%%
\section{Introduction}
\label{sec:intro}

The decay $B\to K\pi \ell^+\ell^-$  remains one of the most important modes in the investigation of  the $b\to s\ell^+\ell^-$ flavour-changing neutral current (FCNC)
transition. This decay occurs predominantly through the vector-resonance channel $B\to K^*(892) \ell^+\ell^-$, which has received most of the experimental and theoretical focus (see~Refs.~\cite{LHCb:2020gog,LHCb:2020lmf} for the latest LHCb results and Refs.~\cite{Bifani:2018zmi,Albrecht:2021tul} for recent reviews).
Several interesting tensions have been observed in the muon mode, concerning the branching ratio and some of the angular observables~\cite{Descotes-Genon:2013wba,Descotes-Genon:2015uva,Gubernari:2022hxn}, in particular the so-called $P_5'$ observable~\cite{Descotes-Genon:2012isb,Descotes-Genon:2013vna}.

However, while the $K^*(892)$ is the most prominent resonance, it is only one of the many possibilities for the $K\pi$ system in a $P$-wave state. Based on the same approach as in \Reff{Cheng:2017smj}, the contribution of excited vector $K^*$ resonances and also of the non-resonant $P$-wave $K\pi$ state  have been studied in \Reff{Descotes-Genon:2019bud}, where the $P$-wave $B\to K\pi$ form factors were obtained from QCD light-cone sum rules (LCSRs). This study  allowed to uncover two important effects.
First, a notable impact of the non-vanishing $K^*$ total width was found, leading to  an $\op(10\%)$ increase of the $B\to K^*$ form factors compared to the narrow-width limit.
Second, large contributions from higher resonances were found to be constrained by existing experimental measurements performed outside the $K^*$ window~\cite{LHCb:2016eyu}. 
These findings thus illustrated the usefulness of investigating the LCSRs for form factors beyond the well-known case of final states with a single narrow resonance.

In this paper we  concentrate on another important part of the  $B\to K\pi \ell\ell$ decay amplitude in which the $K\pi$ pair is produced in an $S$-wave. This requires the knowledge of
the $S$-wave $B\to K\pi$ form factors.
Our main goal is to study these form factors within the same LCSR approach as in \Reff{Descotes-Genon:2019bud}. 
There are several important motivations for this study:

\begin{enumerate}

\item The $S$-wave $K\pi$ state represents a potentially important
background for the $B\to K^*\ell\ell $ channel. The LHCb collaboration has indeed identified a non-negligible $S$-wave fraction of 10\% under the $K^*$ peak~\cite{LHCb:2016ykl}. There are also hints that the interference between the $S$-wave and the other components is important in the higher-resonance region around the $K^*_{0,2}(1430)$~\cite{LHCb:2016eyu}. It was stressed in \Reff{Becirevic:2012dp} that
the scalar component could affect the accurate extraction of $B\to K^*\mu\mu$ angular observables. 
Current LHCb analyses for $B\to K^*\mu\mu$ include this component by treating the $S$-wave fraction and the additional angular coefficients arising from the interference between $S$ and $P$ waves as nuisance parameters~\cite{LHCb:2020lmf}. 

\item It is notoriously difficult to describe the $S$-wave $K\pi$ state at low invariant masses. More specifically, there is a scalar resonance ($K^{*}_0(700)$) in the same region as the $K^*(892)$~\cite{ParticleDataGroup:2022pth}, but it is known to elude a Breit-Wigner (BW) description due to its large width. 
A better description of the $S$-wave component of the $K\pi$ state would thus contribute to a better understanding of its interference with the $P$-wave in the $B\to K\pi\ell\ell$ decay. As described in detail in \Reff{Alguero:2021yus}, angular observables associated with these components can be extracted experimentally (rather than treated as background/nuisance terms). Besides providing useful cross-checks of the experimental analyses, these angular observables could in be principle used to constrain New Physics (NP), provided that a solid theoretical description of the hadronic dynamics is available~\footnote{This description should also include the non-local or ``charm loop'' effects specifically in the $S$ wave, a problem which remains beyond our scope here.}.

\item The $B\to K\pi$ form factors are not only relevant for semileptonic decays, but appear also in factorization theorems for non-leptonic multi-body $B$ decays~\cite{Krankl:2015fha,Virto:2016fbw,Klein:2017xti,Huber:2020pqb,Mannel:2020abt}. The $S$-wave $B\to K\pi$ form factors are also relevant for a complete phenomenological analysis of these decay modes. 

\end{enumerate}

We will consider LCSRs for the $S$-wave $B\to K\pi$ form factors based on the  operator-product expansion (OPE) of the vacuum-to-$B$ correlation function of two currents. One of them is  a $b\to s$ transition current, whereas the other one is a quark-antiquark current interpolating the final $K\pi$  hadronic state. In order to isolate the $K\pi$ $S$-wave, the chosen interpolating current is the scalar light-quark current with strangeness.
The version of the LCSR method with the $B$-meson distribution amplitudes (DAs) used here originated in \Reff{Khodjamirian:2005ea} 
and was used in several other $B$-meson  form factor calculations
(e.g.~\cite{Khodjamirian:2006st,Khodjamirian:2010vf,Wang:2015vgv,Cheng:2017smj,Gubernari:2018wyi,Descotes-Genon:2019bud,Gubernari:2020eft}).
The generalization to the case of two mesons in the final state was proposed in \Reff{Cheng:2017smj}~\footnote{$B\to M_1M_2$ form factors have also been addressed within the LCSRs with dimeson DAs~\cite{Hambrock:2015aor,Cheng:2017sfk}.}.
As for any QCD sum rule, we rely on the dual nature of the underlying correlation function. On the one hand, it is cast into a hadronic dispersion representation with a spectral density 
saturated by the intermediate states with strangeness and  spin-parity $J^P=0^+$. On the other hand, the same correlation function is computed, employing a light-cone OPE in terms of $B$-meson DAs convoluted with perturbatively computed short-distance kernels. 
In this respect, the way the LCSRs are obtained in this paper largely follows~\Reff{Descotes-Genon:2019bud}. Most importantly, we can use 
the same non-perturbative input in the form of $B$-meson DAs, taken into account up to twist four.  

An essential novelty concerns the hadronic part of the LCSRs
obtained here. In the case of $P$-wave $B\to K\pi$ 
form factors,  a set of Breit-Wigner (BW) resonances -- the $K^*(892) $ and its radial excitations -- described reasonably  well the spectral density.
This was supported by measurements of the $\tau\to K \pi \nu_\tau$ decay distribution, where the vector part of the hadronic spectral density is determined by the same $P$-wave $K\pi$ state. However,
for the scalar $K\pi$ state, a simple BW ansatz would constitute an oversimplification. We thus pay special attention to 
this issue and employ a more realistic model for the hadronic spectral density based on the dispersive analysis of~\Reff{VonDetten:2021rax}.
The same spectral density emerges in the auxiliary two-point QCD sum rule for two scalar currents with strangeness. The latter sum rule is used to estimate the quark-hadron duality threshold, in analogy with~Refs.~\cite{Cheng:2017smj,Descotes-Genon:2019bud}.

The rest of the article is organized as follows. In~\Sec{sec:definitions} we define the $S$-wave $B\to K\pi$ 
form factors and discuss the related kinematics. In~\Sec{sec:BDAs} the LCSRs for these form factors 
are derived. In~\Sec{sec:ffmodel} we discuss the coupled-channel model for the scalar $K\to \pi$ form factor and 
introduce the corresponding ansatz for the $B\to K\pi$ form factors. \Sec{sec:analysis} contains a numerical analysis. In~\Sec{sec:BKpiellell} we use the form factors to analyse 
the role of the $K\pi$ $S$-wave in the $B\to K\pi \ell\ell$ decay. Finally, \Sec{sec:Conclusions} contains our concluding discussion.
In~\App{app:OPE} we collect the results for the OPE parts of the LCSRs. In~\App{app:resmodel}
we present the various models for the $B\to K\pi$ form factors. \App{app:2ptSR} contains the analysis of the two-point QCD sum rules in the scalar $K\pi$ channel.

%%%%%%%%%%%%%%%%%%%%%%%%%%%%%%%%%%%%%%%%%%%%%%%
\section{$S$-wave Form Factors and Kinematics}
\label{sec:definitions}

The complete definitions of all $B\to K\pi$ form factors and their partial wave expansions have been presented in \Reff{Descotes-Genon:2019bud}. In this paper we use the same conventions, which we repeat here for convenience and reference. The form factors $F_i^{(T)}(k^2,q^2,q\cdot \overline{k})$ are defined by the following Lorentz decomposition:
\eqa{
i\langle K^-(k_1)\pi^+(k_2)|\bar{s}\gamma^\mu b|\bar{B}^0(q+k)\rangle
&=& F_\perp \, k^\mu_\perp
\,,
\nonumber\\
-i\langle K^-(k_1)\pi^+(k_2)|\bar{s}\gamma^\mu \gamma_5 b|\bar{B}^0(q+k)\rangle
&=&  F_t \, k^\mu_t + F_0 \, k^\mu_0 + F_\parallel \, k^\mu_\|\,, 
\nonumber\\
\langle K^-(k_1)\pi^+(k_2)|\bar{s}\sigma^{\mu\nu} q_\nu b|\bar{B}^0(q+k)\rangle
&=&  F_\perp^T \, k^\mu_\perp \,, 
\label{eq:defFF}\\
\langle K^-(k_1)\pi^+(k_2)|\bar{s} \sigma^{\mu\nu}q_\nu \gamma_5  b|\bar{B}^0(q+k)\rangle
&=& F_0^T \, k^\mu_0 + F_\parallel^T \, k^\mu_\|\,,
\nonumber
}
in terms of the following set of orthogonal Lorentz vectors:
\begin{align}
k^\mu_\perp &= \frac{2}{\sqrt{k^2} \sqrt{ \lambda}} \,
i\epsilon^{\mu\alpha\beta\gamma} \, q_\alpha \, k_{\beta} \, \bar{k}_{\gamma}\ ,   &
k^\mu_t &= \frac{q^\mu}{\sqrt{q^2}}\ , \nonumber \\
k^\mu_0 &= \frac{2\sqrt{q^2}}{\sqrt{\lambda}} \, \Big(k^\mu - \frac{k \cdot q}{q^2} q^\mu\Big)\ ,  &
k^\mu_\| &= \frac{1}{\sqrt{k^2}} \,
\Big(\overline{k}^\mu - \frac{4 (q\cdot k) (q \cdot \overline{k})}{\lambda} \, k_\mu
+ \frac{4 k^2 (q\cdot \overline{k})}{\lambda} \, q_\mu \Big)\ .
\label{eq:kpar}
\end{align}
Here $\lambda\equiv\lambda(m_B^2,q^2,k^2) = m_B^4 + q^4 + k^4 - 2(m_B^2 q^2 + m_B^2 k^2 + q^2 k^2 )$ is the kinematic K\"all\'en function.
The total dimeson momentum is $k=k_1+k_2$, and
\eq{
\overline k^\mu =
\bigg(1- \frac{\Delta m^2}{k^2}\bigg)\ k_1^\mu
- \bigg(1+ \frac{\Delta m^2}{k^2}\bigg)\ k_2^\mu\ ,
\label{eq:kbar}
} 
with $\Delta m^2 \equiv k_1^2 - k_2^2 = m_1^2 - m_2^2$, such that $k\cdot \overline k=0$.
Some useful relations are:
\begin{align}
q\cdot k &=\frac12(m_B^2-q^2-k^2)\ ,   &
q\cdot \overline k &=
\frac{\sqrt{\lambda\,\lambda_{K\pi}}\ \cos \theta_K}{2k^2}\ ,  \nonumber \\
\lambda &= 4(q\cdot k)^2 - 4q^2 k^2\ ,   &
k^2 \overline k^2  &=  -\lambda_{K\pi}  \ ,
\label{eq:kinematicrels}
\end{align}
where $\lambda_{K\pi} \equiv \lambda(k^2,m_K^2,m_\pi^2)$, and $\theta_K$ is the angle between the 3-momenta of the  pion and the
$B$-meson in the $(K\pi)$ rest frame. 

The dependence on $\theta_K$ (i.e. on~$q\cdot \bar k$) can be separated by partial-wave expansion.
\Reff{Descotes-Genon:2019bud} focuses on the $P$-wave ($\ell=1$) components, while here we focus on the $S$-wave ($\ell=0$):
\eq{
\label{eq:PWE0t}
F_{0,t}(k^2,q^2, q\cdot \bar k) =F_{0,t}^{(\ell=0)}(k^2,q^2)
+\sum_{\ell=1}^{\infty} \sqrt{2 \ell+1}\ F_{0,t}^{(\ell)}(k^2,q^2)\ 
P_{\ell}^{(0)}(\cos\theta_K)\, ,
}
where $P^{(0)}_0=1$ has been used.
The same expansion is valid for the tensor form factor $F_{0}^T$, while $F_\perp^{(T)}$ and $F_\|^{(T)}$ contain no $S$-wave components.
Our main task is to find LCSR relations for the three $S$-wave $B\to (K\pi)_S$ form factors $F_{0}^{(\ell=0)}$, $F_{t}^{(\ell=0)}$ and $F_0^{T(\ell=0)}$, referred to, respectively,
as the longitudinal, timelike-helicity  and tensor $S$-wave form factors.
In order to simplify the notation along the paper, hereafter the $S$-wave tensor form factor will be denoted as
\eq{
F_0^{T(\ell=0)} \equiv F_T^{(\ell=0)}\ .
}

For definiteness, we consider the $\bar{B}^0\to K^-\pi^+$ transition.
Isospin symmetry allows one to relate the form factors
of all four $\bar{B}\to K\pi$ transitions:
\begin{eqnarray}
-\langle \bar{K}^0(k_1) \pi^0(k_2)|j_b|\bar{B}^0(p)\rangle=
\langle K^-(k_1) \pi^0(k_2)|j_b|B^-(p)\rangle
\nonumber\\
=\frac{1}{\sqrt{2}}\langle \bar{K}^0(k_1) \pi^-(k_2)| j_b|B^-(p)\rangle
=
\frac{1}{\sqrt{2}}\langle K^-(k_1) \pi^+(k_2)|j_b|\bar{B}^0(p)\rangle\,,
\label{eq:iso}
\end{eqnarray}
where $j_b$ is any one of the $b\to s$ transition currents in~\Eq{eq:defFF}. For brevity we denote the relevant axial-vector and pseudotensor currents by
\eq{
j_A^\mu=\bar{s}\gamma^\mu \gamma_5 b\ ,
\qquad
j^\mu_T=\bar{s} \sigma^{\mu\nu}q_\nu \gamma_5  b\ .
\label{eq:jb}
}
In the sum rules we will also need the form factor of the scalar strange current interpolating the $S$-wave of the $K\pi$ state.
Starting from the standard definition for the vector strange current
in terms of the vector and scalar form factors:
\eqa{
\langle K^-(k_1)\pi^+(k_2) |\bar{s}\gamma_\mu d | 0 \rangle =
f_+(k^2) \ \overline k_\mu
+\frac{m_K^2 - m_\pi^2}{k^2}  f_0(k^2) \ k_\mu\ \,,
\label{eq:KpivectFF} }
and multiplying both sides by $k_\mu$, we recover
the divergence of the vector current on l.h.s.  and relate
the scalar form factor $f_0$ with the hadronic matrix element
\eqa{
\langle K^-(k_1)\pi^+(k_2) |j_S| 0 \rangle =
(m_K^2 - m_\pi^2) f_0(k^2)\equiv F_S(k^2)\,,
\label{eq:KpiscalFF} }
where the scalar strange current is defined as
\eq{
j_S=(m_s-m_d)\bar{s}d\ ,
\qquad
j_S^\dagger= (m_s-m_d)\bar{d}s\ .
\label{eq:scalcurr}
}
The corresponding isospin relations for the $K\pi$  form factors are:
\eqa{
\langle \bar{K}^0(k_1) \pi^0(k_2)|\bar{s}d|0\rangle &=&
-\langle K^-(k_1) \pi^0(k_2)|\bar{s}u|0\rangle
\nonumber\\
&=& \frac{1}{\sqrt{2}}\langle \bar{K}^0(k_1) \pi^-(k_2)| \bar{s}u|0\rangle
=
-\frac{1}{\sqrt{2}}\langle K^-(k_1) \pi^+(k_2)|\bar{s}d|0\rangle\,.
\label{eq:isoKpi}
}
%

%%%%%%%%%%%%%%%%%%%%%%%%%%%%%%%%%%%%%%%%%%%%%%%
\section{LCSRs with $B$-meson Distribution Amplitudes}
\label{sec:BDAs}

\noindent We follow the method proposed in \Reff{Khodjamirian:2006st} and consider a correlation function
\eq{
\S^{\mu}_b(k,q) = i\int d^4x\, e^{i k \cdot x} \langle 0 | \text{T} \{ j^\dagger_{S}(x), j^{\mu}_b(0) \} | \bar{B}^0(q+k) \rangle
= L^{\mu}(k,q)\, \S(k^2, q^2)+\cdots\ ,
\label{eq:corrB}
}
in which the $b\to s$ transition current $j_b$ 
(one of the currents defined in~\Eq{eq:jb}) and the scalar current $j^\dagger_S$  defined in~\Eq{eq:scalcurr} are sandwiched between the on-shell $B$-meson  state and the vacuum, so that $(q+k)^2=m_B^2$. We consider the invariant amplitude 
$\S(k^2,q^2)$ multiplying  a certain Lorentz structure $L^{\mu}$, 
indicating  by dots the other possible structures.

Taking the external momenta
$k$ and $q$  in the region far below the hadronic thresholds,
\eq{
k^2<0,~~ |k^2| \gg \Lambda_{QCD}^2~~ \mbox{and}~~ 
q^2\ll m_b^2\,,
\label{eq:region}
}
the function $\S(k^2,q^2)$ can be calculated by means of a light-cone OPE in terms of $B$-meson LCDAs.
We then employ the dispersion relation in the variable $k^2$, relating the OPE result for the invariant amplitude in the region given by~\Eq{eq:region} to the integral over its imaginary part,
\eq{
\S^\text{OPE}(k^2, q^2) = \frac{1}{\pi} 
\int_{s_\text{th}}^\infty ds \,\frac{\im\, \S(s,q^2)}{s-k^2}\,,
\label{eq:disp1}
}
where $s_\text{th}=(m_K+m_\pi)^2$ is the lowest hadronic threshold.
The spectral density  of the correlation function in~\Eq{eq:corrB} is obtained from unitarity by inserting a full set of hadronic states between the two currents in the $T$-product. The contribution from the $K\pi$ states with the 
lowest threshold $(m_K+m_\pi)^2$ is: 
\begin{eqnarray} 
2\, \im \S_b^{\mu,(K\pi)}(k,q)&=&
\sum\limits_{K\pi}\int d\tau_{K\pi}
\langle 0 |j_S^{\dagger}\,| K(k_1)\pi(k_2) \rangle
\langle K(k_1)\pi(k_2) | j^{\mu}_b| \bar{B}^0(q+k)\rangle
\nonumber\\
&=& L^{\mu}(k,q) \big[2\,\im\, \S^{(K\pi)}(k^2,q^2)\big]\, +\dots,
\label{eq:Kpidisp}
\end{eqnarray}
with the same Lorentz-structure  as in~\Eq{eq:corrB}. Denoting by $\S^{(h)}$
the sum over all other contributions to the invariant amplitude $\S$ with thresholds $s_h>s_\text{th}$, 
we have:
\eq{
\im \S(s,q^2)= \im\,\S^{(K\pi)}(s,q^2)+\im\,\S^{(h)}(s,q^2)\theta(s-s_h) \,.
\label{eq:corr2}
}
We then apply the quark-hadron duality approximation for the dispersion integral over the spectral density of the heavier-threshold states: 
\begin{eqnarray}
\int_{s_h}^\infty ds \,\frac{\im\,\S^{(h)}(s,q^2)}{s-k^2}=
\int_{s_0}^\infty ds \,\frac{\im\,\S^{\text{OPE}}(s,q^2)}{s-k^2}\,,
\label{eq:dual}
\end{eqnarray}
where the integral over imaginary part of the OPE expression is taken 
above the {\it effective threshold} $s_0$.

Performing a Borel transformation in the variable $k^2$ on both  sides of~\Eq{eq:disp1} and using~Eqs.~(\ref{eq:corr2})-(\ref{eq:dual}),
we obtain the following sum rule:
\eq{
\frac1\pi \int_{s_\text{th}}^{s_0} ds \,e^{-s/M^2}\, \im\,\S^{(K\pi)}(s,q^2)
= \frac1\pi \int_{m_s^2}^{s_0}  ds \,e^{-s/M^2}\, \im\, \S^\text{OPE}(s,q^2)\equiv 
\S^\text{OPE}(q^2,s_0,M^2)
\ .
\label{eq:sr}
}
Note that in the OPE expression we neglect the $u$ and $d$ quark masses, hence the lower integration limit on the r.h.s.

\bigskip

Having outlined the method in general, we apply it now to the form factors of the axial-vector $b\to s$ current. We thus start from the correlation function 
\eqa{
\S_A^{\mu}(k,q)&=&i\int d^4x\, e^{i k \cdot x}
\av{ 0 | {\rm T} \{j_S^{\dagger}(x),j^\mu_A(0) \} | \bar{B}^0(q+k) }
\nonumber \\[1mm]
&=&i\,\bigg( k^\mu - \frac{(k\cdot q)}{q^2}q^\mu \bigg)\, \S_{0}(k^2,q^2) + i\,\frac{q^\mu}{q^2} \S_{t}(k^2,q^2)\ .
\label{eq:corrA}
}
The Lorentz decomposition in two independent four-vectors allows one to obtain the two sum rules for the longitudinal and timelike-helicity  form factors from the two invariant amplitudes $\S_0$ and $\S_t$,
respectively. To proceed, we derive the $K\pi$-state contribution to the 
hadronic spectral density: 
\eqa{ 
&&
2\, \im \S_A^{\mu,(K\pi)}(k,q)= \sum_{K\pi}\int d\tau_{K\pi}
\langle 0 |j_S^\dagger\,| K(k_1)\pi(k_2) \rangle
\langle K(k_1)\pi(k_2) | j_A^\mu| \bar{B}^0(q+k)\rangle
\label{eq:KpidispA}\\
&&
=\frac{3\sqrt{\lambda_{K\pi} q^2}}{32\pi k^2}F^{*}_S(k^2)\int_{-1}^1 d \cos\theta_K
\bigg[\frac{2}{\lambda}\Big( k^\mu-\frac{k\cdot q}{q^2}q^\mu\Big)F_0(k^2,q^2, q\cdot \bar k)
+\frac{q^\mu}{q^2}F_t(k^2,q^2, q\cdot \bar k)\bigg]\,,
\nonumber
}
where the isospin-related $\bar{K}^0\pi^0$ state is included, the phase space integral is reduced to the 
angular integration  and the definitions of both $K\pi$ and $B\to K\pi$ form factors are used.
We then use partial wave expansions and integrate over the angle $\theta_K$,
employing the orthogonality of the Legendre polynomials,
\eq{
\int_{-1}^1 d \cos\theta_K\, F_{0,t}(k^2,q^2, q\cdot \bar k)= 2F^{(\ell=0)}_{0,t}(k^2,q^2)\ .
}
Matching the coefficients of the Lorentz structures 
in~Eqs.~(\ref{eq:corrA}) and~(\ref{eq:KpidispA}), we obtain the  $S$-wave $K\pi$ state contributions to the imaginary parts of the invariant amplitudes:
\eqa{
\im\,\S_0^{(K\pi)}(s,q^2)&=&\frac{3\sqrt{\lambda_{K\pi}(s)}}{16\pi s \sqrt{\lambda(s)}} F_S^{*}(s)
\sqrt{q^2}F^{(\ell=0)}_{0}(s,q^2)\ ,
\nonumber
\\
\im\,\S_t^{(K\pi)}(s,q^2)&=&\frac{3\sqrt{\lambda_{K\pi}(s)}} {32\pi s} 
F_S^{*}(s)
\sqrt{q^2} F^{(\ell=0)}_{t}(s,q^2)\ ,
\label{eq:KpiIm}
}
where $\lambda(s)=\lambda(m_B^2,q^2,s)$ and $\lambda_{K\pi}(s) = \lambda(s,m_K^2,m_\pi^2)$.
The resulting LCSRs take the form:

\eqa{
\frac{3}{16\pi^2} \int_{s_\text{th}}^{s_0} ds \,e^{-s/M^2}
\frac{\sqrt{\lambda_{K\pi}(s)}}{s\sqrt{\lambda(s)}} F_S^{*}(s)
\sqrt{q^2}F^{(\ell=0)}_{0}(s,q^2)&=&
\S_0^\text{OPE}(q^2,s_0,M^2)\ ,
\label{eq:LCSRF0}\\
\frac{3}{32\pi^2} \int_{s_\text{th}}^{s_0} ds \,e^{-s/M^2}
\frac{\sqrt{\lambda_{K\pi}(s)}}{s} F_S^{*}(s)
\sqrt{q^2}F^{(\ell=0)}_{t}(s,q^2)&=&
\S_t^\text{OPE}(q^2,s_0,M^2)\ .
\label{eq:LCSRFt}
}

Following the same procedure for the tensor form factor, and starting from the correlation function
\eq{
{\cal S}_T^{\mu}(k,q) = i\int d^4x\, e^{i k \cdot x}
\langle 0 | {\rm T} \{j^\dagger_S(x),j^\mu_T(0) \} | \bar{B}^0(q+k) \rangle 
= \bigg( q^2k^\mu - (k\cdot q)q^\mu \bigg)\, {\cal S}_T (k^2,q^2)\,,
}
with the pseudotensor transition current $j^\mu_T$ defined in~\Eq{eq:jb}, we obtain the following sum rule for the tensor form factor:
\eq{
\frac{3}{16\pi^2}\int_{s_\text{th}}^{s_0} ds \,e^{-s/M^2}
\frac{\sqrt{\lambda_{K\pi}(s)}}{s\sqrt{\lambda(s)}} F_S^{*}(s)
\frac{F^{(\ell=0)}_{T}(s,q^2)}{\sqrt{q^2}}=
\S_T^{\text{OPE}}(q^2,s_0,M^2)\,.
\label{eq:LCSRFT}
}

We can also derive a separate sum rule for the timelike-helicity
form factor, as done in~\mbox{Refs.\cite{Cheng:2017smj,Descotes-Genon:2019bud}}, starting from a different correlation function,
\eqa{
\S_5(k^2,q^2)&=& i\int d^4x\, e^{i k \cdot x}
\langle 0 | {\rm T} \{ j^\dagger_S(x),j_5(0) \} | \bar{B}^0(q+k) \rangle\,,\quad 
\label{eq:corr-Ft}
}
where the pseudoscalar $b\to s$  current $j_5 = (m_b+m_s)\bar{s}i\gamma_5 b~$ is used and the correlation function itself represents an invariant amplitude.
We use the definition of the $B\to K\pi$ form factor generated by the pseudoscalar current:
\eq{
\av{ K^-(k_1) \pi^+(k_2)|(m_b+m_s)\bar{s}i\gamma_5 b|\bar{B}^0(p) }
=  \sqrt{q^2}F_t (k^2,q^2,q\cdot k)\ .
}
The resulting LCSR reads
\eq{
\frac{3}{32\pi^2}
\int_{s_\text{th}}^{s_0} ds \,e^{-s/M^2}
\frac{\sqrt{\lambda_{K\pi}(s)}}{s} F_S^{*}(s)
\sqrt{q^2}F^{(\ell=0)}_{5}(s,q^2)
=
\S^\text{OPE}_5(q^2,s_0,M^2)\ .
\label{eq:LCSRF5}
}
We use the notation $F_5^{(\ell=0)} = F_t^{(\ell=0)}$ here, in order to distinguish the timelike-helicity form factor derived from the sum rules with the axial and pseudoscalar transition currents.
The LCSRs in~Eqs.(\ref{eq:LCSRFt}) and~(\ref{eq:LCSRF5}) are identical except for the functions $\S_t^{\text{OPE}}$ and $\S_5^{\text{OPE}}$, which have a different form within the adopted accuracy of the OPE (see~\App{app:OPE}). Thus a numerical comparison of these OPE functions will determine to which extent $F_5^{(\ell=0)} = F_t^{(\ell=0)}$, which will constitute a useful test of the LCSR approach. As argued below, both LCSRs are identical in the heavy-quark limit.

\bigskip

The four light-cone sum rules derived in this section,~Eqs.~(\ref{eq:LCSRF0}),~(\ref{eq:LCSRFt}),~(\ref{eq:LCSRFT}) and~(\ref{eq:LCSRF5}),
can be written compactly as:
\eq{
\int_{s_{\rm th}}^{s_0} ds \ e^{-s/M^2}
\omega_i(s,q^2) \,F^*_S(s)\, F_i^{(\ell=0)}(s,q^2)
= \S_i^{\text{OPE}}(q^2,s_0,M^2) \ ,
\label{eq:AllLCSRs}
}
for
$i = \{0,t,5,T\}$, with the functions $\omega_i(s,q^2)$ given by
\eq{\label{eq:omega}
\omega_0(s,q^2)
= q^2\,\omega_T(s,q^2)
= \frac{2\,\omega_t(s,q^2)}{\sqrt{\lambda(s)}}
= \frac{2\,\omega_5(s,q^2)}{\sqrt{\lambda(s)}}
=\frac{3\sqrt{\lambda_{K\pi}(s)\,q^2}}{16\pi^2s\sqrt{\lambda(s)}} \ .
}
The functions $\S^{\text{OPE}}_i(q^2,s_0,M^2)$, 
computed within the OPE following~\Reff{Descotes-Genon:2019bud}, are all collected in~\App{app:OPE}.
The sum rules given by~\Eq{eq:AllLCSRs} together with the OPE functions, 
represent the first main results of this paper.

\bigskip

It is also instructive to consider the heavy-quark limit of the sum rules obtained here. As already discussed in detail in \Reff{Khodjamirian:2006st}, applying the LCSR method with $B$ meson DAs defined in HQET, we are implicitly leaving some $O(1/m_b)$ corrections unaccounted, which are to be regarded as "systematic" uncertainties of the method. These effects 
still lack their systematic study by expanding the heavy-light current and $B$ state in the correlation function and retaining the terms beyond the leading order in HQET.
The fact that all "standard" heavy-light form factors, such as the ones in $B\to \pi$ and $B\to \rho$ transitions, calculated from $B$-DA sum rules are in a good agreement 
with the results of an alternative light-meson LCSRs, ensures that the inverse heavy-mass corrections to the correlation function are small. Apart from that, the OPE part of the sum rules (see e.g. the expression written in a compact form in \Eq{eq:FOPE}) contains additional $O(1/m_b$) terms originating from different sources. First, we have the standard expansion of the $B$-meson decay constant and mass:
\eq{
f_B=\frac{\hat{f}}{\sqrt{m_b}} +O(1/m_b), \quad m_B=m_b+\bar{\Lambda}.
}
Furthermore, there are $O(s/m_B^2)$ terms of kinematical nature in the coefficients of OPE. Their origin is discussed in \Reff{Khodjamirian:2006st}.
Finally, the $B$-meson DAs depend on the variable $s/m_B$ bounded in the sum rules by the threshold $s_0$ which does not scale with $m_b$.
Hence, the heavy quark limit of LCSRs is determined by the behavior of the DAs near $\omega=s/m_B \sim 0$. Keeping in mind the above considerations, by comparing the coefficients $I^{(2)}_{5,n}$ and $I^{(2)}_{t,n}$
in~\App{app:OPE}, it is easy to notice that 
the heavy mass limits of the LCSRs in~Eqs.(\ref{eq:LCSRFt}) 
and~(\ref{eq:LCSRF5}) determining one and the same form factor 
$F_t^{(\ell=0)}$, are equal. 
To avoid confusion, we remind that the heavy mass expansion plays a secondary role in LCSRs, because the main hierarchy of power corrections in the light-cone OPE is determined by the Borel scale $M$ in the channel of the light-meson interpolating current. This scale, chosen much larger than $\Lambda_\text{QCD}$, is independent of $m_b$.

%%%%%%%%%%%%%%%%%%%%%%%%%%%%%%%%%%%%%%%%%%%%%%%%%%%%%%
\section{Two-channel model of form factors}
\label{sec:ffmodel}

The generalized LCSR approach used here implies that the sum rules for $B\to K\pi$ form factors only determine  weighted integrals of the form factors over the $K\pi$ invariant mass, contrarily to $B\to P$ or $B\to V$ form factors that can be directly extracted from LCSRs (see e.g \Reff{Khodjamirian:2006st}).
Thus, we first need to model the $B\to K\pi$ form factors and only then use the LCSRs obtained here to constrain the parameters of the model. This was the procedure followed in Refs.~\cite{Cheng:2017smj,Descotes-Genon:2019bud} for the $P$-wave form factors.
We could, in principle, follow the same approach and consider a sum of Breit-Wigner~(BW) resonances, as explained in~\App{app:BWparam}. However, this description is certainly insufficient in the present case, as the $S$-wave of the $K\pi$ system is known to exhibit a more complicated structure than the $P$-wave. Indeed, in addition to well-identified scalar resonances such as the $K^*_0(1430)$, more elusive ones have been detected, in particular the $K^*_0(700)$ (also known as $\kappa$)~\cite{Buettiker:2003pp,Descotes-Genon:2006sdr,Pelaez:2016klv,Pelaez:2020uiw,Pelaez:2020gnd,Pelaez:2021dak}. These resonances correspond to poles in the second Riemann sheet located far from the real axis in scattering amplitudes or form factors, and they are therefore difficult to distinguish from  slow variations of the 
nonresonant background in these amplitudes.

The scalar $K\pi$ form factor has been extensively studied in the literature with various parametrisations, by means of e.g. the $K$-matrix~\cite{Jamin:2001zq} or dispersion relations~\cite{Moussallam:2007qc,Bernard:2013jxa,Escribano:2014joa}.
 For our purposes, we need a model with the same appealing features as the Breit-Wigner ansatz used in 
our previous work~\cite{Descotes-Genon:2019bud}. Specifically, it should possess appropriate analytical properties, with poles corresponding to known resonances and cuts for the relevant open channels. And it should also be easily generalised to $B\to K\pi$ form factors, with a simple dependence on the parameters to be constrained by the sum rules. 

A recent description matching the above requirements was provided in \Reff{VonDetten:2021rax}, using a two-channel dispersive model to include both elastic scattering at low energies and inelastic effects and resonances at higher energies. We will briefly discuss the main features of this model for the $K\pi$ scalar form factor, before adopting an extension to the $B\to K\pi$ form factors of interest.

%%%%%%%%%%%%%%%%%%%%%%%%%%%%%%%%%%%%%%
\subsection{The $K\pi$ scalar form factor}

We start by recalling salient features of the two-channel dispersive model of \Reff{VonDetten:2021rax} for
the scalar form factor $F_S$, or equivalently, for the related form factor $f_0$, see~\Eq{eq:KpiscalFF}. Inspired by the Bethe-Salpeter approach, this formalism reproduces the elastic Omn\`es parametrisation at low energies and includes inelastic effects through resonances similarly to the isobar model, and it has been used to study both the scalar $K\pi$ scattering amplitude and the $K\pi$ scalar form factors. 

Due to the small impact of the $K\eta$ channel, only the $K\pi$ and $K\eta'$  channels are considered in \Reff{VonDetten:2021rax}. The scalar form factors $f_0^{K\pi}\equiv f_0$ and $f_0^{K\eta'}$ for both channels are collected in a two-component vector $\mathbf{f_0} = (f_0,f_0^{K\eta'})^T$ modeled as:\,\footnote{
In \Reff{VonDetten:2021rax}, $f_0$ is defined as the matrix element of the state $K^0 \pi^-$, while here we use the $K^-\pi^+$ as defined in~\Eq{eq:KpiscalFF}, which are equivalent up to an overall (-1) normalisation factor.
}
\eq{
\mathbf{f_0}(s)=\Omega(s)[{\bf 1}-V_R(s)\Sigma(s)]^{-1}M(s) \equiv B(s)M(s)\ ,
\label{eq:FSmodel}
}
where $\Omega$ is the Omn\`es function, $\Sigma$ is the dressed loop operator and $V_R$ is the interaction potential. \App{app:dispmodel} provides the definition of these $2\times 2$ matrices, with the index $a=1,2$ indicating the $K\pi$ and $K\eta'$ channels, respectively.

The source term $M(s)$ describes the resonances, making it possible to obtain a description of the form factor above the elastic region:
\eq{
\label{eq:Mrewrite}
M_a (s)=\sum_{k=0}^{k_{\rm max}} c_a^{(k)}s^k
-\sum_r g_a^{(r)} \frac{s-s_{K\eta}}{(s-\tilde{M}^2_{(r)})(s_{K\eta}-\tilde{M}^2_{(r)})} \alpha^{(r)} \ .
}
The coefficients $c^{(k)}$ and the resonance couplings $\alpha^{(r)}$ are process-dependent, as well as the order of the polynomial $k_{\rm max}$. By tuning $k_{\rm max}$, the description at intermediate energies can be improved at the expense of changing the high-energy behaviour. The masses $\tilde{M}_{(r)}$ of the resonances and their couplings $g_a^{(r)}$ to the $K \pi$ and $K\eta'$ channels can then be determined from a fit to the $K\pi$ scattering data~\cite{Pelaez:2016tgi}.  Based on this knowledge of the $K\pi$ scattering,
the description of the $K\pi$ form factor $f_0$ in \Eq{eq:FSmodel} can be obtained by fitting the $\tau^-\to K_S\pi^-\nu_\tau$ spectrum from the Belle experiment~\cite{Belle:2007goc}. In fact, a joint fit of the scalar and vector $K\pi$ form factors is performed as described in detail in \Reff{VonDetten:2021rax}.

\begin{figure}[t]
\centering
\includegraphics[width=0.475\textwidth]{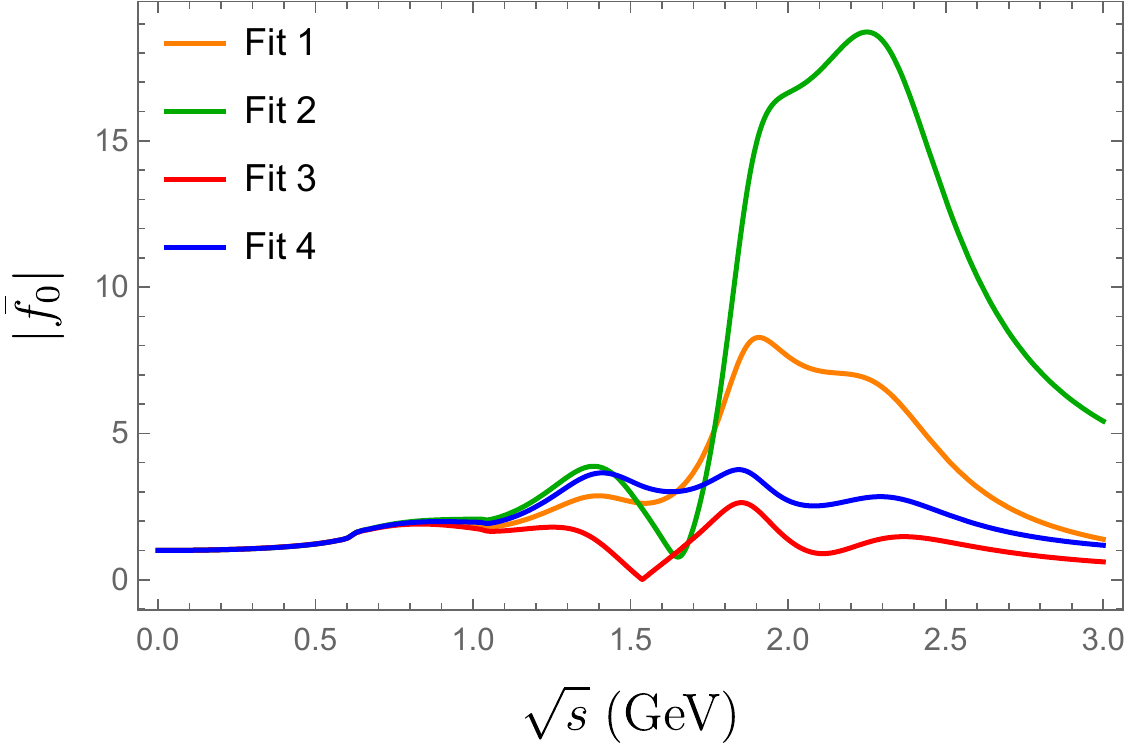}
\hspace{5mm}
\includegraphics[width=0.475\textwidth]{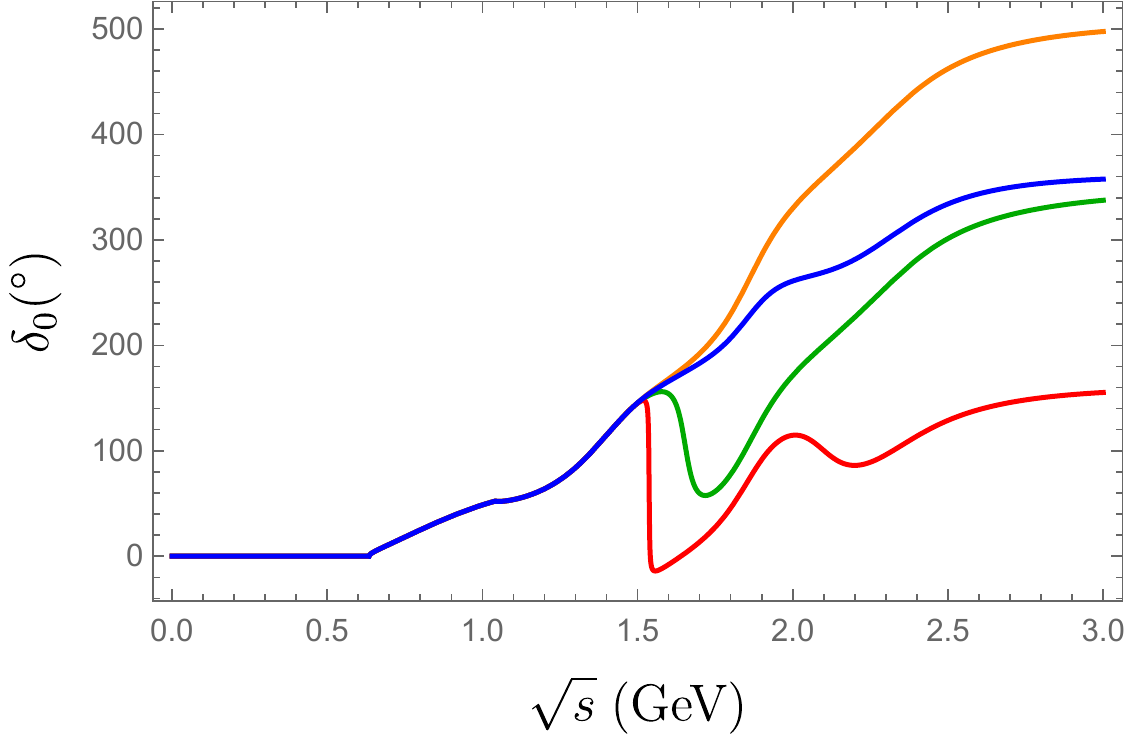}
\caption{\it Modulus of the normalized scalar form factor $|\bar{f}_0|$ and its strong phase $\delta_0$ obtained from the four different fit scenarios of~\Reff{VonDetten:2021rax}. }
\label{fig:ffs}
\end{figure}

As a result, in \Reff{VonDetten:2021rax}, four different descriptions of the scalar $K\pi$ form factor are obtained, all fitting the data equally well. All four models contain the resonance $K^*_0(1430)$ in the interaction potential. Models 1 and 2 also contain the $K^*_0(1950)$ resonance.
In~\Fig{fig:ffs} we plot the normalized form factor
\eq{
\bar{f}_0(s) \equiv \frac{f_0(s)}{f_0(0)} \equiv |\bar{f}_0| e^{i\delta_0} \ ,
\label{eq:f0norm}
}
for the four models, using the outcome of~\Reff{VonDetten:2021rax}. At $q^2=0$ we use the model-independent condition $f_0(0) = f_+(0)$ to have 
a more precise value of the vector $K\pi$ form factor.
The large variations above $\sqrt{s}>2$ GeV are caused by the different  assumptions chosen concerning the polynomial terms $c_a$ in \Eq{eq:Mrewrite}, as well as the presence or the absence of an additional term for the $K^*_0(1950)$ resonance. We notice that three models (1,2,3) yield a similar contribution from the $K^*_0(1430)$ whereas model 4 is much lower. This provides an illustration of the weak constraints on the parameters of this dispersive model in the intermediate energy region around~$1.5 - 2.5 \GeV$.

In the following, we will use all these four models to determine the $B\to K\pi$ form factors, interpreting the variation between the models as a qualitative measure of systematic uncertainty.

%%%%%%%%%%%%%%%%%%%%%%%%%%%%%%%%%%%%%%%%%
\subsection{$B\to K\pi$ form factors for the $K\pi$ $S$-wave}

We can generalise the above parametrization quite easily to the $B\to K\pi$ form factors with the $K\pi$ system in the $S$ wave. Specifically, for each $B\to K\pi$ form factor $F_i^{(\ell=0)}$ a two-component vector $\mathbf{F}_i$ is defined including the $B\to K\pi$ and $B\to K\eta'$ form factors as components with $a=1$ and $a=2$.
Following the previous discussion, we write
\eq{
\mathbf{F}_i(s,q^2)=\Omega(s)[{\bf 1}-V_R(s)\Sigma(s)]^{-1}N_i(s,q^2) \equiv B(s) N_i(s,q^2)\ ,
}
with the source term for a given form factor
\eq{
N_{i,a}(s,q^2)=\sum_{k=0}^{k_{\rm max}} d_{i,a}^{(k)}(q^2)s^k-\sum_r g_a^{(r)} \frac{s-s_{K\eta}}{(s-\tilde{M}^2_{(r)})(s_{K\eta}-\tilde{M}^2_{(r)})} \beta_i^{(r)}(q^2)\ .
}
Compared to our parametrisation for $P$-wave form factors with BW resonances in \Reff{Descotes-Genon:2019bud}, and to the equivalent description given for the $S$ wave in~\App{app:BWparam}, we can see that there is an additional channel to be considered $(K\eta')$ which doubles the number of parameters.
Moreover, there is an additional polynomial term for each of the two channels, with an order which is not determined a priori. 
We constrain these parameters by assuming that $\mathbf{F}_i$ and $\mathbf{f}_0$ have the same phase for each channel, leading to the constraints ${\rm Im}[(BM)_a^* (BN_i)_a]=0$ for $a=1,2$. One solution is provided by $N_{i,a}(s,q^2) = \hat{\rho}_i(s,q^2) M_a(s)$, leading to
\eq{
F_i^{(\ell=0)}(s, q^2)=\hat{\rho}_i(s,q^2) f_0(s)\ .
}
We then further assume that the only $s$-dependence in $\hat{\rho}_i(s,q^2)$ arises from kinematic effects.
The latter can be identified, noticing that the alternative model with Breit-Wigner line shapes discussed in~\App{app:BWparam} must feature similar kinematic structures. In particular, from~\Eq{eq:FFmodels1}, we expect the form factors $F_0^{(\ell=0)}$ and $F_T^{(\ell=0)}$ to have a kinematic factor $\sqrt{\lambda(s)}$ (coming from their definition in terms of $k_0^\mu$) which will not be  present for $F_t$ and $F_5$. In addition, we may factor out the kinematic $q^2$ dependence to simplify the analysis of the sum rules. To this extent, we define
\eq{
\hat{\omega}_i(s) \equiv \kappa_i(s,q^2)\,{\omega}_i(s,q^2)
\ ,
}
where
\eq{
\kappa_0(s,q^2) = \frac{\sqrt{\lambda(s)}}{\sqrt{q^2}} \ ,\quad \kappa_{5,t}(s,q^2) = \frac{2}{\sqrt{q^2}} \ , \quad \kappa_T(s,q^2) = \sqrt{\lambda(s) q^2}\ ,
}
such that the factors $\kappa_i(s,q^2)$ cancel out the entire kinematic $s$ and $q^2$ dependence in $\omega_i(s,q^2)$ defined in \Eq{eq:omega}, leading to
\begin{equation}
\hat{\omega}\equiv \hat\omega_0=\hat\omega_T=\hat\omega_5=\hat\omega_t
=\frac{3}{16\pi^2}\frac{\sqrt{\lambda_{K\pi}(s)}}{s}\ .
\end{equation}
Taking into account these elements, we obtain 
\eq{
\label{eq:LCSRszpar21}
F_i^{(\ell=0)}(s,q^2)=
\kappa_i(s,q^2)\, \rho_i(q^2)\,f_0(s)\ , 
}
where $\rho_i(q^2)$ is a real-valued function (independent of the channel $a$) that, by assumption, only depends on $q^2$. 
As a result, the sum rules given in \Eq{eq:AllLCSRs} become constraints on the functions $\rho_i$,
\eqa{
\label{eq:rhodef}
\S_i^{\rm OPE}(q^2,s_0,M^2) 
&=& \rho_i(q^2)  \int_{s_{\rm th}}^{s_0}ds\ e^{-s/M^2}(m_K^2-m_\pi^2)\hat{\omega}(s) |f_0(s)|^2
\nonumber \\
&\equiv& \rho_i(q^2) \;I_{\rm SR}(s_0,M^2)\ ,
}
where the integral $I_{\rm SR}$ only depends on $s_0$, $M^2$ and the form factor model for $f_0$.
This leads to
our final expression for the $S$-wave $B\to K\pi$ form factors,
\begin{equation}
\label{eq:FSRfinal}
F_i^{(\ell=0)}(s,q^2)=
\frac{\kappa_i(s,q^2) f_0(s)\,\S_i^{\rm OPE}(q^2,s_0,M^2)}{I_{\rm SR}(s_0, M^2)}
\ ,
\qquad i=\{0,t,5,T\}
\ .
\end{equation}
At this stage one could perform a $z$-expansion on both sides of the sum rules in \Eq{eq:AllLCSRs}, as done in \Reff{Descotes-Genon:2019bud}.
Since the model for the form factors does not obey a simple parametrization in this variable, we refrain from doing so and work with~\Eq{eq:FSRfinal} directly. 

A comment is in order concerning the comparison with the $P$-wave case of \Reff{Descotes-Genon:2019bud}, where both the $K\pi$ and $B\to K\pi$ form factors were modelled as a superposition of Breit-Wigner resonances with relative phases depending on $s$. The reality of the product of the $B\to K\pi$ form factors with the vector $K\pi$ form factor could be easily implemented there for each of the resonance contributions, by choosing the corresponding relative phase equal to that of the vector form factor. Here, we consider a rather different model for the $B\to K\pi$ scalar form factors, as the $s$-dependent phase is encoded in the overall  matrix $B(s)$, together with the phase in the $K\eta'$ channel and involving all resonances at once. Satisfying the reality constraint is therefore harder than in the $P$-wave case, which explains that a lesser number of parameters are fixed by the sum rules: only two per $B\to K\pi$ form factor in the scalar case, rather than two per $B\to K\pi$ form factor and per resonance in the vector case.

\begin{table}
\centering
\setlength{\tabcolsep}{10pt}
\begin{tabular}{@{}cc c @{\hspace{1cm}} cc c@{}}
\toprule[0.7mm]
Parameter &Value & Ref & Parameter & Value & Ref \\
\midrule[0.7mm]
$m_{\pi^\pm}$           &  $140\MeV$  		        & \cite{ParticleDataGroup:2022pth}			& $m_{K^\pm}$ 			&  $494 \MeV$ 			& \cite{ParticleDataGroup:2022pth}\\
$m_{B^0}$ 				&  $5.28 \GeV$  			& \cite{ParticleDataGroup:2022pth}
& $f_B$              	&  $207^{+17}_{-9} \MeV$  	&  \cite{Gelhausen:2013wia} \\
%\midrule
$\overline m_b (\overline m_b)$ 		&   $4.18^{+0.03}_{-0.02}\GeV$ 
& \cite{ParticleDataGroup:2022pth}&
$\overline m_c (\overline m_c)$ 		&  $1.27\pm 0.02\GeV$  & \cite{ParticleDataGroup:2022pth}\\
$m_s(2\,\mbox{GeV})$ & $93.4^{+8.6}_{-3.4}~\mbox{MeV}$  
& \cite{ParticleDataGroup:2022pth} &
$m_d(2\,\mbox{GeV})$ & $4.67^{+0.48}_{-0.17}~\mbox{MeV}$ &
\cite{ParticleDataGroup:2022pth} \\
\midrule
$\lambda_B$  	     	     &    $460\pm 110 \MeV$	&   	 \cite{Braun:2003wx}		&  $R$ 	&     	$0.4^{+0.5}_{-0.3}$	&  \cite{Braun:2017liq} \\
\bottomrule[0.7mm]
\end{tabular}
\caption{\it Compendium of input values used in the numerical analysis.
}
\label{tab:inputs}
\end{table}

%%%%%%%%%%%%%%%%%%%%%%%%%%%%%%%%%%%%%%%%%%%%%%%%
%%%%%%%%%%%%%%%%%%%%%%%%%%%%%%%%%%%%%%%%%%%%%%%%
\section{Numerical analysis}
\label{sec:analysis}

%%%%%%%%%%%%%%%%%%%%%%%%%%%%%%%%%%%%%%%%%%%%%%%%
\subsection{Numerical input}

We follow the strategy outlined in \Reff{Descotes-Genon:2019bud} (see also
\Reff{Cheng:2017smj} for further illustration). The inputs used in the numerical analysis and their sources are collected in~\Tab{tab:inputs}. 
Despite the fact that the OPE for LCSRs is computed in HQET, the $b$-quark 
mass parameter still explicitly enters the LCSR for the heavy-light pseudoscalar current. 
We also need the $c$-quark mass for an estimate of nonlocal contributions 
in the analysis of $B\to K\pi\ell \ell$ decays. In both cases, we adopt  
typical $\overline{\rm MS}$ heavy quark masses, respectively: 
$\overline m_b =4.18\GeV$ and $\overline m_c =1.27\GeV$. The scale of the OPE 
is around $\mu=1\GeV $, hence we renormalize the 
$s$-quark mass value given in \Tab{tab:inputs} to $\overline m_s(1\GeV)=123(14) \MeV$.

Furthermore, we use the QCD sum rule estimate for the inverse moment $\lambda_B \equiv \lambda_B(1.0\, {\rm GeV})$ of the $B$-meson DA from~\Reff{Braun:2003wx}, consistent with a more recent estimate of~\Reff{Khodjamirian:2020hob} obtained with the same method.
This moment is the most important parameter determining the two- and three-particle $B$-meson DAs up to twist-four. 
For these DAs, we use the so-called ``Model I''~from~\Reff{Braun:2017liq} specified in~Appendix~B of \Reff{Descotes-Genon:2019bud}, and based on the exponential model proposed originally in \Reff{Grozin:1996pq}
(see also \Reff{Khodjamirian:2006st}).
The only additional parameter needed in this model
and within the adopted approximation of the light-cone OPE is the ratio $R=\lambda_E^2/\lambda_H^2$ of the two normalization parameters $\lambda_E$ and $\lambda_H$ determining the vacuum-to-$B$ matrix elements of the quark-antiquark-gluon HQET currents. The choice of $R$ 
is discussed in detail in \Reff{Descotes-Genon:2019bud}. 
For consistency, we also use the QCD sum rule result for 
the $B$-meson decay constant quoted in~\Tab{tab:inputs}, which is close to (but less accurate than) the most recent lattice QCD average $f_B =190.0(1.3)$ MeV in \Reff{FLAG:2021npn}.

For the $K\pi$ form factor, we use the four models obtained from the fits in \Reff{VonDetten:2021rax}. The corresponding numerical
values of the normalized form factor $\bar{f_0}(s)$ are presented in the ancillary files attached to this paper. For the normalization $f_0(0)$ in \Eq{eq:f0norm}, we use $f_0(0)=f_+(0)=1.0$, which agrees with the analysis in \Reff{Descotes-Genon:2019bud} using the Belle data 
\cite{Belle:2007goc} on the $\tau\to K\pi\nu_\tau$ decay. For comparison, the current lattice QCD average at $N_f=2+1+1$ given  in \Reff{FLAG:2021npn}
is $f_+(0) = 0.9698(17)$. We note that a different normalization would simply correspond to a rescaling of our form factors. 

Following \Reff{Descotes-Genon:2019bud}, we use the two-point QCD sum rule to fix the effective threshold parameter $s_0$ and the Borel mass squared $M^2$ appearing in the LCSRs. This sum rule and the procedure to fix $s_0$ and $M^2$ are described in~\App{app:2ptSR}.
As a result, we use the following values
\eq{
s_0 = 1.8 \; {\rm GeV}^2\ , \quad
M^2 = 1.25 \;{\rm GeV}^2\ ,
\label{eq:s0M}
}
independently of the $K\pi$ form factor model.
These values satisfy the two-point sum rule for all such models, and at the same time, render the LCSRs well behaved. On the one hand, they lead to a reasonable convergence of the 
light-cone OPE, measured by the relative size of the contribution from three-particle DAs to the functions $S_i^\text{OPE}$\,:
\eq{
\frac{\S_{i, 3p}^\text{OPE}}{\S_{i, 2p}^\text{OPE}} < 30 \% \ ,
}
in the range $q^2= [0,6.0]\GeV^2$.
On the other hand, the integral of the spectral density above the effective threshold is less than 40\% of the total integral, sufficiently suppressing the sensitivity of the LCSRs to the quark-hadron duality approximation.

\begin{figure}
    \centering
\includegraphics[width=0.7\textwidth]{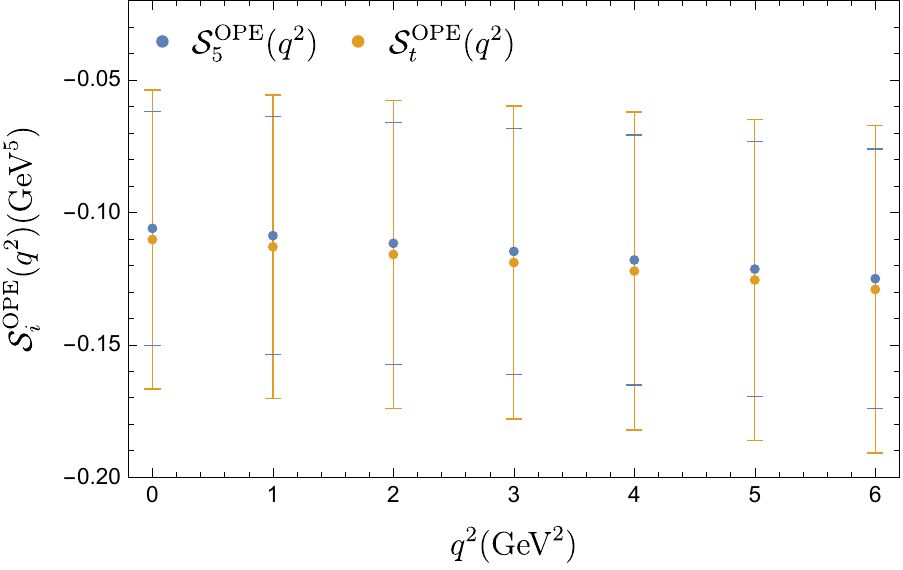}
    \caption{\it Comparison between the OPE contributions $\S_5^{\text{OPE}}$ and $\S_t^{\text{OPE}}$ as a function of $q^2$, for the values of $s_0$ and $M^2$ given in \Eq{eq:s0M}.}
    \label{fig:S5versusSt}
\end{figure}

%%%%%%%%%%%%%%%%%%%%%%%%%%%%%%%%%%%%%%%%%%%%%%%%
\subsection{Results for $S$-wave $B\to K\pi$ form factors}

Using the above inputs, we find the following central values for the integral $I_{\rm SR}$ defined in \Eq{eq:rhodef},
\eq{
I_{\rm SR}(s_0=1.8 \;{\rm GeV}^2,M^2=1.25\; {\rm GeV}^2) =
\big \{6.9,\,9.7,\,5.3,\,8.4 \big\} \cdot 10^{-3} \;{\rm GeV}^4 \ ,
}
respectively from the fits $\{1,2,3,4\}$ for $f_0$, discussed in~\Sec{sec:ffmodel}.
Using these values, we can calculate the functions $\rho_i$, and hence 
determine the form factors $F_i^{(\ell=0)}(s,q^2)$ from~\Eq{eq:FSRfinal}. 
We first comment on the numerical difference between $\S_5^{\text{OPE}}$ and $\S_t^{\text{OPE}}$ shown in~\Fig{fig:S5versusSt} for different values of $q^2$. Within  uncertainties, the two OPE functions agree.
Therefore, in what follows the numerical results for $i=5$ will not be used as they would lead to very similar results.

\begin{figure}[t]
\centering
\includegraphics[height=0.31\textwidth]{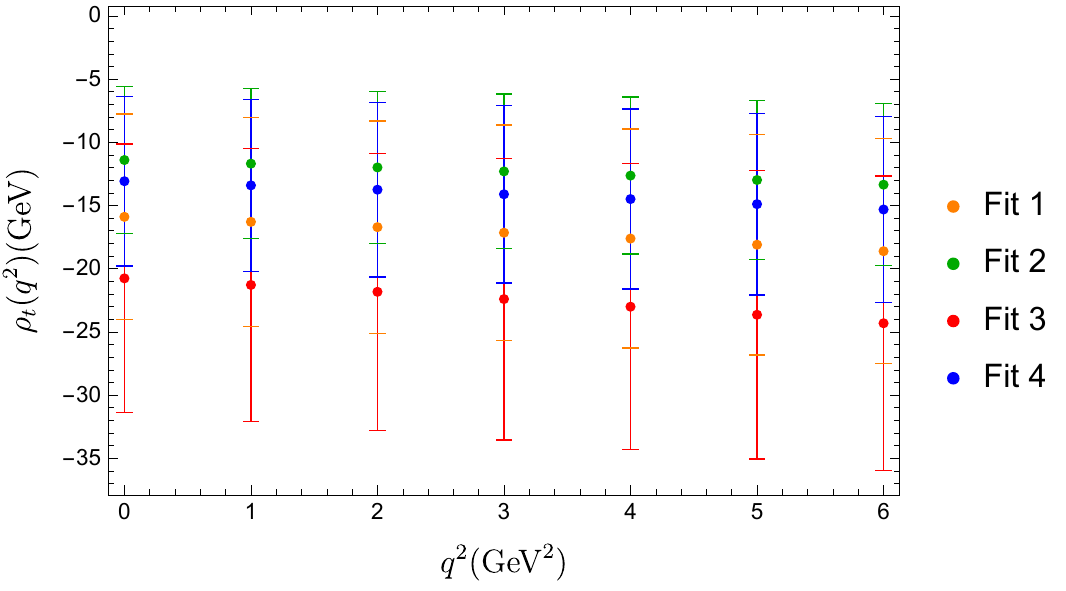}
\\[10mm]
\includegraphics[height=0.31\textwidth]{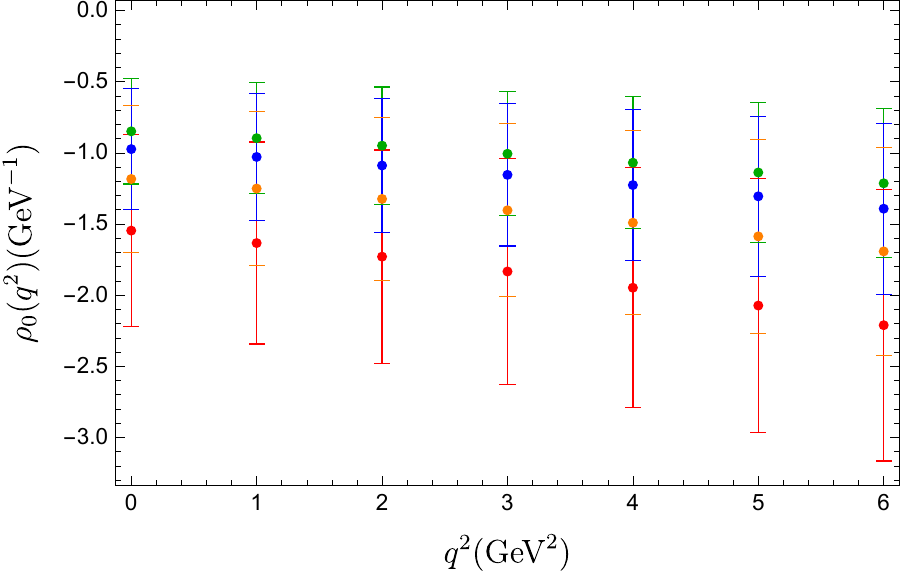}
\hspace{3mm}
\includegraphics[height=0.31\textwidth]{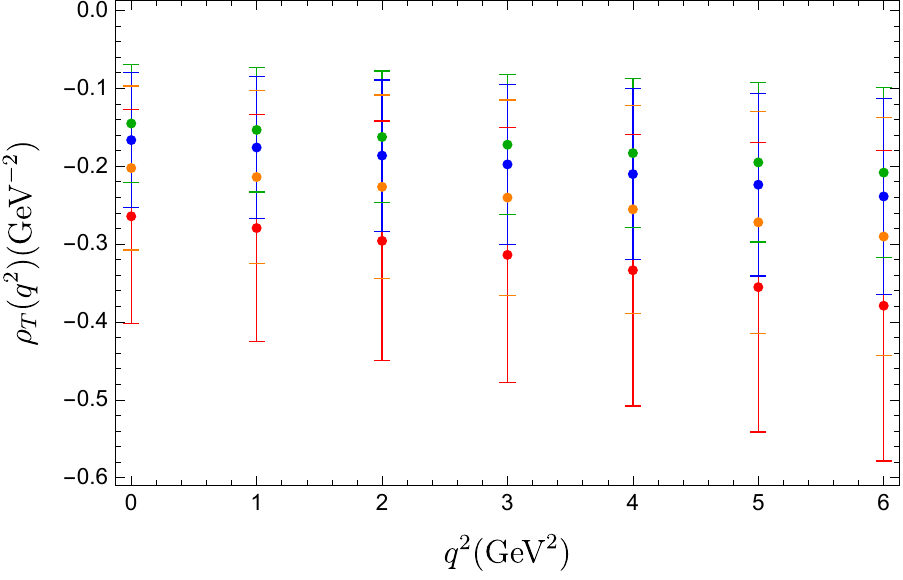}
\caption{\it Results for the functions $\rho_i(q^2)$ defined in \Eq{eq:rhodef} for all four fit models.}
\label{fig:rhodep}
\end{figure}

We also note that the functions $\S_i^\text{OPE}$ have a mild dependence on $q^2$ in the region of 
LCSR validity, yielding rather constant functions $\rho_i(q^2)$.
These functions are shown in~\Fig{fig:rhodep}.
\begin{figure}
\subfloat{\includegraphics[height=0.29\textwidth]{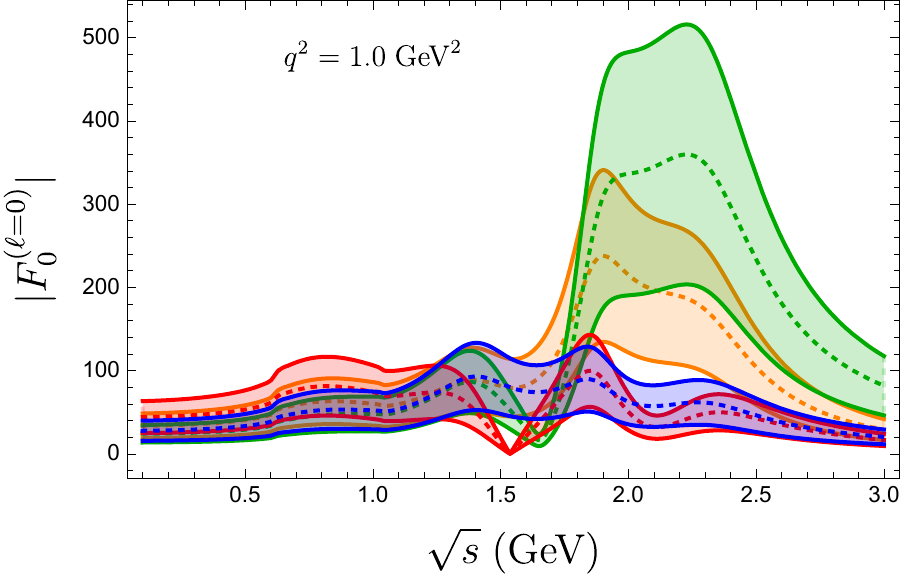}}
\hspace{5mm}
\subfloat{\includegraphics[height=0.29\textwidth]{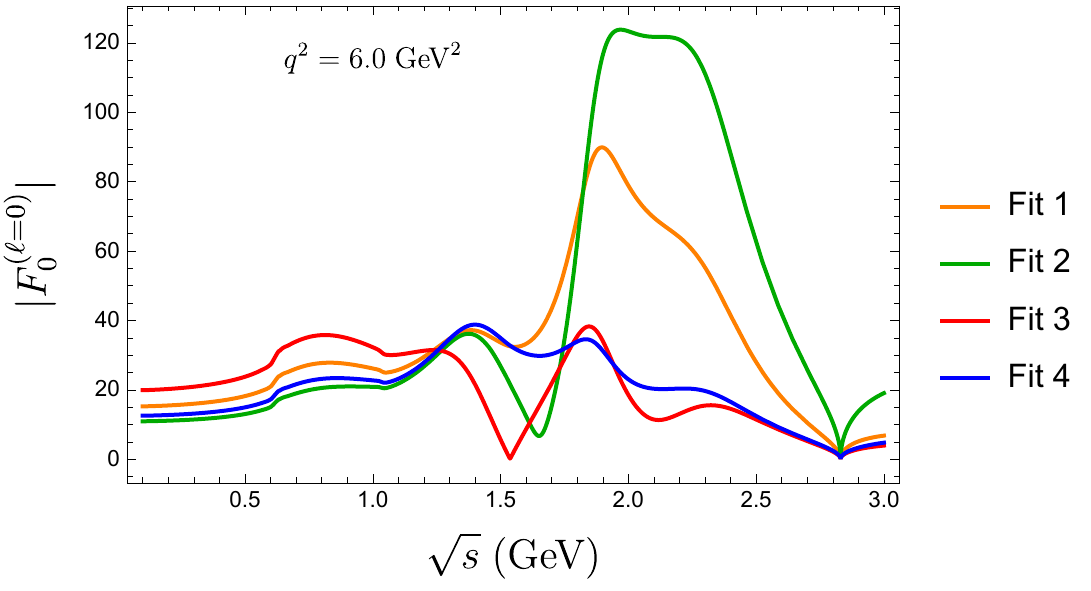}}\\ 

\vspace{5mm}
\subfloat{\includegraphics[height=0.29\textwidth]{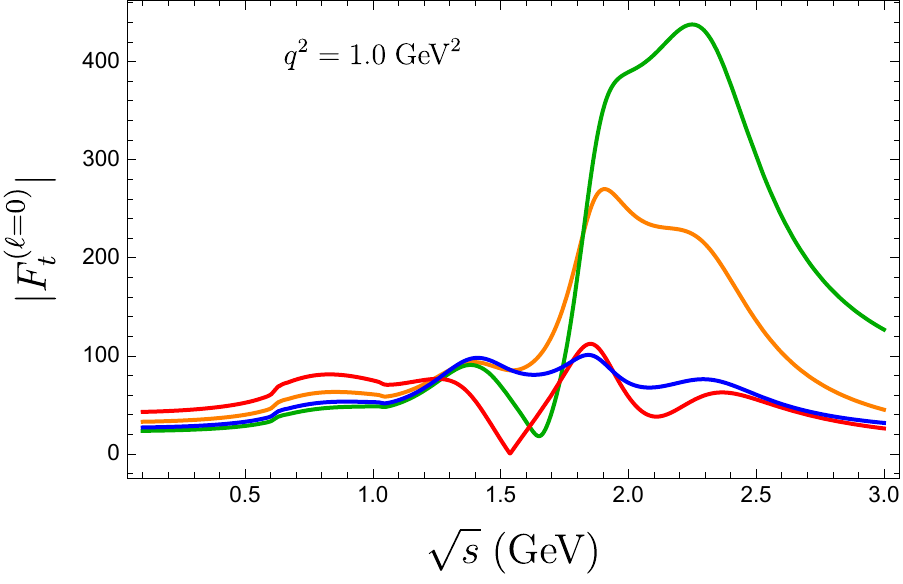}}
\hspace{5mm}
\subfloat{\includegraphics[height=0.29\textwidth]{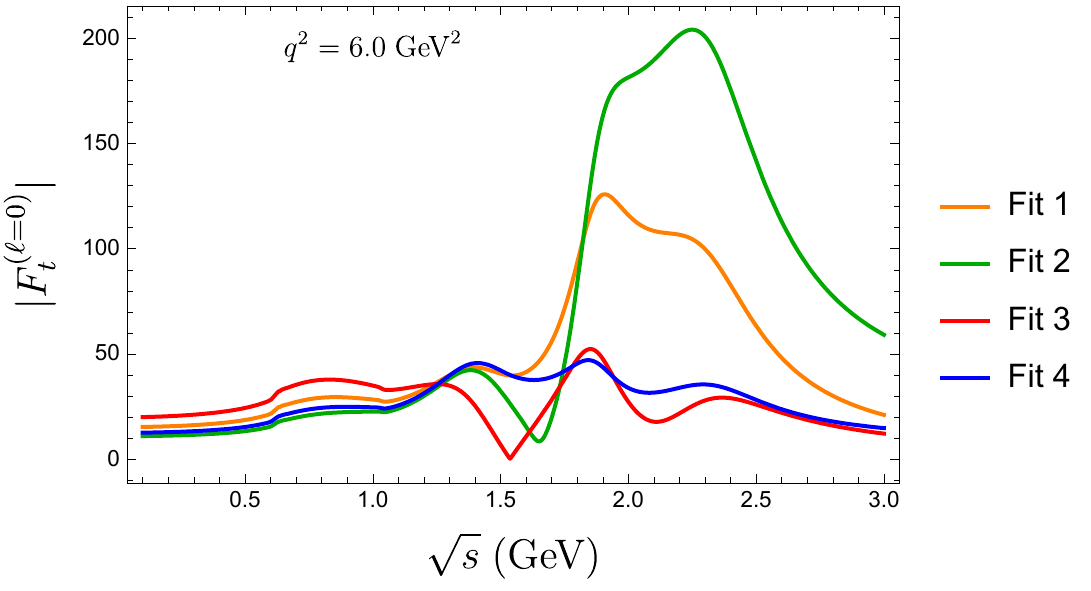}}\\

\vspace{5mm}
\subfloat{\includegraphics[height=0.29\textwidth]{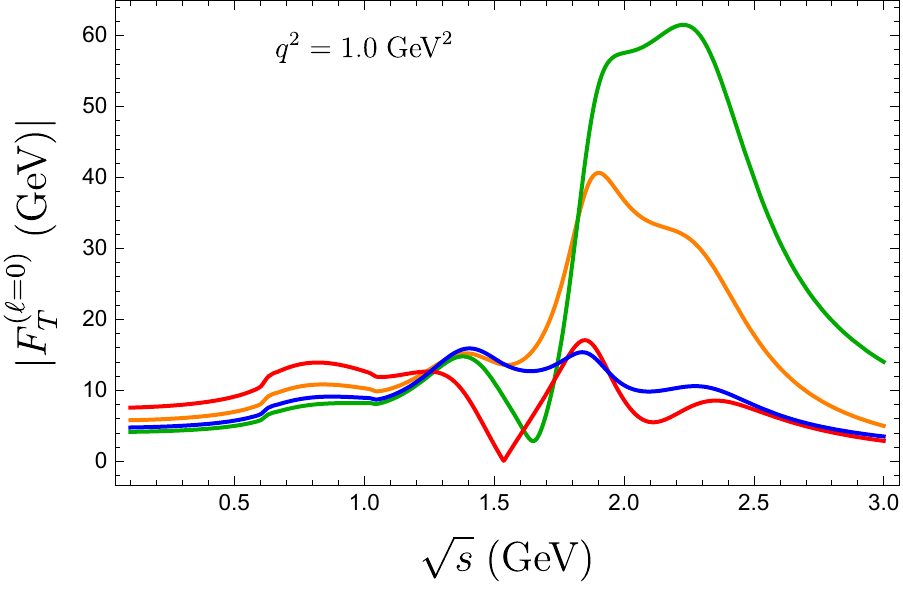}}
\hspace{5mm}
\subfloat{\includegraphics[height=0.29\textwidth]{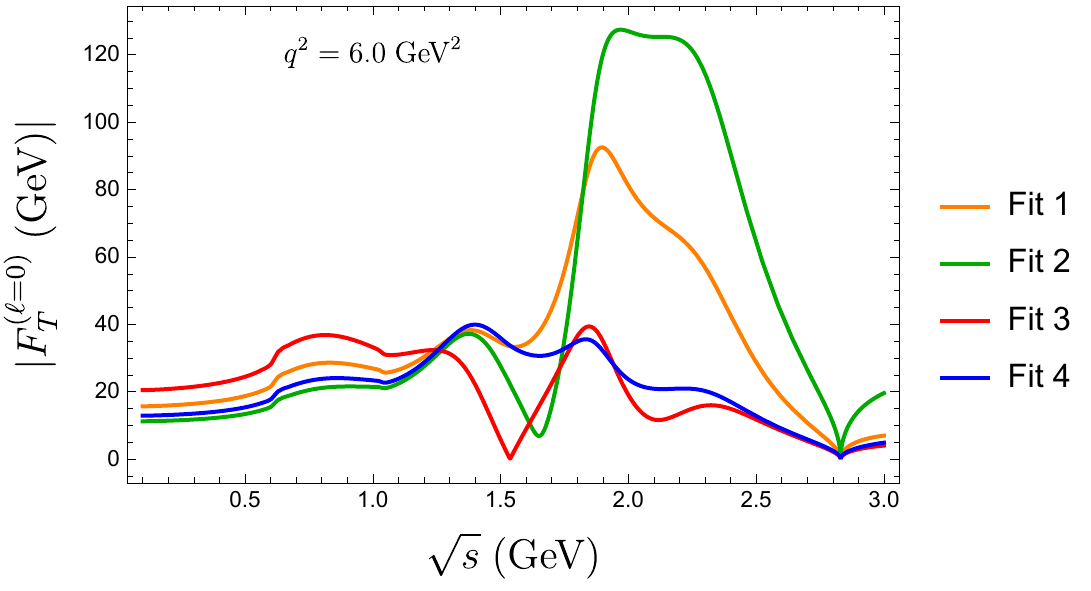}}
\caption{\it The form factors $F^{(\ell=0)}_0, F_t^{(\ell=0)}, 
F_T^{(\ell=0)}
\equiv F_0^{T{(\ell=0)}}$ for the different fit models at $q^2=1.0\ \GeV^2$ (left) and $q^2=6.0\ \GeV^2$ (right). }
\label{fig:formfactors}
\end{figure}

In~\Fig{fig:formfactors}, we present our results for the S-wave form factor 
$F_{0,t}^{(\ell=0)}$ and $F_T^{(\ell=0)}\equiv F_0^{T(\ell=0)}$ at $q^2=1.0$ and $6.0$ GeV$^2$ as a function of the $K\pi$ invariant mass $\sqrt{s}$. We display the uncertainties coming from the OPE calculation in~\Fig{fig:formfactors}(a). For other values of $q^2$, as well as for the form factors 
$F_t^{(\ell=0)}$ and $F_T^{(\ell=0)}$, the
corresponding uncertainties are in the 
same ballpark, hence we omit them in the other panels of \Fig{fig:formfactors}.

By definition, the $q^2$ dependence of the form factors in~\Eq{eq:FSRfinal} is, apart from the factors $\kappa_i$, determined by $\mathcal{S}^{\rm OPE}_i$, which are parameterized by $\rho_i$. As the latter are rather constant functions of $q^2$, the resulting $q^2$ dependence of the form factors is almost entirely given by the kinematical factors $\kappa_i$. This is similar to the $P$-wave case discussed in terms of Breit-Wigner model in \Reff{Descotes-Genon:2019bud}. For completeness, we show the explicit $q^2$ dependence of the $B\to K\pi$ 
$S$-wave form factors in the next subsection, but we can already notice that these $B\to K\pi$ form factors strongly resemble the $K\pi$-form factor, revealing large deviations between different S-wave models for values of $s\gtrsim (1.8\GeV)^2$. 

\begin{figure}[t]\centering
\subfloat[]{\includegraphics[height=0.29\textwidth]{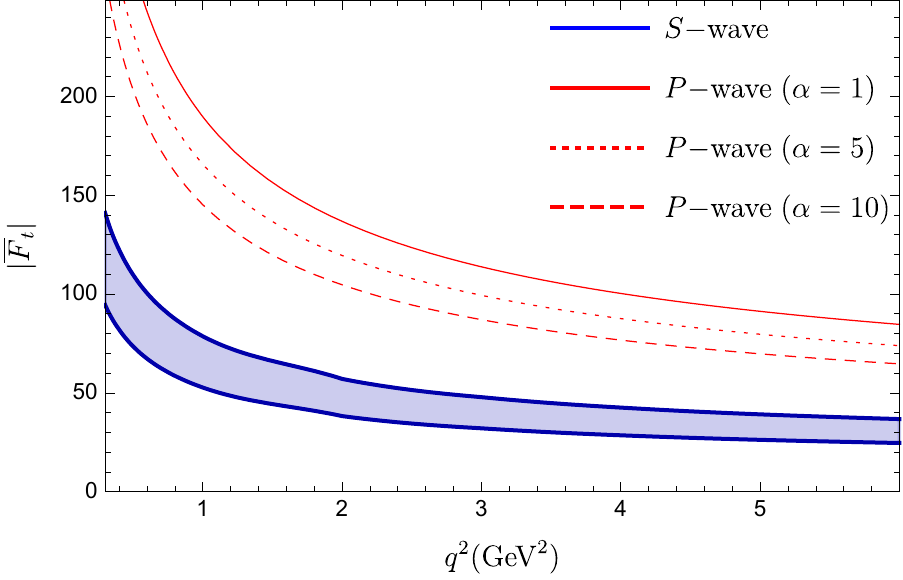}}\\
\subfloat[]{\includegraphics[height=0.29\textwidth]{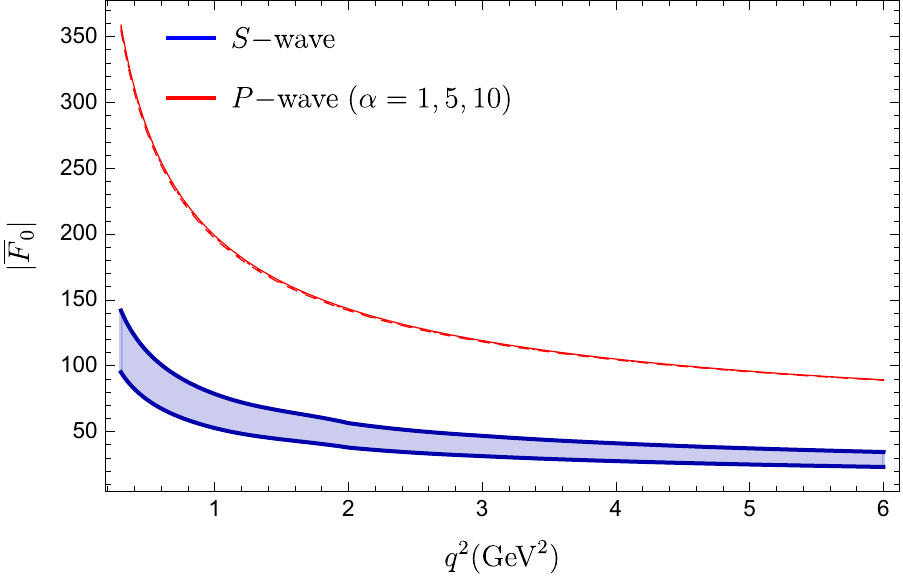}} \hspace{5mm}
\subfloat[]{\includegraphics[height=0.29\textwidth]{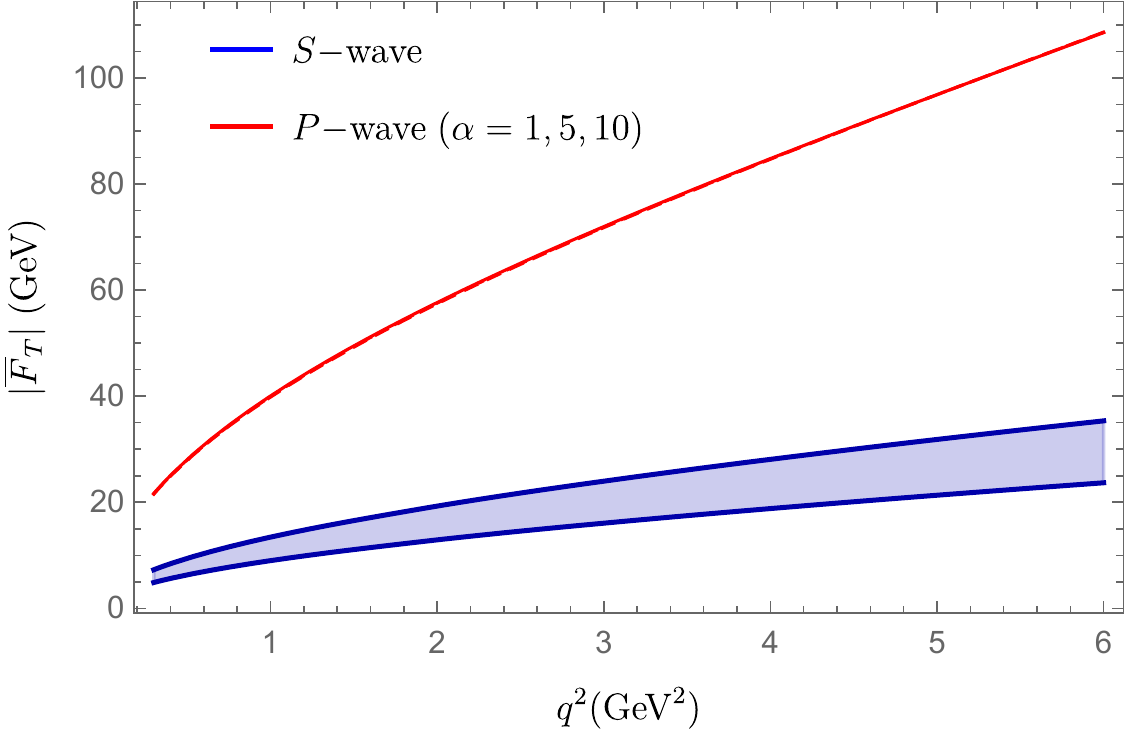}}
\caption{\it The $P$-wave form factors $F_{0,t,T}^{(\ell=1)}$ from \Reff{Descotes-Genon:2019bud} for $\alpha = 1,~5$ and 10 compared to the $S$-wave form factors $F_{0,t,T}^{(\ell=0)}$ (we define $F_T^{(\ell=0,1)} \equiv F_0^{T(\ell=0,1)}$).
The shaded bands indicate the full range of the $S$-wave models. The form factors are integrated over a $100$ {\rm MeV} region around the $K^*(892)$ resonance: $(0.796 \;{\rm GeV})^2 < s < (0.996\; {\rm GeV})^2$.  }
\label{fig:q2depPwave2}
\end{figure}

%%%%%%%%%%%%%%%%%%%%%%%%%%%%%%%%%%%%%%%%%%%%%%%%%%%%%%%%%%
\subsection{Interplay of $S$- and $P$-wave form factors}

We will combine our results with the $P$-wave $B\to K\pi$ form factors studied in \Reff{Descotes-Genon:2019bud} where the LCSRs were used to constrain the contributions of both $K^*(892)$ and $K^*(1410)$ $P$-wave resonances. In~\Reff{Descotes-Genon:2019bud} a floating parameter $\alpha$ was introduced to vary the relative size of the $K^*(1410)$ contribution to the form factors, defined by
\eq{
\F_{K^*(1410)}(q^2) = \alpha\, \F_{K^*(892)}(q^2) \, .
}
Upper bounds on $\alpha$ were derived from LHCb measurements in the $K^*(1430)$ region in ~\Reff{Descotes-Genon:2019bud}: the consideration of $P$-wave moments led to the bound $\alpha\lesssim 10$ whereas the branching ratio (neglecting $S$-wave contributions) led to $\alpha \lesssim 3$. We will show this in more detail later.

\begin{figure}[t]\centering
\subfloat{\includegraphics[height=0.29\textwidth]{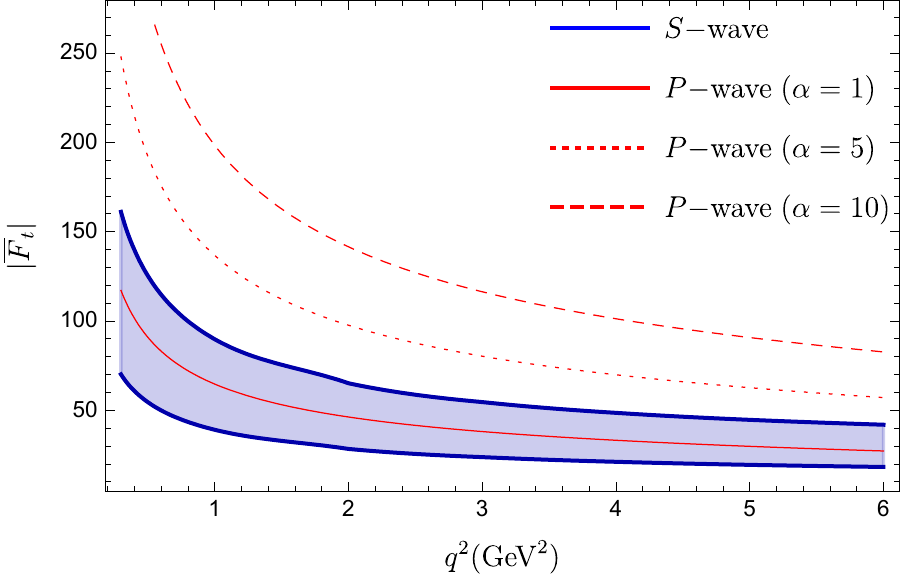}}\\
\subfloat{\includegraphics[height=0.29\textwidth]{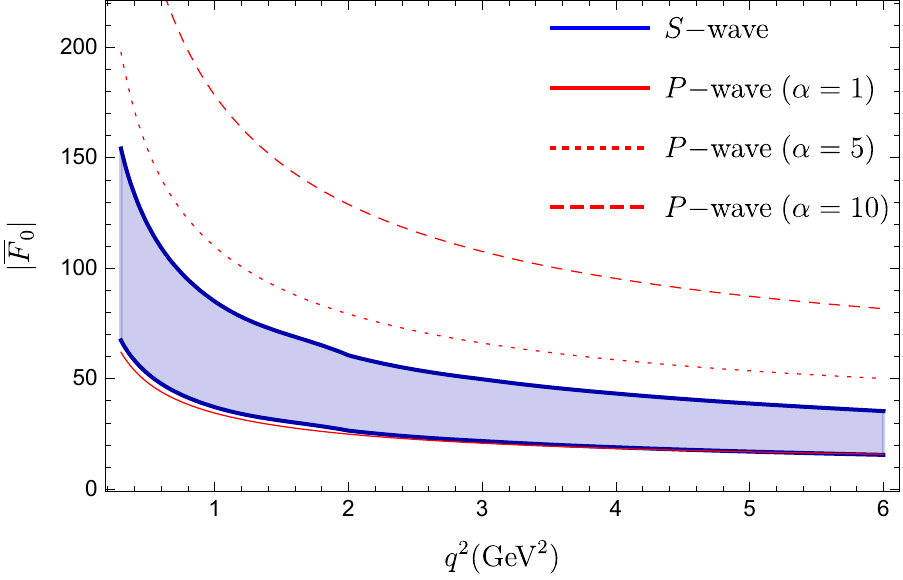}} \hspace{5mm}
\subfloat{\includegraphics[height=0.29\textwidth]{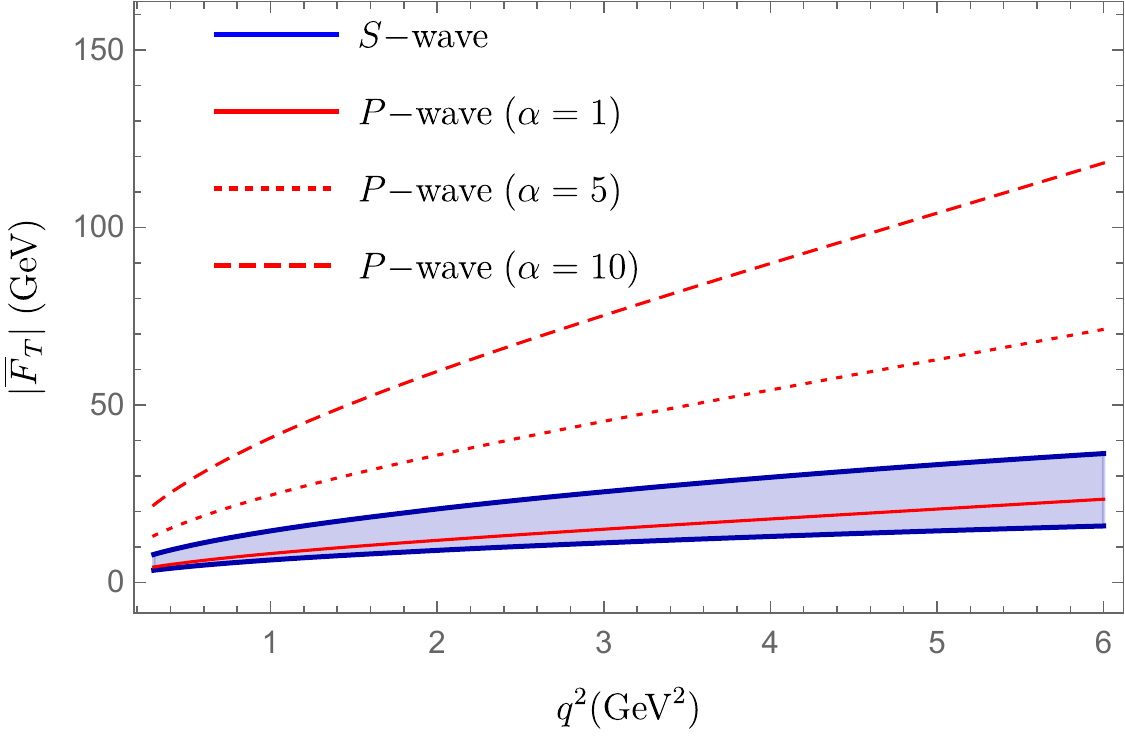}}
\caption{\it $P$- and $S$-wave form factors as in~\Fig{fig:q2depPwave2}, but in the higher $s$ region containing the resonances
$K^*(1410)$ and $K^{*}_0(1430)$: $(1.33\;{\rm GeV})^2< s <(1.53\;{\rm GeV})^2$.}
\label{fig:q2depPwave3}
\end{figure}

We consider the models for the $P$-wave form factors $F_{0,t,T}^{(\ell=1)}$ with $\alpha=1,5,10$ (we will focus on the case $\alpha=1$ later). In~\Fig{fig:q2depPwave2} and \Fig{fig:q2depPwave3}, they are compared to the corresponding $S$-wave form factors, for which the full range of variations between the different fit models is interpreted as a systematic uncertainty. In practice, the model 3 (model 4) always yields the lowest (highest) value for the $S$-wave form factor and we show the corresponding range of variation. We define normalized binned form factors through 
\begin{equation}
 \overline{F}_{0,t,T}^{(\ell)}(q^2)= \frac{1}{s_{2}-s_{1}}\int\limits_{s_{1}}^{s_{2}}\!ds\, F_{0,t,T}^{(\ell)}(s,q^2)\, ~~~(\ell=0,1)\,.
\end{equation}

In~\Fig{fig:q2depPwave2} we show the normalized form factors by integrating the form factors in a $100\MeV$ bin\,\footnote{
This bin is inspired by the LHCb analysis in \Reff{LHCb:2020lmf} where a similar region was chosen.
} around the $K^*(892)$ resonance.
We can see that the  form factors $F_{0,t,T}^{(\ell=0)}$  and 
$F_{0,t,T}^{(\ell=1)}$ have a similar $q^2$ dependence, while the magnitude 
 of each form factor depends  on the specific $S$-wave model or on the value of $\alpha$ in the $P$-wave case. For $F_0$ and $F_T$, the variation of $\alpha$ has only a tiny effect, indistinguishable in the plots. As expected, the magnitudes of the $S$-wave form factors, though noticeable in this region, are smaller than their $P$-wave counterparts. We add that the $F_t$ form factor does not contribute to $B\to K^* \ell \ell$ in the limit of massless leptons, but this form factor plays an important role in non-leptonic decays \cite{Krankl:2015fha,Virto:2016fbw,Klein:2017xti,Huber:2020pqb}.

In \Fig{fig:q2depPwave3}, the same comparison is shown for the region $1.33<\sqrt{s}<1.53$ GeV, which is dominated by the $P$-wave resonance $K^*(1410)$ 
and the $S$-wave resonance $K^*(1430)$. As expected, varying $\alpha$ has a much more significant impact on the $P$-wave model in this region. The interplay between the $P$ and $S$ waves in the $B\to K\pi$ form factors is also more substantial, so that both partial waves contribute at the same level, (and they are very close numerically for $\alpha\simeq 1$).

%%%%%%%%%%%%%%%%%%%%%%%%%%%%%%%%%%%%%%%%%%%%%%%%%%%%%%%%%%
%%%%%%%%%%%%%%%%%%%%%%%%%%%%%%%%%%%%%%%%%%%%%%%%%%%%%%%%%%
\section{Application to the $B\to K\pi\ell\ell$ decay}
\label{sec:BKpiellell}

In \Reff{Descotes-Genon:2019bud}, we applied  LCSRs 
to the $B\to K\pi\ell\ell$ decay with the $K\pi$ system in 
the $P$ wave.
 We are now in a position to extend this analysis by adding the $S$-wave contribution. The discussion is aimed at clarifying two different issues: the pollution from the $S$-wave component under the $K^*(892)$ peak, and the exploitation of the LHCb measurements in the $K^*(1410)$ region. Before discussing a few applications, we will recall elements already presented in \Reff{Descotes-Genon:2019bud}, adapting them to include the $S$ wave.

%%%%%%%%%%%%%%%%%%%%%%%%%%%%%%%%%%%%%%%%%%%%%%%%%%%%%%%%%%
\subsection{Formalism}

The amplitude $\A \equiv \A(\bar B^0 \to K^-(k_1) \pi^+(k_2) \ell^-(q_1) \ell^+(q_2) )$ is given by:
\eq{
i\A = g_F \frac{\alpha}{4\pi} \bigg[ (C_9 \,L_{V\mu} + C_{10} \,L_{A\mu})\  \F_L^\mu 
+  \frac{L_{V\mu}}{q^2} \Big\{  2 m_b C_7\, \F^{T\mu}_R  -i\, 16\pi^2 \,\H^\mu \Big\}   \bigg]
\label{AmplitudeBKpill}
}
with $g_F \equiv 4G_F/\sqrt{2} \,V^*_{ts} V_{tb}$, $L_{V(A)}^\mu \equiv \bar u_\ell(q_1) \gamma^\mu(\gamma_5) v_\ell(q_2)$,
and the local and non-local hadronic matrix elements:
\eqa{
\F_L^\mu &\equiv& i\, \langle K^-(k_1) \pi^+(k_2)|\bar{s}\gamma^\mu P_L\, b|\bar{B}^0(p)\rangle
= \frac12 \big( F_\perp \, k_\perp^\mu + F_\| \, k_\|^\mu + F_0 \, k_0^\mu +F_t \, k_t^\mu \big) \ , \\
\F^{T\mu}_R &\equiv& \langle K^-(k_1) \pi^+(k_2)|\bar{s}\sigma^{\mu\nu} q_\nu P_R\, b|\bar{B}^0(p)\rangle
= \frac12 \big( F^T_\perp \, k_\perp^\mu + F^T_\| \, k_\|^\mu + F^T_0 \, k_0^\mu \big) \ , \\
\H^\mu  &\equiv& i \int dx\,e^{i\,q\cdot x } \langle K^-(k_1) \pi^+(k_2)| T\{ j_{\rm em}^\mu(x) \,\op_{\rm 4q}(0) \} |\bar{B}^0(p)\rangle
= \H_\perp  k_\perp^\mu + \H_\|  k_\|^\mu + \H_0 \, k_0^\mu \ ,\ \quad
}
with $q=q_1+q_2$ and $p=q+k$. In addition to the form factors $F^{(T)}_i$, the decay amplitude
involves the functions $\H_i(k^2,q^2,q\cdot \bar k)$ describing the non-local effects which appear when the lepton pair couples to the electromagnetic current, through a penguin contraction of the four-quark operators $\op_{\rm 4q} \sim \bar s b \bar q q$.

We define the decomposition in terms of transversity amplitudes $\A^{L,R}_i$ :
\eq{
i \A = \frac{\alpha \,g_F}{8\pi \N} \bigg\{
L_\mu \big( \A^L_\perp \, k_\perp^\mu + \A^L_\| \, k_\|^\mu + \A^L_0 \, k_0^\mu + \A^L_t \, k_t^\mu \big) +
R_\mu \big( \A^R_\perp \, k_\perp^\mu + \A^R_\| \, k_\|^\mu + \A^R_0 \, k_0^\mu + \A^R_t \, k_t^\mu \big) 
\bigg\} \ ,
\label{eq:iA1}
}
where $L^\mu \equiv \bar u_\ell(q_1) \,\gamma^\mu P_L\, v_\ell(q_2)$ and $R^\mu \equiv \bar u_\ell(q_1) \,\gamma^\mu P_R\, v_\ell(q_2)$,
and $P_{L,R} = (1\mp \gamma_5)/2$ are the left- and right-chirality components of the lepton current. 
 The normalization constant $N$ is set to the value 
\begin{equation}\label{Eq:normN}
{\cal N}= \alpha G_F V_{tb} V_{ts}^* \sqrt{\frac{ \sqrt{\lambda\lambda_{K\pi}\lambda_q}}{3\cdot 2^{13}\pi^7 m_B^3\,k^2}}\,,
\end{equation}
for easier comparison with the $P$-wave results in the narrow-width limit for the $K^*$ meson~\cite{Descotes-Genon:2019bud}.

Comparing with~\Eq{AmplitudeBKpill} one can see that
\eq{
\A_i^{L,R} = \N \bigg[
(C_9 \mp C_{10}) F_i + \frac{2m_b}{q^2} \Big\{  C_7 F_i^T - i \frac{16\pi^2}{m_b} \H_i  \Big\}
\bigg]\ ,
\quad i=\{\perp,\|,0,t\}\ ,
}
keeping in mind that $\A_i^{L,R} \equiv \A_i^{L,R}(k^2,q^2,q\cdot \bar k)$, etc. For $\A_t^{L,R}$ only the first term is present  due to $F_t^T = \H_t = 0$.
In addition, since we consider two leptons of equal masses, one has $L_\mu k_t^\mu = - R_\mu k_t^\mu$ and the timelike-helicity amplitude depends only on the $C_9$\,-\,independent combination
$\A_t\equiv \A_t^L-\A_t^R$.

Concerning the non-local form factors $\H_i$, we will use the 
operator product expansion (OPE) at leading power, which allows us to express the functions $\H_i$ in terms of the local form factors. Using the notation of~\Reff{Asatrian:2019kbk}, we have
\eq{
\H_i(k^2,q^2,q\cdot\bar k) = \frac{i\,q^2}{32\pi^2} \Delta C_9(q^2) F_i(k^2,q^2,q\cdot\bar k) + \op(\alpha_s) + \cdots
}
where the ellipses denote higher OPE contributions, and the function $\Delta C_9(q^2)$ is given by
\begin{align}
\Delta C_9(q^2)= \frac49 (C_F C_1+C_2) \bigg[
\frac23 + \frac{4m_c^2}{q^2} - \log\frac{m_c^2}{m_b^2}
-\bigg(2+\frac{4m_c^2}{q^2}\bigg) \sqrt{\frac{4m_c^2-q^2}{q^2}}
\arctan \sqrt{\frac{q^2}{4m_c^2-q^2}}
\bigg]\,,
\end{align}
keeping only the leading contributions from the current-current $(\bar s c)( \bar c b)$ operators. The definitions used here are the same as in
\Reff{Asatrian:2019kbk}, 
where $C_1(m_b)\simeq -0.29$ and $C_2(m_b)\simeq 1.01$. The resulting transversity amplitudes in this approximation are given by
\eq{
\label{eq:calABtoKll}
\A_i^{L,R} = \N \bigg[
\big(C_9 +\Delta C_9(q^2) \mp C_{10}\big) F_i +
\frac{2m_b}{q^2} C_7 F_i^T
\bigg]\ ,
\quad i=\{\perp,\|,0,t\}\ .
}
For the numerical inputs, we use $C_9(m_b) = 4.3$, $C_{10}(m_b)=-4.2$, $C_7^{\rm eff}(m_b) = -0.3$
and $\alpha_{\rm em}(m_b) = 1/129$ (see e.g.~\Reff{Huber:2005ig}).
The transversity amplitudes $\A_i^{L,R}$ in~\Eq{eq:calABtoKll} can be expanded in partial waves $\A_i^{L,R(\ell)}(k^2,q^2)$ in the same way as the form factors. 
The form factors $F_{0,t}$ contain the $S$-wave, as described in~\Eq{eq:PWE0t}, whereas $F_{\perp,\|}$ start at the $P$-wave only (see~\Reff{Descotes-Genon:2019bud} for their partial-wave expansion).

Following the same steps as in \Reff{Descotes-Genon:2019bud} and considering the decay chain $B\to V^*(\to \ell\ell) K\pi$, we may rewrite the amplitude $\A$ in terms of the helicity amplitudes $H_\lambda^{L,R}$:
\eq{
\label{eq:Mpolarisations}
i \A
   =\frac{\alpha g_F}{8\pi {\mathcal {N}}}
      \sum_\lambda g_{\lambda\lambda} [(\epsilon_\lambda \cdot L)\ H^L_\lambda + (\epsilon_\lambda \cdot R)\  H^R_\lambda]\ ,
}
with $\lambda=\{0,t,+,-\}$ and $g_{tt}=1$, $g_{00}=g_{++}=g_{--}=-1$. 
The polarisations of the virtual intermediate gauge boson $V^*$ defined in the $B$-meson rest frame are
\eq{
\epsilon_\pm^\mu=(0,1,\mp i,0)/\sqrt{2}\quad 
\epsilon_0^\mu=(-q_z,0,0,-q_0)/\sqrt{q^2}\quad
\epsilon_t^\mu=(q_0,0,0,q_z)/\sqrt{q^2}\ ,
}
where $q^\mu=(q_0,0,0,q_z)$. We can then define transversity amplitudes,  performing the partial-wave expansion up to the $P$-wave:
\begin{align}
H_+^{L,R}&=\sqrt{3}\,\frac{\widehat{A}_{\|}^{L,R}+\widehat{A}_\perp^{L,R}}{\sqrt{2}}(-\sin\theta_K) + \cdots\ ,
&
H_-^{L,R}&=\sqrt{3}\,\frac{\widehat{A}_{\|}^{L,R}-\widehat{A}_\perp^{L,R}}{\sqrt{2}}(-\sin\theta_K) + \cdots\ ,
\label{eq:Hpm}
\\[2mm]
H_0^{L,R}&=\sqrt{2} (\widehat{S}_0^{L,R}+\sqrt{3}\widehat{A}_0^{L,R} \cos\theta_K + \cdots)\ ,
&
H_t&=-\sqrt{2} (\widehat{S}_t+\sqrt{3}\widehat{A}_t \cos\theta_K + \cdots)\ ,
\label{eq:Ht0}
\end{align}
with $H_t \equiv H_t^L - H_t^R$. Here $\widehat{S}^{L,R}_i$ and 
$\widehat{A}^{L,R}_i$ denote~\footnote{
We have changed the normalisation of the $S_i$ amplitudes compared to \Reff{Descotes-Genon:2019bud}, to be consistent with the partial-wave expansions of the longitudinal and time-like components in~\Eq{eq:PWE0t}.} 
the amplitudes with $\ell_{K\pi}=0$ and $\ell_{K\pi}=1$ respectively,
 and the ellipsis indicates the $D$-wave as well as higher partial waves. 
The amplitudes entering Eqs.~(\ref{eq:Hpm}) and~(\ref{eq:Ht0}) are related  to the transversity amplitudes introduced in~\Eq{eq:iA1}:  
\begin{align}
\widehat{A}_{\perp}^{L,R} &= -\frac{\sqrt{\lambda_{K\pi}}}{k^2}\A_{\perp}^{L,R(1)}\ ,
&
\widehat{A}_{\|}^{L,R} &= \frac{\sqrt{\lambda_{K\pi}}}{k^2}\A_{\|}^{L,R(1)}\ ,
\nonumber\\[2mm]
\widehat{A}_0^{L,R}&= -\A_0^{L,R(1)}/\sqrt{2}\ , 
&
\widehat{A}_t &= -\A_t^{(1)}/\sqrt{2}\ ,
\nonumber\\[2mm]
\widehat{S}_0^{L,R}&= -\A_0^{L,R(0)}/\sqrt{2}\ , 
&
\widehat{S}_t &= -\A_t^{(0)}/\sqrt{2}\ ,
\label{eq:helampl}
\end{align}
where the first two lines were already shown in 
\Reff{Descotes-Genon:2019bud}, but the 
last line is new, following from our consideration of $S$-wave contributions.\footnote{
We also corrected a typo in Eq.~(6.22) of~\Reff{Descotes-Genon:2019bud}
(Eq.~(124) in the arXiv version) regarding the sign in the relation between $\widehat{A}_\perp$ and ${\cal A}_\perp$.
}

%%%%%%%%%%%%%%%%%%%%%%%%%%%%%%%%%%%%%%%%%%%
\subsection{Differential decay rate}

The differential decay rate  for $\bar{B}\to K^-\pi^+\ell\ell$ is given by
\begin{equation}
\frac{d\Gamma}{dq^2\, ds\, d\cos\theta_\ell \, d\cos\theta_K\, d\phi}
  =\frac{1}{2^{15}\pi^6m_B}
  \frac{\sqrt{\lambda\lambda_q\lambda_{K\pi}}}{m_B^2q^2 s}
   \sum_{s_1,s_2} |\A|^2\ ,
\end{equation}
where $s=k^2$ and
$\lambda_q \equiv \lambda(q^2,m_{\ell}^2,m_{\ell}^2)$. According to~\Eq{eq:Mpolarisations},  $|\A|^{2}$ involves
the products of the hadronic amplitudes $\widehat{A}_i^{L,R}$ 
(known in terms of the form factors $F_i,F_i^{T}$ and non-local contributions neglected here) and 
  the leptonic amplitudes $L_\lambda$ and $R_\lambda$ (which 
can be easily evaluated in the $B$-meson rest frame). 
Summing over the spins of the outgoing leptons yields the final expression:
\begin{equation} \label{eq:differentialdecayrate}
\frac{d\Gamma}{dq^2\, ds\, d\cos\theta_\ell \, d\cos\theta_K\, d\phi}
  =\frac{9}{32\pi}\bar{I}(q^2,s,\theta_\ell,\theta_K,\phi)\,,
\end{equation}
containing  the following decomposition in terms of angular observables:
\begin{align} \label{eq:angulardist}
 \bar{I}(q^2, s, \theta_\ell, \theta_K, \phi)& = 
      \bar{I}_1^s \sin^2\theta_K + \bar{I}_1^c \cos^2\theta_K
      + (\bar{I}_2^s \sin^2\theta_K + \bar{I}_2^c \cos^2\theta_K) \cos 2\theta_\ell
\nonumber \\       
    & + \bar{I}_3 \sin^2\theta_K \sin^2\theta_\ell \cos 2\phi 
      + \bar{I}_4 \sin 2\theta_K \sin 2\theta_\ell \cos\phi 
\nonumber \\       
    & - \bar{I}_5 \sin 2\theta_K \sin\theta_\ell \cos\phi
\nonumber \\      
    & - (\bar{I}_6^s \sin^2\theta_K +
      {\bar{I}_6^c \cos^2\theta_K})  \cos\theta_\ell 
      + \bar{I}_7 \sin 2\theta_K \sin\theta_\ell \sin\phi
\nonumber \\ 
    & - \bar{I}_8 \sin 2\theta_K \sin 2\theta_\ell \sin\phi
      - \bar{I}_9 \sin^2\theta_K \sin^2\theta_\ell \sin 2\phi
      \nonumber\\
&+\bar{\tilde{I}}_{1b}^c\cos(\theta_K) 
+ \bar{\tilde{I}}_{2b}^c \cos(\theta_K)\cos(2 \theta_\ell) \nonumber\\
& 
 +\bar{\tilde{I}}_4\sin(\theta_K)\sin(2 \theta_\ell)\cos(\phi)
 - \bar{\tilde{I}}_5\sin(\theta_K)\sin(\theta_\ell)\cos(\phi)\nonumber\\
&  
 +\bar{\tilde{I}}_7\sin(\theta_K)\sin(\theta_\ell)\sin(\phi)
 - \bar{\tilde{I}}_8\sin(\theta_K)\sin(2 \theta_\ell)\sin(\phi)
\,. 
\end{align}
The expressions for the angular observables in which the contributions of the $S$ and $P$ waves
are separated are\,\footnote{
We use the same classification as in \Reff{Alguero:2021yus},  but we use
the same definition of the angles as in \Reff{Altmannshofer:2008dz} for $\bar{B}\to K^-\pi^+\ell\ell$. Moreover,
we enforce the same normalisation for the $S$ and $P$-wave angular coefficients, recasting $\bar{\tilde{I}}_{1a}^c$ and $\bar{\tilde{I}}_{2a}^c$ as contributions to $\bar{I}_1^{s,c}$ and $\bar{I}_2^{s,c}$, in order to avoid any ambiguity in the definition of the differential decay rate. We  recall that  only SM operators in the weak effective Hamilonian are taken into account in our study.
}:
\begin{align}
  \bar{I}_1^s & = \frac{(2+\beta_\ell^2)}{4} \left[|\widehat{A}_{\perp}^L|^2 + |\widehat{A}_{\parallel}^L|^2 + (L\to R) \right] 
            + \frac{4 m_\ell^2}{q^2} \re(\widehat{A}_{\perp}^L(\widehat{A}_{\perp}^R)^* + \widehat{A}_{\parallel}^L(\widehat{A}_{\parallel}^R)^*)\nonumber\\
& \qquad +\frac{1}{3}\left[|\widehat{S}_0^L|^2 +  (L\to R)  + (1-\beta_\ell^2)(|\widehat{S}_t|^2 +
      2\re(\widehat{S}_0^L(\widehat{S}_0^R)^*)\right] ,
\\
  \bar{I}_1^c & =  |\widehat{A}_{0}^L|^2 +|\widehat{A}_{0}^R|^2  + \frac{4m_\ell^2}{q^2} 
               \left[|\widehat{A}_t|^2 + 2\re(\widehat{A}_{0}^L(\widehat{A}_{0}^R)^*) \right] \nonumber\\
& \qquad +\frac{1}{3}\left[|\widehat{S}_0^L|^2 + (L\to R)  + (1-\beta_\ell^2)(|\widehat{S}_t|^2 +
      2\re(\widehat{S}_0^L(\widehat{S}_0^R)^*)\right] ,
\\
  \bar{I}_2^s & = \frac{ \beta_\ell^2}{4}\left[ |\widehat{A}_{\perp}^L|^2+ |\widehat{A}_{\parallel}^L|^2 + (L\to R)\right] -\frac{1}{3}\beta_\ell^2\left[|\widehat{S}_0^L|^2 + (L\to R) \right]
  ,
\\
  \bar{I}_2^c & = - \beta_\ell^2\left[|\widehat{A}_{0}^L|^2 + (L\to R)\right]-\frac{1}{3}\beta_\ell^2\left[|\widehat{S}_0^L|^2 + (L\to R) \right],
\\
  \bar{I}_3 & = \frac{1}{2}\beta_\ell^2\left[ |\widehat{A}_{\perp}^L|^2 - |\widehat{A}_{\parallel}^L|^2  + (L\to R)\right],
\\
  \bar{I}_4 & = \frac{1}{\sqrt{2}}\beta_\ell^2\left[\re (\widehat{A}_{0}^L(\widehat{A}_{\parallel}^L)^*) + (L\to R)\right],
\\
  \bar{I}_5 & = \sqrt{2}\beta_\ell\left[\re(\widehat{A}_{0}^L(\widehat{A}_{\perp}^L)^*) - (L\to R) 
\right],
\\
\label{eq:I6c}
  \bar{I}_6^s  & = 2\beta_\ell\left[\re (\widehat{A}_{\parallel}^L(\widehat{A}_{\perp}^L)^*) - (L\to R) \right],
\\
   \bar{I}_6^c  &  = 0,
\\
  \bar{I}_7 & = \sqrt{2} \beta_\ell \left[\im (\widehat{A}_{0}^L(\widehat{A}_{\parallel}^L)^*) - (L\to R) 
\right],
\\
  \bar{I}_8 & = \frac{1}{\sqrt{2}}\beta_\ell^2\left[\im(\widehat{A}_{0}^L(\widehat{A}_{\perp}^L)^*) + (L\to R)\right],
\\
\label{eq:AC-last}
  \bar{I}_9 & = \beta_\ell^2\left[\im (\widehat{A}_{\perp}^L(\widehat{A}_{\parallel}^L)^{*}) + (L\to R)\right]\,,
\end{align}
with $\beta_\ell=\sqrt{\lambda_q}/q^4=\sqrt{1-4m_\ell^2/q^2}$.
The angular observables containing interferences between the $S$ and $P$-waves are:
\begin{align}
\bar{\tilde{I}}_{1b}^c &= \frac{2}{3}\sqrt{3}
 \re\left[\widehat{S}_0^L (\widehat{A}_0^L)^* + (L\to R)  + (1 - \beta_\ell^2)(\widehat{S}_0^L (\widehat{A}_0^R)^* + 
      \widehat{S}_0^R (\widehat{A}_0^L)^* + \widehat{S}_t \widehat{A}_t^*)\right],\\
\bar{\tilde{I}}_{2b}^c &= -\frac{2}{3}\sqrt{3}\beta_\ell^2
 \re\left[\widehat{S}_0^L(\widehat{A}_0^L)^* (L\to R) \right],\\ 
\bar{\tilde{I}}_4 &= \frac{2}{3}\sqrt{\frac{3}{2}}\beta_\ell^2
 \re\left[\widehat{S}_0^L (\widehat{A}_\parallel^L)^* +(L\to R) \right],\\  
\bar{\tilde{I}}_5 &= \frac{4}{3}\sqrt{\frac{3}{2}}\beta_\ell
 \re\left[\widehat{S}_0^L (\widehat{A}_\perp^L)^* -(L\to R) \right],\\  
\bar{\tilde{I}}_7 &= \frac{4}{3}\sqrt{\frac{3}{2}}\beta_\ell
 \im\left[\widehat{S}_0^L (\widehat{A}_\parallel^L)^* -(L\to R) \right],\\  
\bar{\tilde{I}}_8 &= \frac{2}{3}\sqrt{\frac{3}{2}}\beta_\ell^2
 \im\left[\widehat{S}_0^L (\widehat{A}_\perp^L)^* +(L\to R) \right]\,. 
 \end{align}
As already indicated in \Reff{Descotes-Genon:2019bud}, the choice of normalisation in~\Eq{Eq:normN} yields a very simple expression for~\Eq{eq:differentialdecayrate}. If we neglect $S$-wave contributions,
setting $\widehat{S}_0^{L,R}=\widehat{S}_t=0$, we can see that $\bar{I}$ is formally the same expression as the one obtained in Eqs~(3.10) and~(3.21) of \Reff{Altmannshofer:2008dz},  with the angular coefficients  $\bar{I}_{1s,1c,2s,2c,3,4,5,6s,6c,7,8,9}$ given by Eqs.~(3.34)-(3.45) of the same reference, as long as the transversity amplitudes $A_i$ of Eqs.~(3.28)-(3.31) in \Reff{Altmannshofer:2008dz} are replaced by the transversity amplitudes $\widehat{A}_i$ given in~\Eq{eq:helampl}. We also agree with the structure of the differential decay rate in terms of transversity amplitudes given in \Reff{Alguero:2021yus} for the terms involving the $S$ wave.

%%%%%%%%%%%%%%%%%%%%%%%%%%%%%%%%%%%%%%%%%%%%%%%%%%%%%%%%%%%%%%%%%%%%%%%%%%%%%%
\subsection{Predictions for the differential rate in the $K^*(892)$ region}

We start by considering the case where the $K\pi$ invariant mass is close to the $K^*(892)$ $P$-wave resonance. At such low invariant masses, the $S$ and $P$ waves are dominant and one can neglect higher partial waves. The differential rate, integrated over all angles, is then
\begin{equation}
\label{eq:BRBtoKpillSP}
 \frac{d\Gamma}{dq^2ds}=
 |\widehat{S}^L|^2+|\widehat{S}^R|^2+|\widehat{A}_{\|}^L|^2+|\widehat{A}_{\|}^R|^2+|\widehat{A}_{\perp}^L|^2+|\widehat{A}_{\perp}^R|^2+|\widehat{A}_{0}^L|^2+|\widehat{A}_{0}^R|^2 \,.
\end{equation}
We note that the $F_t^{{(\ell=0,1)}}$ form factors do not enter because we are assuming massless leptons. 

\begin{figure}\centering
\subfloat{\includegraphics[height=0.33\textwidth]{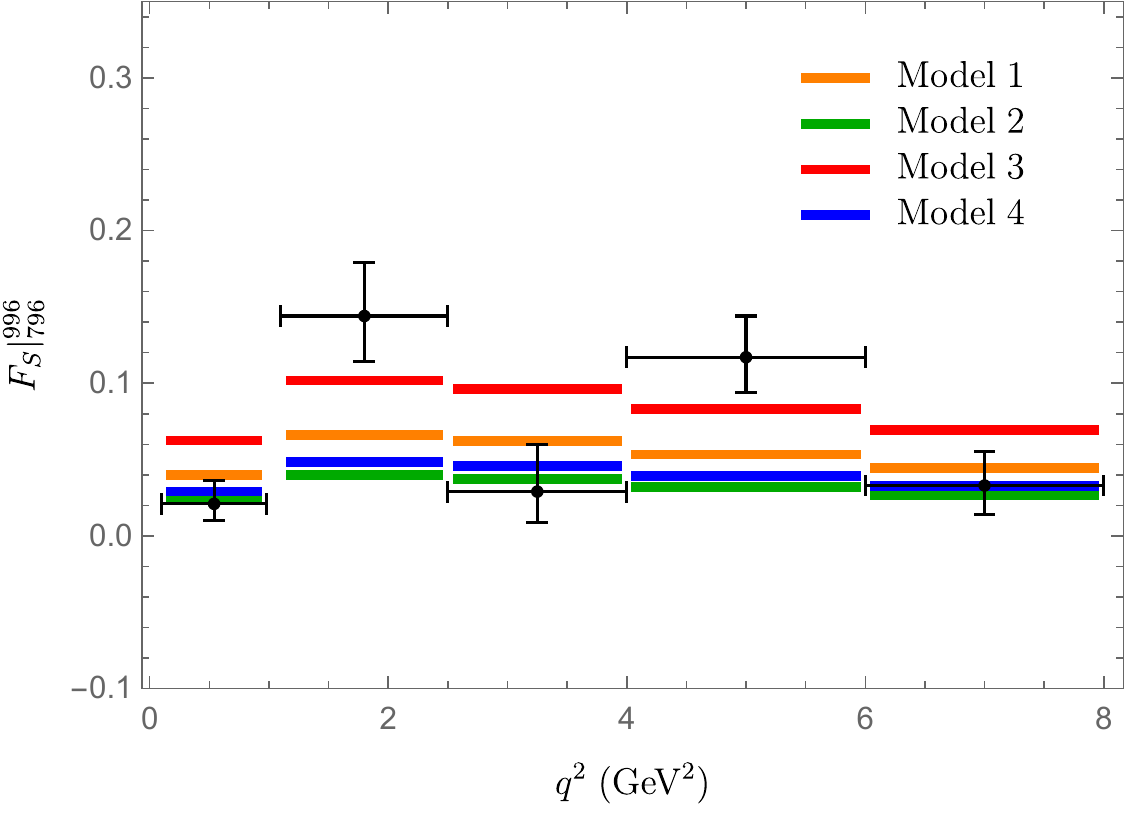}}\hspace{5mm}
\subfloat{\includegraphics[height=0.33\textwidth]{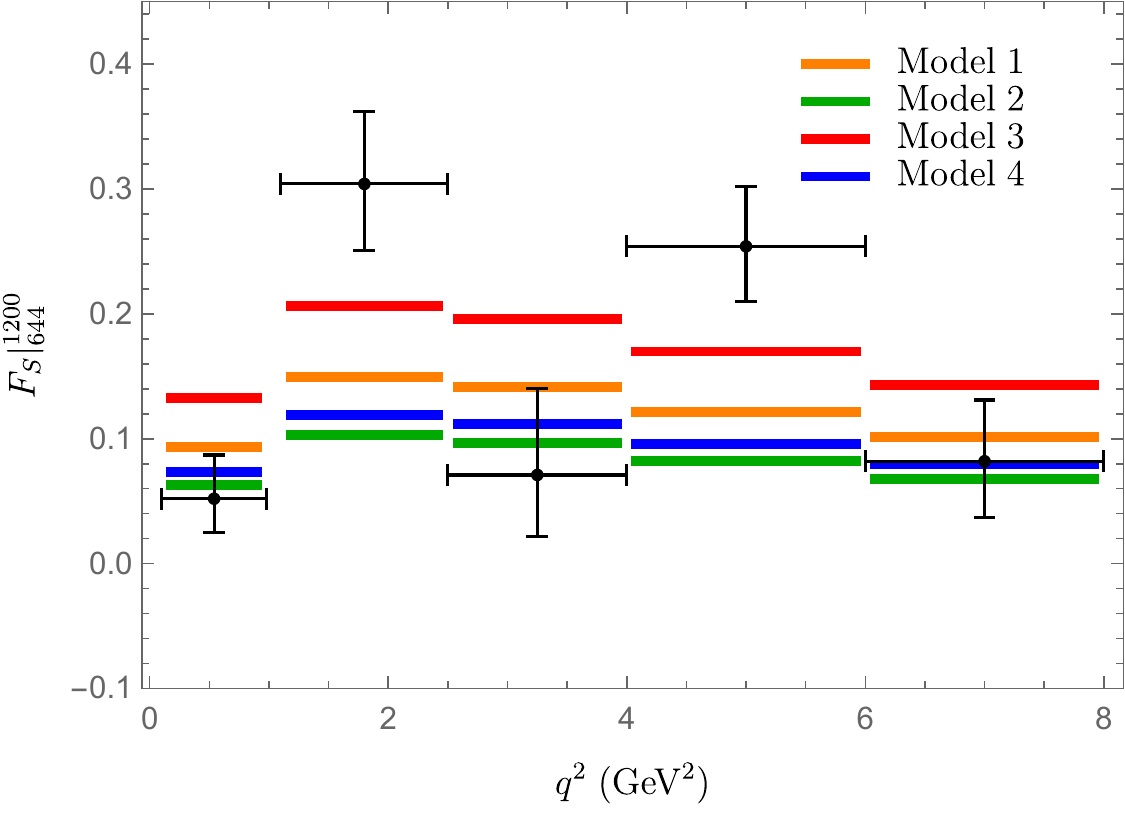}}
\caption{\it $S$-wave fration $F_S$ in bin 1 (left) and bin 2 (right) for different $q^2$ bins compared with the LHCb data points from \Reff{LHCb:2016ykl}.}
\label{fig:FS}
\end{figure}

In~\Reff{LHCb:2016ykl}, the LHCb collaboration presented measurements of the $B\to K \pi \mu\mu$ differential decay rate in the $K^*(892)$ region. To quantify the $S$-wave contribution, they also measured the $S$-wave fraction in several $q^2$ bins and in two different ranges of the $K\pi$ invariant mass around the $K^*$ resonance. 
In analogy with the expression of $F_S$ in Eq.~(5) of~\Reff{LHCb:2016ykl}, we define:
\begin{equation}
F_S \equiv \frac{\int_\text{bin} ds \,(|\widehat{S}^L|^2+|\widehat{S}^R|^2)}
{\int_\text{bin} ds\, (|\widehat{S}^L|^2+|\widehat{S}^R|^2+ |\widehat{A}_{\|}^L|^2+|\widehat{A}_{\|}^R|^2+|\widehat{A}_{\perp}^L|^2+|\widehat{A}_{\perp}^R|^2+|\widehat{A}_{0}^L|^2+|\widehat{A}_{0}^R|^2)} \,,
\end{equation}
and we compute this fraction in 
\begin{align}
    {\rm bin\; 1}:&  \quad\quad (0.796 \; {\rm GeV})^2<s< (0.996 \; {\rm GeV})^2 \,,\\ 
      {\rm bin\; 2}:& \quad\quad (0.644\; {\rm GeV})^2 <s < (1.200 \; {\rm GeV})^2 \,.
\end{align}
Our results for both bins are presented in~\Fig{fig:FS}, using the $P$-wave model of~\Reff{Descotes-Genon:2019bud} for $\alpha=1$, compared with the LHCb data from \Reff{LHCb:2016ykl}. We predict a rather small $S$-wave contribution in this region,  in agreement with some of the LHCb data points. 
 It would be very difficult to achieve a  full agreement, as the data features very rapid changes in $F_S$ as $q^2$ varies. Since this observable 
 is driven by form factors with a monotonous $q^2$ behavior, we cannot propose any plausible theoretical explanation for such rapid variations.

\begin{figure}\centering
\subfloat{\includegraphics[height=0.55\textwidth]{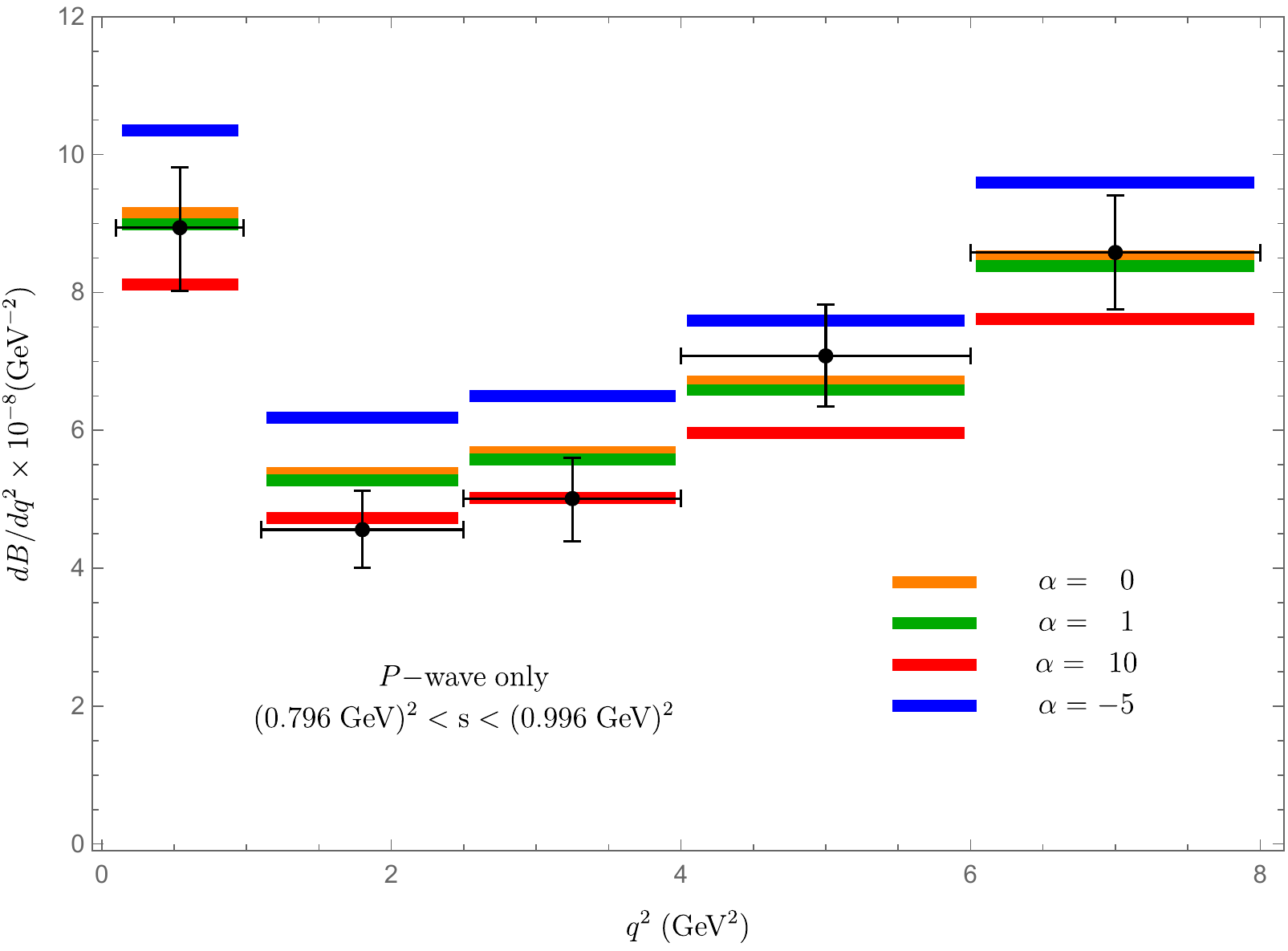}}
\caption{\it Theory predictions for the $B\to (K\pi)_P \ell^+\ell^-$ branching ratio within the $K\pi$ invariant mass bin $(0.796 \; {\rm GeV})^2 <s < (0.996 \; {\rm GeV})^2$, for different values of $\alpha$, compared to the LHCb measurements of $B\to K^*\mu^+\mu^-$ in~\Reff{LHCb:2016ykl}.}
\label{fig:brlowbin}
\end{figure}

The LHCb collaboration then presented a measurement of the $B\to K^* \ell\ell$ branching ratio by subtracting the $S$-wave contribution $F_S$ from the data. In~\Fig{fig:brlowbin}, we compare this experimental data with our predictions for the branching ratio restricted to the $P$-wave component, calculated at various values of $\alpha$ in the region $(0.796 \; {\rm GeV})^2 < s < (0.996 \; {\rm GeV})^2$ and in different $q^2$ bins. We normalize the branching ratio to the 
$q^2$-bin size in the same way as in the experimental analysis:  
\begin{equation}\label{eq:normbr}
    \frac{dB}{dq^2} = \frac{1}{q_2^2- q_1^2} \int_{q_1^2}^{q_2^2} dq^2 \int_{s_{\rm min}}^{s_{\rm max}} ds\;\tau_B\; \frac{d\Gamma}{dq^2 ds} \ .
\end{equation}
We observe that higher values of $\alpha$ push down the predictions for the branching ratios in the $K^*(892)$ region, while lower values push them up. From now on, we will  set $\alpha=1$ for the $P$-wave contribution, as it yields a good agreement with the LHCb measurements of the branching ratios.

Finally, we note that our results could allow us to predict angular observables associated with different moments of the $S$-wave contribution in the $K^*(892)$ region~\cite{Becirevic:2012dp,Alguero:2021yus}. Comparing such predictions with data could thus give more insight into the 
dynamics of the $S$-wave component. However,  we are not aware of corresponding experimental data on the $S$-wave in this region.  The $S$-wave contribution has been discussed for branching ratios~\cite{LHCb:2016ykl} 
but it was treated as a nuisance parameter in the context of angular moments ~\cite{LHCb:2020lmf}. 
We will thus leave  this study for further work, focusing on the $K^*(1410)$ and $K^*_0(1430)$ region from now on.

%%%%%%%%%%%%%%%%%%%%%%%%%%%%%%%%%%%%%%%%%%%%%%%%%%%%%%%%%%%%%%%%%%%%%%%%%%%%%%
\subsection{Differential decay rate in the $K^*(1410)$ and $K^*_0(1430)$ region} \label{sec:Br1430region}

\begin{figure}
\centering
\subfloat{\includegraphics[height=0.55\textwidth]{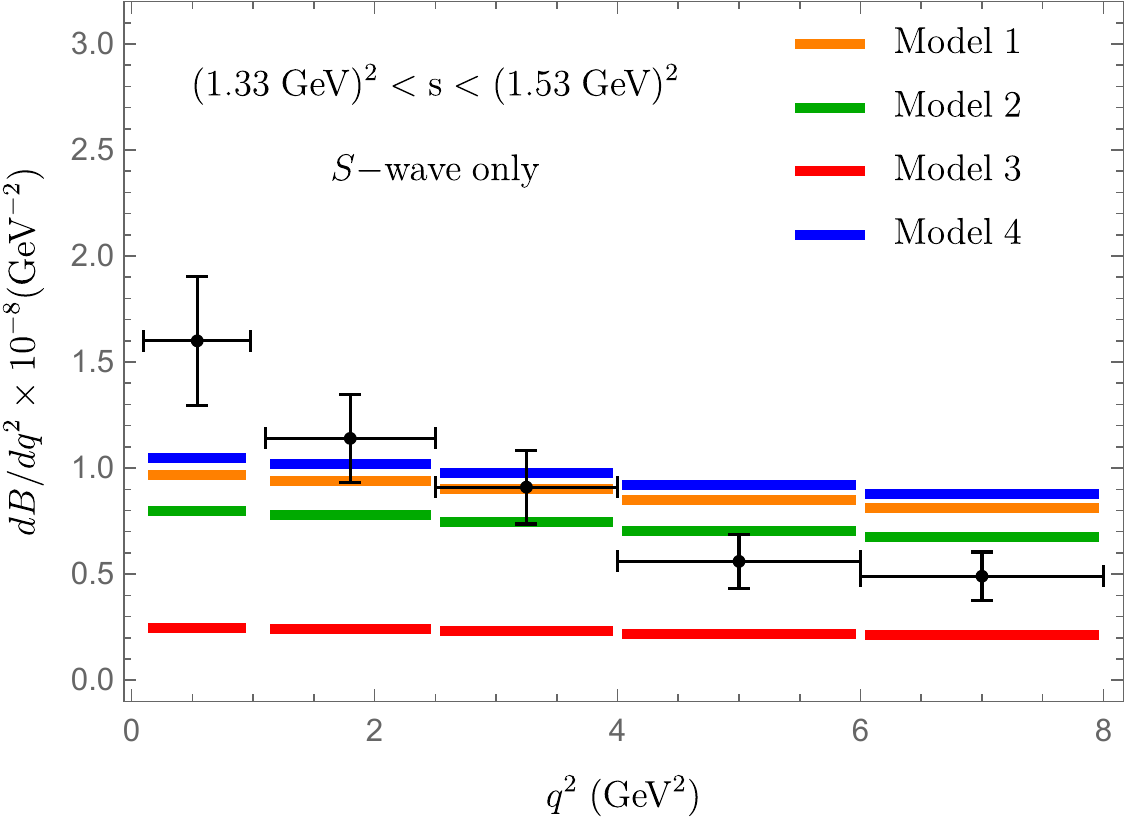}}
\caption{\it LHCb data and the $S$-wave only contribution to the differential rate in different $q^2$ bins integrated over $s$ in the high bin.}
\label{fig:difrateSonly}
\end{figure}

The LHCb collaboration also measured the $B\to K \pi \mu\mu$ differential decay rate in the region
$(1.33\;{\rm GeV})^2<s<(1.52\; \rm{GeV})^2$  and in different $q^2$ bins in
\Reff{LHCb:2016eyu}. Taking~\Eq{eq:BRBtoKpillSP}, we compute this rate 
using the four different $S$-wave models and with different values of $\alpha$ for the $P$-wave contribution. The $S$-wave 
is substantially more important here than in the $K^*(892)$ region. In~\Fig{fig:difrateSonly}, we show our results considering only the $S$-wave contribution for the four different models (normalized following~\Eq{eq:normbr}). We see that in the higher $q^2$ bins, some of the models yield already too large value compared to the data, even before including the (positive) $P$-wave contribution. Moreover the sum of the $S$- and $P$-wave contributions to the branching ratios should be smaller than the experimental value, since the latter also includes a (positive) contribution of the $D$ wave that we are not able to estimate at this stage, but which is not necessarily negligible~\cite{LHCb:2016eyu}.

\begin{figure}\centering
\subfloat{\includegraphics[height=0.55\textwidth]{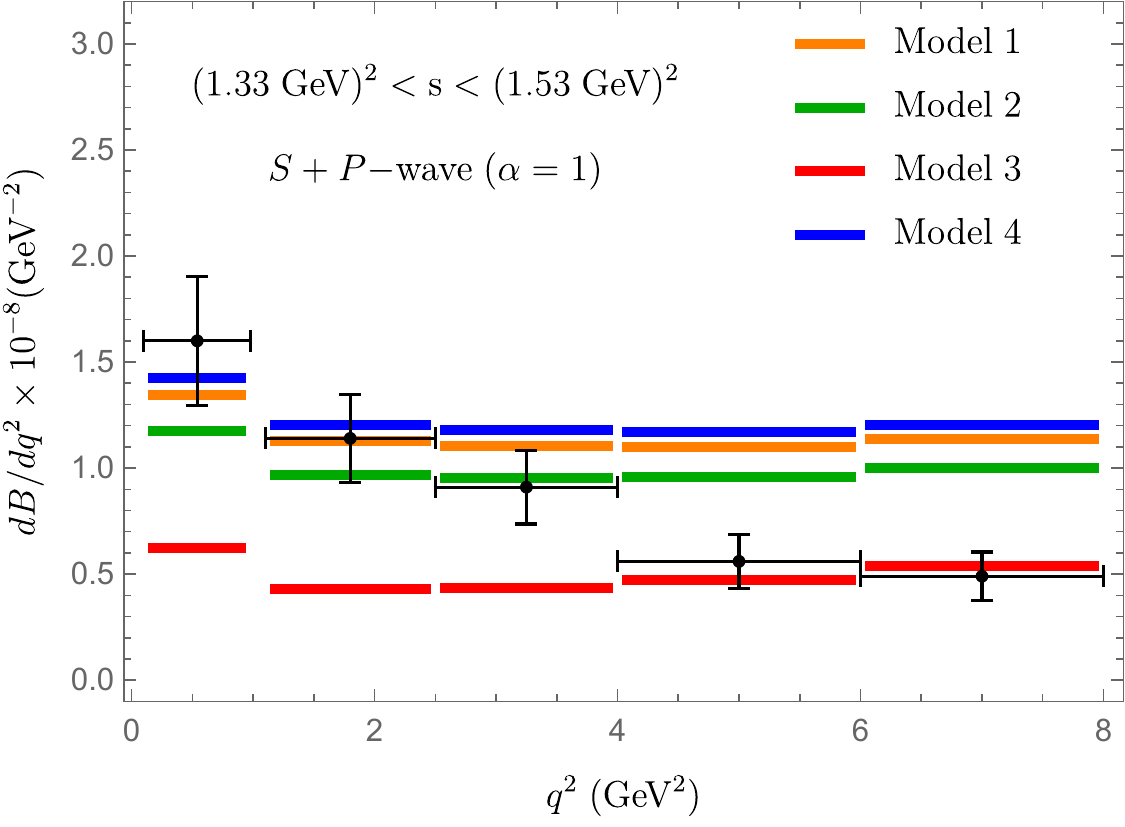}}
\caption{\it LHCb data and $P$ ($\alpha=1$) and $S$-wave contributions to the differential rate in different $q^2$ bins integrated over $s$ in the high bin.}
\label{fig:difratetot}
\end{figure}

Adding the $P$-wave (with $\alpha=1$) gives the predictions 
shown in~\Fig{fig:difratetot}. We observe good agreement for the lower $q^2$ bins. 
At larger $q^2$, we cannot reproduce the measured $q^2$ dependence, as our result combines the increasing
$P$-wave contribution with an almost constant $S$-wave contribution (for all four models).

It seems difficult to improve the situation significantly by changing the parameter $\alpha$ of the $P$-wave contribution. Indeed, this $P$-wave contribution is responsible for the 
satisfactory agreement for the $B\to K\pi\ell\ell$ branching ratio at low $q^2$ in the two regions of $K\pi$ invariant mass that we have considered here.

The $D$-wave contribution constitutes a possible missing element.
It is not included in our analysis but would yield a further positive contribution to the branching ratio. This contribution might change the $q^2$-dependence of the branching ratio, but at the same time, it will increase the overall prediction and thus worsen the agreement with the data.
 
We thus expect that the origin of the disagreement 
encountered with data at large $q^2$
might be related to the overall normalisation and the $q^2$-dependence of the $S$-wave contribution around the scalar resonance $K^*_0(1430)$.
At this stage, one should remember that the four versions of the  $B\to K\pi$ $S$-wave form factors adopted here originate from the models for the scalar $K\pi$ form factor of \Reff{VonDetten:2021rax}.
These versions were not meant to exhaust all the possible models for this form factor, but rather to show the possible range of variation at intermediate $K\pi$ invariant mass allowed by the dispersive approach and the limited amount of data available to fix the free parameters of the models.
The description of the $K\pi$ scalar form factor, and consequently, of the models
for the $B\to K\pi$ $S$-wave form factors, can certainly be explored further, 
in particular, concerning the impact of 
the so-called source term (the polynomial terms describing the high-energy behaviour), and the presence of additional resonances around 2~GeV. Therefore, one should interpret the fact that we get only a partial agreement with the data (although in the right ballpark regarding the prediction for the branching ratio) as the indication that the $S$-wave models considered here may serve as a good starting point requiring further tuning. This in turn could yield a larger range of possibilities regarding the contribution from the $K^*_0(1430)$, which is fairly similar in the models 1,2,3 (see~\Fig{fig:ffs}), leading to rather close predictions for the branching ratio (see~\Fig{fig:difratetot}). We will refrain from entering such an investigation here, as we want mainly to highlight the possibilities given by our framework.

%%%%%%%%%%%%%%%%%%%%%%%%%%%%%%%%%%%%%%%%%%%%%%%%%%%%%%%%%%%%%%%%%%%%%%%%%%%%%%
\subsection{Angular observables in the $K^*(1410)$ and $K^*_0(1430)$ region}

We can now turn to the analysis of angular observables
 performed in \Reff{Descotes-Genon:2019bud} and extend it to include the $S$ wave.  In \Reff{LHCb:2016eyu} the LHCb experiment has analysed the moments $\Gamma_i$ ($i=1\ldots 41$) of the angular distribution of $B\to K\pi\mu^+\mu^-$ in the region of $K\pi$ and dilepton invariant masses $\sqrt{k^2}\in [1.33,1.53]\GeV$ and $q^2\in [1.1,6]\GeV^2$, respectively~\footnote{
We neglect lepton masses in line with the analysis of Refs.~\cite{LHCb:2016eyu,Dey:2015rqa}.
}.
This region of $K\pi$ masses contains contributions from $K^*$ resonances in the $S$, $P$ and $D$ waves, and the moments analysed in \Reff{LHCb:2016eyu} contain contributions from all partial waves, following the analysis in \Reff{Dey:2015rqa}. The corresponding expansion can be written as
\eq{
\frac{d\Gamma}{dq^2\, ds\, d\Omega}
=\frac{1}{4\pi} \sum_{i=1}^{41} f_i(\Omega)\ \tilde{\Gamma}_i(q^2,k^2) \,,
}
where $d\Omega=d\cos\theta_\ell\,d\cos\theta_K\,d\phi$. Since the decomposition takes into account the possibility of $S$, $P$ and $D$-wave contributions, it features many different angular structures $f_i(\Omega)$. The normalisations chosen are such that  
\begin{equation}
\label{eq:BRBtoKpill}
 \frac{d\Gamma}{dq^2dk^2}=
\tilde{\Gamma}_1=|\widehat{S}^L|^2+|\widehat{S}^R|^2+
|\widehat{A}_{\|}^L|^2+|\widehat{A}_{\|}^R|^2+
|\widehat{A}_{\perp}^L|^2+|\widehat{A}_{\perp}^R|^2+
|\widehat{A}_{0}^L|^2+|\widehat{A}_{0}^R|^2+
\ldots\,,
\end{equation}
where the ellipsis denotes higher partial waves.
The other moments can be obtained from Table~5 of \Reff{LHCb:2016eyu} with $\tilde\Gamma_i=\bar\Gamma_i \tilde\Gamma_1$.
We recall that \Reff{LHCb:2016eyu} uses the same definition of the kinematics as in \Reff{Dey:2015rqa}, whereas we follow a prescription for the angles in agreement with \Reff{Altmannshofer:2008dz}: the comparison requires us to perform the redefinition
$\theta_\ell \to \pi-\theta_\ell$
leading to a change of sign for $\Gamma_i$ for $i$ from 11 to 18 and 29 to 33 between our definition and the one used in \Reff{Aaij:2016kqt}.

We can determine combinations of the moments $\tilde\Gamma$ involving only $S$- and $P$-wave amplitudes.
In addition to the relations already given in \Reff{Descotes-Genon:2019bud} involving only $P$-wave amplitudes, we have the following relations~\footnote{We recall that there are degeneracies among the moments, so that these relations can be rewritten in terms of other moments, which are equivalent theoretically but may lead to slightly different results experimentally. The list of such degeneracies is given in \Reff{Descotes-Genon:2019bud}.} free from $D$-wave contributions
\begin{eqnarray}\label{eq:rels}
    |\widehat{S}^L|^2+|\widehat{S}^R|^2+
    |\widehat{A}_0^L|^2+|\widehat{A}_0^R|^2 &=&
\frac{1}{54} (4 \tilde\Gamma_{1} - 14 \sqrt{5} \tilde\Gamma_{3} - 63 \tilde\Gamma_{5} - 
  50\sqrt{5} \tilde\Gamma_{6} - 70 \tilde\Gamma_{8})\ ,
\quad\\
\label{eq:rels2}\re(\widehat{A}_0^L \widehat{S}^{L*}+\widehat{A}_0^R \widehat{S}^{R*})  &=&
\frac{1}{54} (-5\sqrt{21} \tilde\Gamma_{4} - 27 \sqrt{5} \tilde\Gamma_{7} + \sqrt{105} \tilde\Gamma_{9})\ , 
\end{eqnarray}
The other interferences between the $P$-wave $\widehat{A}_i^{L,R}$ amplitudes and the $S$-wave $\widehat{S}^{L,R}$ amplitudes involve also the $D$ waves and we will consider them only at a later stage.

%%%%%%%%%%%%%%%%%%%%%%%%%%%%%%%%%%%%%%%%%%%%%%%%%%%%%%%%%%%%%%%%%%%%%%%%%%%%%%
\subsection{Predictions for the moments involving only $S$ and $P$ waves}

Using~\Eq{eq:rels}, we define two $S-P$ moments:
\begin{eqnarray}\label{eq:momdef}
\av{M_0} &\equiv& \tau_B\, \av{   |\widehat{S}^L|^2+|\widehat{S}^R|^2+
    |\widehat{A}_0^L|^2+|\widehat{A}_0^R|^2} \,,\\
  \av{M_{0\re}} &\equiv& 
\tau_B\, \av{\re(\widehat{A}_0^L \widehat{S}^{L*}+\widehat{A}_0^R \widehat{S}^{R*})}\, .
\end{eqnarray}
Taking the experimental values and correlations of the moments given in \Reff{LHCb:2016eyu},
we obtain from \eqref{eq:rels} and \eqref{eq:rels2} the following values in the ranges $\sqrt{k^2}\in [1.33,1.53]\GeV$ and $q^2\in [1.1,6.0]\GeV^2$:
\begin{eqnarray}
\label{eq:momexp}
\av{M_0}_{\rm exp} &=
(0.03\pm 1.86)\times 10^{-8}\ ,\\
\label{eq:mom2exp}
\av{M_{0\re}}_{\rm exp} &=
(0.34 \pm 0.70)\times 10^{-8}\ ,
\end{eqnarray}
Using now our $S$-wave form factors and the $P$-wave form factors 
(at $\alpha=1$) from \Reff{Descotes-Genon:2019bud}, we find for the first moment 
\begin{equation}
\av{M_0} = \left\{5.1, 4.4,1.9, 5.5  \right\}\times 10^{-8} \, ,
\end{equation}
where the values correspond to the four $S$-wave models under consideration. The spread in the values for the $S$-wave models can be understood from~\Fig{fig:ffs}, as these models have different behaviour around  $\sqrt{s}=1.5$ GeV, which is within our integration region. 
As discussed in \Sec{sec:ffmodel}, 
these differences stems from different assumptions on the polynomial (non-resonant) part of the models, as well as the inclusion (or not) of the $K^*_0(1950)$ resonance in the source terms of the model.
 
Comparing with~\Eq{eq:momexp}, we observe that our predictions overshoot the measurements by $(1-2)\sigma$. For higher values of $\alpha$ the moment $\av{M_0}$ becomes even larger. Actually, our predictions for most of the models overshoot the measurement even without a $P$-wave contribution (which would only increase the value of the moment). Indeed, we find for the $S$-wave only:
\begin{equation}\label{eq:MS}
\av{M_S}\equiv \tau_B\langle|\widehat{S}^L|^2+|\widehat{S}^R|^2 \rangle = \left\{4.37, 3.62, 1.12, 4.73  \right\}\times 10^{-8} \,. 
\end{equation}
For the interference of the amplitudes $\av{M_{0\re}}$ we find, for $\alpha=1$,
\begin{equation}
\av{M_{0\re}} =
\left\{-1.20, -1.08, -0.53, -1.24  \right\}\times 10^{-8} \ ,
\end{equation}
which lay somewhat below the experimental value in~\Eq{eq:mom2exp}.

Given the experimental measurements, our results for these moment seems to favour fit Model~3 together with $\alpha\approx1$, which is in agreement with the results found for the branching ratio for $4.0<q^2<6.0~\GeV^2$ in the previous section (although not at lower $q^2$).

%%%%%%%%%%%%%%%%%%%%%%%%%%%%%%%%%%%%%%%%%%%%%%%%%%%%%%%%%%%%%%%%%%%%%%%%%%%%%%
\subsection{Neglecting the $D$-wave contributions}

From the results of \Reff{LHCb:2016eyu}, it remains unclear if one can assume that the $D$-wave is negligible for $B\to K\pi \ell\ell$ in the $K^*_0(1430)$ 
region. On the one hand, the LHCb collaboration indicate that they expect a large $D$-wave contribution in this region, and on the other hand they obtain only a rather weak bound on the $D$-wave fraction 
of the branching ratio, $F_D<0.29$ (and compatible with zero). 

Assuming that the $D$-wave contributions are indeed negligible, we get 26 constraints, corresponding to the vanishing of some moments:
 \eq{
\tilde\Gamma_{4,5,9,10,13,14,17,18,20,22,23,25,27,28,30,32,33,36,37,40,41} 
=0\,,
}
and some linear combinations:
\begin{eqnarray}
%0&=& 
&&\tilde\Gamma_1 + \sqrt{5}\tilde\Gamma_3 + \sqrt{5}\tilde\Gamma_6 + 5\tilde\Gamma_8 
=\tilde\Gamma_2 + \sqrt{5} \tilde\Gamma_7=
\nonumber\\
&&\tilde\Gamma_{19} + \sqrt{5} \tilde\Gamma_{21}=
\tilde\Gamma_{24} + \sqrt{5} \tilde\Gamma_{26}
=\tilde\Gamma_{29}+\sqrt{5}\tilde\Gamma_{31}=0\,.
\end{eqnarray}
All these constraints are satisfied at 1.5 $\sigma$ or less, apart from $\tilde\Gamma_5=0$ and $\tilde\Gamma_{22}=0$, which are only satisfied at 2$\sigma$ and 1.7$\sigma$, respectively. This suggests indeed that the data in \Reff{LHCb:2016eyu} are compatible with the assumption of negligible $D$-wave contributions. Our study of the branching fraction in~\Sec{sec:Br1430region} does not suggest the need for a large $D$-wave component either.

\begin{table}
\centering
\def\arraystretch{1.5}
\setlength{\tabcolsep}{7pt}
\begin{tabular}{@{}ccr@{\hspace{8mm}}c@{}}
\toprule[0.7mm]
Moment$/\tau_B$ & Amplitude & Exp. Value $\times 10^{8}$ & Theory $\times 10^{8}$ \\
\midrule[0.7mm]
$-\frac{\sqrt{5}}{2}(\tilde\Gamma_{3} + 2\tilde\Gamma_{6}) $ 
& $\tau_B\langle|\widehat{S}^L|^2+|\widehat{S}^R|^2 \rangle = \av{M_{S}}$ 
& $2.16\pm 1.62$ \hspace{4mm} 
& $[1.12, 4.73]$\\
$\frac{1}{2}\tilde\Gamma_{2} $
& $\tau_B\langle\re(\widehat{A}_0^L \widehat{S}^{L*}+\widehat{A}_0^R \widehat{S}^{R*}) \rangle = \av{M_{0\re}} $ 
& $ -0.84 \pm 0.29$ \hspace{4mm} 
& $[-0.53, -1.24]$ \\
$-\sqrt\frac{5}{3}\tilde\Gamma_{11}$
& $\tau_B\langle\re(\widehat{A}_{||}^L \widehat{S}^{L*}+\widehat{A}_{||}^R \widehat{S}^{R*}) \rangle $ 
& $ -0.31 \pm 0.69$ \hspace{4mm} 
& $[-0.23, -0.54]$ \\
$\sqrt\frac{5}{3}\tilde\Gamma_{15}$
&  $\tau_B\langle\im(\widehat{A}_\perp^L \widehat{S}^{L*}+\widehat{A}_\perp^R \widehat{S}^{R*}) \rangle $ 
& $ 0.57 \pm 0.69$ \hspace{4mm} 
& $[-0.17, -0.36]$  \\
 $\frac{1}{\sqrt{3}}\tilde\Gamma_{34} $
 & $\tau_B\langle\re(\widehat{A}_{\perp}^L \widehat{S}^{L*}-\widehat{A}_{\perp}^R \widehat{S}^{R*}) \rangle $ 
 & $ 0.35 \pm 0.26$ \hspace{4mm} 
 & $[-0.14, -0.34]$\\
 $-\frac{1}{\sqrt{3}}\tilde\Gamma_{38} $
 & $\tau_B\langle\im(\widehat{A}_{||}^L \widehat{S}^{L*}-\widehat{A}_{||}^R \widehat{S}^{R*}) \rangle $ 
 & $ 0.14 \pm 0.25$ \hspace{4mm} 
 & $[-0.29, -0.61]$\\
\bottomrule[0.7mm]
\end{tabular}
\caption{\it Moments depending on both $S$ and $P$ interference terms obtained setting the $D$-wave contributions to zero.}
\label{tab:noDwavemom}
\end{table}

In~\Tab{tab:noDwavemom}, we list all the moments that have both an $S$ and $P$ wave contribution and their experimental values\,\footnote{
The moment $\av{M_{0\re}}$ was already discussed in the previous section. We give here a different value, obtained by choosing the simplest combination of moments $\tilde\Gamma$ under the assumption that the $D$-wave is negligible. The value quoted in~\Tab{tab:noDwavemom} is different from~\Eq{eq:momexp}, but compatible, given the large uncertainties.
}. 
For the theoretical values, we assume $\alpha=1$ for the $P$-wave and we quote as an uncertainty the spread of values from the different $S$-wave models. It turns out that  the lower and upper values always come from the models 3 and 4, respectively. One finds a good agreement for the first three moments in~\Tab{tab:noDwavemom}, whereas the last three moments are less well reproduced but still compatible within the large uncertainties.

Once we neglect $D$-wave contributions, we can also split $\av{M_0}$ between the $S$-wave only part $\av{M_S}$ defined in~\Eq{eq:MS} and the $P$-wave part: 
\begin{equation}
    \av{M_P} \equiv \tau_B\langle|\widehat{A}_0^L|^2+|\widehat{A}_0^R|^2 \rangle = \tau_B\frac{1}{6} (2\tilde\Gamma_{1} + 3 \sqrt{5}\tilde\Gamma_{3} + 2 \sqrt{5}\tilde\Gamma_{6})  \ ,
\end{equation}
for which we find, using $\alpha=1$,
\begin{equation}
    \av{M_P} = 0.75 \times 10^{-8}\,, 
\end{equation}
and higher predictions for larger $\alpha$ values. Comparing with the experimental value
\begin{equation}
    \av{M_P}_{\rm exp}= (-0.52 \pm 0.87) \times 10^{-8} \ ,
\end{equation}
suggests once again a small value of $\alpha$  if $D$ waves can be neglected. 
For the $S$-wave contribution, we already calculated the moment $\av{M_S}$ in~\Eq{eq:MS}, where we find good agreement with the measurement given in~\Tab{tab:noDwavemom}. 

We conclude by considering the two $P$-wave moments already discussed in \Reff{Descotes-Genon:2019bud}:
\begin{eqnarray}
\av{M_{||}}\equiv \tau_B \langle| \widehat{A}_{||}^L|^2+ |\widehat{A}_{||}^R|^2\rangle &=& 0.21 \times 10^{-8} \ ,\\
\av{M_\perp}\equiv \tau_B\langle|\widehat{A}_\perp^L|^2+|\widehat{A}_\perp^R|^2\rangle  &=& 0.11 \times 10^{-8} \ ,
\end{eqnarray}
where we quote our results using $\alpha=1$. As in \Reff{Descotes-Genon:2019bud}, we compare this with the $S$ and $D$-wave free combination of moments:
\begin{eqnarray}
\av{M_{||, \perp}} =  \tau_B \frac{1}{36} (5\tilde\Gamma_{1} - 7 \sqrt{5}\tilde\Gamma_3 + 5 \sqrt{5}\tilde\Gamma_6 - 35\tilde\Gamma_{8} \mp 5\sqrt{15}\tilde\Gamma_{19} \pm35 \sqrt{3}\tilde\Gamma_{21}) \ ,
\end{eqnarray}
where the upper (lower) sign applies to $|| (\perp)$. Using the experimental data gives
\begin{eqnarray}
\av{M_{||}}_{\rm exp} &=& (1.07 \pm 1.13) \times 10^{-8} \ ,\\
\av{M_\perp}_{\rm exp} &=& (0.94 \pm 1.06) \times 10^{-8} \ .
\end{eqnarray}
On the other hand, when neglecting the $D$-wave contribution, we find also a different combination of moments that probes the same underlying amplitudes: 
\begin{eqnarray}
\av{M_{||}}_{\rm exp} =  \tau_B \frac{1}{3} (\tilde\Gamma_{1} + \sqrt{5}\tilde\Gamma_{6} - \sqrt{15}\tilde\Gamma_{19})&=& (0.61 \pm 0.74) \times 10^{-8} \ ,\\
\av{M_\perp}_{\rm exp} = \tau_B \frac{1}{3} (\tilde\Gamma_{1} + \sqrt{5}\tilde\Gamma_{6} + \sqrt{15}\tilde\Gamma_{19}) &=& (1.76 \pm 0.72) \times 10^{-8} \ .
\end{eqnarray}
We observe that our results agree with both these experimental values, and also that they agree with each other within their still large uncertainties. Again this suggests that at the current level of uncertainty, the $D$ wave contribution can be safely neglected.

%%%%%%%%%%%%%%%%%%%%%%%%%%%%%%%%%%%%%%%%%%%%%%%%%%%%%%%%%%%%%%%%%%%%%%%%%%%%%%
\section{Conclusions}
\label{sec:Conclusions}

Exclusive $B$-meson decays can be used as powerful tests of the Standard Model, provided that accurate theoretical predictions can be made. These predictions require the knowledge of certain non-perturbative hadronic matrix elements, such as form factors. Among the many approaches to the calculation of form factors, LCSRs in various versions have been extensively used, and are currently advantageous in some respects. One such advantage of the LCSRs with $B$-meson distribution amplitudes is that they provide form factors of the $B$-meson transition into dimeson state, as was demonstrated in \Reff{Cheng:2017smj} for the $B\to\pi\pi$ form factors and applied in \Reff{Descotes-Genon:2019bud} to the $B\to K\pi $ form factors, focusing 
in both cases on the $P$-wave final states.

In this article, we have extended the work of~\Reff{Descotes-Genon:2019bud} and derived LCSRs for the $B\to K\pi$ transitions with an $S$-wave $K\pi$ state.
These sum rules provide integral relations between
the convolution of the $K\pi$ scalar form factor with a  $B\to K\pi$ form factor on one side, and the OPE of a specific correlation function expressed in terms of $B$-meson LCDAs on the other side.
On the OPE side of the sum rules, we computed the two- and three-particle contributions up to twist four, and determined the optimal threshold parameter $s_0$ from a separate QCD sum rule.
On the hadronic side, we  considered a consistent dispersive model \cite{VonDetten:2021rax} that takes into account the interference of the $K\pi$ and $K\eta'$ $S$-wave states, and addresses the difficulties of describing the $S$-wave spectrum.

We have studied the implications of the resulting sum rules for the parameters of the $B\to K\pi$ form factors. The form factors inferred from the LCSRs are valid in the 
phenomenologically relevant large-recoil region, i.e. $q^2 \leq 8-10 \GeV^2$.
At the same time, the LCSRs reliably constrain the region 
of the $K\pi$ invariant mass  from the threshold up to $m_{K\pi} \approx 1.4\GeV$, which is the region below $m_{K\pi}^2 < s_0$, 
where the spread between the models of $K\pi$ form factors used in 
our analysis is inessential.

We have then applied our results for the $S$-wave $B\to K\pi$ form factors to the $B\to K\pi\ell\ell$ decay, combining them with the earlier results of \Reff{Descotes-Genon:2019bud} for the $P$-wave $B\to K\pi$ form factors. 
Concerning the impact of our results in the $K^*(980)$ region, we can predict accurately the branching ratio if we use our previous results for the $P$-wave (setting the model parameter $\alpha=1$). The contributions from the $S$-wave in this region, measured by $F_S$ is found rather small for all $q^2$ values, in agreement with some of the LHCb measurements available.
We reiterate that we have focused here on the ``local'' form factors involved in $B\to K\pi\ell\ell$. A dedicated study of the non-local (``charm-loop'') contributions to this decay is required, although recent studies suggest that they are small at least in the $K^*(892)$ case~\cite{Gubernari:2020eft}. In any case, the non-local effect is proportional to the local form factors at the leading order in an Operator Product Expansion, and our numerical analysis has relied on this approximation.

We have then considered the LHCb measurements of the $B\to K\pi\ell\ell$ branching ratio and angular observables  for a $K\pi$ invariant mass around the $K^*(1410)$ and $K^*_0(1430)$ resonances.
The $S$-wave contribution is larger in this region, leading to results for the branching ratio in the right ballpark, but with an unsatisfatory $q^2$-dependence. We understand it as being the sign that the initial model for the scalar $K\pi$ form factor could be further refined to help reproducing the $B\to K\pi$ data more accurately.
In particular, most of the four models yield a similar contribution from the $K^*_0(1430)$ resonance, which could be modified by tuning some of the parameters of the model (presence and characteristics of the $K^*_0(1950)$ resonance, high-energy behaviour of the source term). This would require further data to constrain efficiently our model. We illustrated how we could extract further information from the angular observables, considering first observables that do not involve the $D$ wave, before discussing the larger set of observables that could be predicted if we neglect $D$-wave contributions.
Keeping our $P$-wave model with $\alpha=1$ and the four $S$-wave models inspired by \Reff{VonDetten:2021rax}, we found a good agreement with the data for some of the moments and a reasonable compatibility for the others, given the large experimental uncertainties associated with their measurements. A complete description of the $B\to K\pi\ell\ell$ would obviously require a parametrisation of the $D$-wave contribution, whose size is only loosely constrained by the LHCb data.

Our description of the $B\to K\pi$ form factors begins at the $K\pi$ production threshold, includes the $K^*(892)$ region, and extends to  the vicinity of the first excited resonances $K^*(1410)$ and $K^*_0(1430)$, allowing to make predictions to branching fractions and observables in this entire kinematic region. It would thus be very beneficial to perform a full and detailed angular analysis of the $B\to K\pi\ell\ell$ decay,  not only around the $K^*(892)$ (to understand better the $S$-wave contribution in this region), and  $K^*_0(1430)$ (to confirm the experimental results~\cite{LHCb:2016eyu} that we have used here), but also  for  a broader range of $K\pi$ invariant masses. Such measurements will provide very useful data to restrict our models in a much more precise way,
helping to clarify the questions left open by the existing $B\to K\pi\ell\ell$ measurements. 

One important question is the role of the $P$-wave excited resonances.
According to~\Reff{LHCb:2016ykl}, there is no evidence for a non-resonant $P$-wave component in the region around $K^*(892)$. 
In terms of a hadronic dispersion relation, a non-resonant background in the lower mass region is formed by the contributions of the heavier resonances. 
So far, following~\Reff{Descotes-Genon:2019bud}, we have only included the $K^*(1410)$ in our $P$-wave model.
Hence, the observation by LHCb  suggests a strong suppression of its contribution. Looking at the data in ~\Reff{ParticleDataGroup:2022pth} this suppression can be understood, taking into account the smallness of the  partial width
\eqa{
\Gamma(K^*(1410)\to K\pi)
&=&
\mbox{BR}(K^*(1410)\to K\pi)\times \Gamma^\text{tot}_{K^*(1410)}
\nonumber\\
&\simeq&
6.6\%\times 232~\mbox{MeV}=15.3~\mbox{MeV}\ ,
}
resulting in a suppressed  $K^*(1430)K\pi$ strong coupling\,\footnote{
For comparison, for the scalar resonance 
$\Gamma(K^*(2430)\to K\pi) \simeq  93\%\times 270~\mbox{MeV} =251 \MeV$.
}.
However, according to~\Reff{ParticleDataGroup:2022pth}
there is a heavier vector resonance $K^*(1680)$, with a larger total and partial width: 
\eqa{
\Gamma(K^*(1680)\to K\pi)
&=&
\mbox{BR}(K^*(1680)\to K\pi)\times \Gamma^\text{tot}_{K^*(1680)}
\nonumber\\
&\simeq&
38.7\%\times 322~\mbox{MeV}=
124.6 \MeV \ ,
}
whose influence on both regions of $K^*(890)$ and $K^*(1410)$  still has to be clarified.

All this shows the necessity for a more detailed partial-wave analysis of the $B\to K\pi\ell\ell$ 
differential distribution in the $K\pi$ invariant mass.
This could lead to a consistent picture of the contributions from higher resonances to the $B\to K\pi\ell\ell$ decay, and to a deeper understanding of the dynamics of $b\to s\ell\ell$ transitions that remain under intense theoretical and experimental scrutiny. 

We note that in addition to the FCNC $B\to K\pi\ell\ell$ decays, our method and some of our  results are directly applicable to other modes of current interest.
First, we can obtain LCSRs for the $B\to \pi\pi$ $S$-wave form factors using the OPE expressions derived here and taking the $m_s\to 0$ limit, although a separate dedicated 
model of the pion scalar form factor will be needed to describe  the dynamics of the di-pion state. These form factors are important hadronic inputs for a detailed partial-wave  analysis of the semileptonic $B\to \pi\pi\ell\nu $ decay relevant for $V_{ub}$ extraction and for the Cabibbo-suppressed FCNC $B\to \pi\pi\ell\ell$ 
decays.
Furthermore, our results for $B\to K\pi$ form factors apply to other decay modes of interest, 
including the rare $B\to K\pi\nu\bar\nu$ decays, the non-leptonic $B$ decays to three or more hadrons such as $B\to K\pi\pi$, or searches for  ALPs or dark photons through 
the $B\to K\pi\,a$ and $B\to K\pi \gamma'$ decays. 
We thus conclude that  a combination of QCD-based LCSRs  with a dispersive approach to hadronic interactions 
substantially enlarges the set of exclusive $B$ decays that can be used to probe the Standard Model and to look for New Physics.

%%%%%%%%%%%%%%%%%%%%%%%%%%%%%%%%%%%%%%%%%%%%%%%%%%%%%%%%%%%%%%%%%%%%%%%%%%%%%%
\section*{Acknowledgments}

S.D.G acknowledges supports from the European Union’s Horizon 2020 research and innovation programme under the Marie Sk\l odowska\,-\,Curie grant agreement No 860881-HIDDeN.

The research of A.K. is supported by the Deutsche Forschungsgemeinschaft (DFG, German Research Foundation) under the grant 396021762 - TRR 257 “Particle Physics Phenomenology after the Higgs Discovery''.

J.V. acknowledges funding from the European Union's Horizon 2020 research and innovation programme under the Marie Sk\l odowska-Curie grant agreement No 700525
`NIOBE', from the Spanish MINECO through the ``Ram\'on y Cajal'' program RYC-2017-21870,
the “Unit of Excellence Mar\'ia de Maeztu 2020-2023” award to the Institute of Cosmos Sciences (CEX2019-000918-M)
and from the grants PID2019-105614GB-C21 and 2017-SGR-92, 2021-SGR-249 (Generalitat de Catalunya).

\newpage

%%%%%%%%%%%%%%%%%%%%%%%%%%%%%%%%%%%%%%%%%%%%%%%%%%%%%%%%%%%%%%%%%%%%%%%%%%%%%%%%%%%%
%%%%%%%%%%%%%%%%%%%%%%%%%%%%%%%%%%%%%%%%%%%%%%%%%%%%%%%%%%%%%%%%%%%%%%%%%%%%%%%%%%%%
%%%%%%%%%%%%%%%%%%%%%%%%%%%%%%%%%%%%%%%%%%%%%%%%%%%%%%%%%%%%%%%%%%%%%%%%%%%%%%%%%%%%
%%%%%%%%%%%%%%%%%%%%%%%%%%%%%%%%%%%%%%%%%%%%%%%%%%%%%%%%%%%%%%%%%%%%%%%%%%%%%%%%%%%%

\appendix

\allowdisplaybreaks

%%%%%%%%%%%%%%%%%%%%%%%%%%%%%%%%%%
\section{OPE expressions for the light-cone sum rules}
\label{app:OPE}

We present here the OPE functions appearing on the r.h.s. of the sum rules in~\Eq{eq:AllLCSRs}, including contributions from two- and three-particle $B$-meson DAs up to twist-4. Their definitions and the {\bf Model I} adopted for their shape are presented and discussed in Appendix~B of~\Reff{Descotes-Genon:2019bud}.

The generic form of the OPE function for any form factor is written as
\eqa{
&&\hspace{-15mm} \S_i^{\text{OPE}}(q^2,s_0,M^2) = (m_s-m_d)\times \nonumber\\
&&\sum_{n\ge 0} \ \frac{f_B m_B}{(M^2)^n} \ \bigg\{ \int\limits_0^{\sigma_0} d\sigma~e^{-s(\sigma)/M^2}
I_{i,n}(\sigma)
+ \sum_{\ell\ge 0}
\eta(\sigma_0) \D_\eta^\ell[I_{i,n+\ell+1}](\sigma_0) \ e^{-s_0/M^2}
\bigg\}
\ ,
\label{eq:FOPE}
}
where $i=\{5,0,t,T\}$ and we are using the notation $\S_T^{\text{OPE}}\equiv \S_0^{T,\text{OPE}}$.
The functions $I_{i,n}$ consist of two- and three-particle contributions:
\eq{
I_{i,n}(\sigma) = I_{i,n}^{(2)}(\sigma) + 
\int _0^{m_B \sigma}  \!\!d\omega_1 \int_{m_B\sigma -\omega_1}^\infty \!\! \frac{d\omega_2}{\omega_2}\, I_{i,n}^{(3)}(\sigma,\omega_1,\omega_2)\,,
}
with $n\leq 3$ in the adopted twist-4 approximation. The 
variable $\sigma$ used in~\Eq{eq:FOPE} is related to the invariant
 $s=k^2$ via:
\eqa{
\hat s(\sigma) = \sigma - \frac{\sigma \hat q^2 -\hat m_s^2}{\bar \sigma}\ , 
\quad
\sigma(s) = \frac12 \bigg\{ 1+\hat s - \hat q^2 - \sqrt{(1-\hat s+\hat q^2)^2 - 4(\hat q^2-\hat m_s^2)}  \bigg\}\ ,
}
where $\bar{\sigma}\equiv 1-\sigma$, 
$\hat s \equiv s/m_B^2$, $\hat q^2 \equiv q^2/m_B^2$, $\hat m_s \equiv m_s/m_B$ and $\sigma_0\equiv \sigma(s_0)$.
The operator  $\D^{\ell}_\eta$ in~\Eq{eq:FOPE} is defined by acting $\ell=0,1,2,..$ times on a generic function $F(\sigma)$:
\eqa{
\D_\eta^0 [F](\sigma_0) = F(\sigma_0)\ ;
\quad
\D^{1}_\eta [F](\sigma_0) = \frac{d}{d\sigma} \big[\eta(\sigma) F(\sigma)\big] \bigg|_{\sigma=\sigma_0} \ ;
\nonumber \\
\D_\eta^2 [F](\sigma_0) = \frac{d}{d\sigma} \bigg[\eta(\sigma) \frac{d}{d\sigma} [\eta(\sigma) F(\sigma)]\bigg] \bigg|_{\sigma=\sigma_0}\ ;
\quad \text{etc},
\label{eq:Deta}
}
with
\eq{
\eta(\sigma) = \frac{\bar\sigma^2}{\bar\sigma^2 m_B^2-(q^2-m_s^2)}\ . 
}
The full expressions for the coefficients $I_{i,n}^{(2)}(\sigma)$ and
$I_{i,n}^{(3)}(\sigma,\omega_1,\omega_2)$ are given in 
the ancillary Mathematica file named `{\tt OPEcoefficientsSwave.m}' (see below for more details).
As a sample, we present here only the results for the two-particle coefficients $I_{i,n}^{(2)}(\sigma)$ for $m_s = 0$ and $q^2=0$: 
\eqa{
I_{5,0}^{(2)}(\sigma) &=& m_b\bigg(- \frac{m_B^2 \phi_+}2 - \frac{2\, g_+}{\bar\sigma^2} + \frac{m_B \bar\Phi_\pm}{\bar\sigma} \bigg) \ , \quad
I_{5,1}^{(2)}(\sigma) = \frac{2 m_b m_B^2 g_+}{\bar\sigma}\ , \quad
I_{5,2}^{(2)}(\sigma) = I_{5,3}^{(2)}(\sigma) = 0\ ,
\nonumber\\
I_{0,0}^{(2)}(\sigma) &=& -m_B \phi_+
+\frac{\bar\Phi_\pm}{\bar\sigma} \ , \quad
I_{0,1}^{(2)}(\sigma) = \frac{4 m_B g_+}{\bar\sigma}\ , \quad
I_{0,2}^{(2)}(\sigma) = I_{0,3}^{(2)}(\sigma) = 0\ ,
\nonumber\\
I_{t,0}^{(2)}(\sigma) &=&  - \frac{m_B^3 \bar\sigma \phi_+}{2} + \frac{2m_B\, g_+}{\bar\sigma} + \frac{m_B^2 \bar\Phi_\pm}{2} \ , \quad
I_{t,1}^{(2)}(\sigma) =  2 m_B^3\,g_+\ , \quad
I_{t,2}^{(2)}(\sigma) = I_{t,3}^{(2)}(\sigma) = 0\ ,
\nonumber\\
I_{T,0}^{(2)}(\sigma) &=& - \frac{\phi_+}{\bar\sigma} \ , \quad
I_{T,1}^{(2)}(\sigma) =  \frac{4\, g_+}{\bar\sigma^2} \ , \quad
I_{T,2}^{(2)}(\sigma) = 
I_{0,3}^{(2)}(\sigma) = 0\ ,
\label{eq:OPEfunctions}
}
where for brevity we have omitted the arguments of the LCDAs, i.e. $\phi_+\equiv \phi_+(m_B\sigma)$, etc.
These results can be easily extracted from the ancillary file. 
For example, the expression for $I_{t,1}^{(2)}(\sigma)$ given in~\Eq{eq:OPEfunctions} is obtained by typing in a Mathematica notebook:

\bigskip

{\tt ISWt[2,1]/.(<<"OPEcoefficientsSwave.m")/.\{ms -> 0, q2 -> 0\}}

\bigskip

\noindent The arguments in brackets are such that, for example, $I^{(k)}_{t,n}$={\tt ISWt[k,n]}.
The expressions for the three-particle contributions contain an additional combination of variables denoted as $u= (\sigma m_B-\omega_1)/\omega_2$ and  $\bar u \equiv 1-u$.

%%%%%%%%%%%%%%%%%%%%%%%%%%%%%%%%%%%%%%%%%%%%%%%%%%%%%%

\section{Models of form factors} 
\label{app:resmodel}

In this appendix, we discuss models which could be considered for the 
S-wave $K\pi$ and $B\to K\pi$ form factors. In the first subsection,
 we find it illustrative to consider a resonance model 
 similar to the one employed in \Reff{Descotes-Genon:2019bud}, even though the Breit-Wigner description fails to give an accurate description of   the strange scalar sector at low masses. In the second subsection, we provide further information concerning the two-channel dispersive model
 that we chose.

%%%%%%%%%%%%%%%%%%%%%%%%%%%%%%%%%%%%%%%%%%%%%%%%%%%%%%%%%%
\subsection{Breit-Wigner parametrization} \label{app:BWparam}

\subsubsection{The $K\pi$ scalar form factor}

The resonance ansatz yields the following description for the matrix element leading to the $K\pi$ scalar and vector form factors:
\eq{
\langle K^-(k_1) \pi^+(k_2) | \bar s \gamma^\mu d | 0 \rangle = 
\sum_R  BW_R(k^2) \langle K^-(k_1) \pi^+(k_2) | R(k) \rangle \langle R(k) | \bar s \gamma^\mu d | 0 \rangle
\ .
\label{eq:ResSumKpi}
}
In the following, we will focus on the scalar form factor $f_0$ in the Lorentz decomposition of this matrix elements, so that the relevant part of the  sum includes the scalar resonances $R = \{K^\star_0(700),K^\star_0(1430)\}$.
The third factor in the right-hand side is related to the $R$ decay constants $f_R$:
\eq{
\langle R(k) | \bar s \gamma^\mu d | 0 \rangle = f_R\,k_\mu 
\label{eq:decayconstant}
}
and the phases of the states $\langle R(k) |$ are defined so that they are real and positive.
The second factor in (\ref{eq:ResSumKpi}) is related to the strong coupling of the resonances to the $K^- \pi^+$  state:
\eq{
g_{RK\pi} \,e^{i\varphi_R}  = \ \langle K^- \pi^+ | R(k) \rangle = -\sqrt2\ \langle \bar K^0 \pi^0 | R(k) \rangle\ ,
\label{eq:gRKpi}
}
where  we include a phase $\varphi_R$ related to the normalization of  the hadronic states. Later on,
this phase will be merged with the relative phases between the separate resonance contributions to the form factors. We neglect any
$k^2$-dependence of the strong couplings although this assumption, well-founded for narrow resonances, might prove more debatable for broad ones.
The first factor in (\ref{eq:ResSumKpi}) is an energy-dependent Breit-Wigner function:
\eq{
BW_R(s) = \frac1{m_R^2 - s -i\sqrt{s}\, \Gamma_R(s)}\ ,
}
with
\eq{
\Gamma_{R}(s)
=\Gamma^{\rm tot}_{R} \left[\frac{\lambda_{K\pi}(s)}{\lambda_{K\pi}(m_{R}^2)}
\right]^{1/2}\frac{m_{R}^3}{s^{3/2}} \,\theta\big(s- s_{\rm th} \big)\,.
\label{eq:Gamma(s)}
}
The strong coupling $g_{RK\pi}$ is determined by the total width of the resonance $R$,
\eq{
\Gamma^{\rm tot}_{R}  = \frac{g_{RK\pi}^2}{16\pi} \frac{\lambda_{K\pi}^{1/2}(m_{R}^2)}{m_{R}^3} \, \frac{1}{\B(R\to K^- \pi^+)}\, .
\label{eq:Gammatot}
}

Plugging Eqs.~(\ref{eq:gRKpi})~and~(\ref{eq:decayconstant}) into~\Eq{eq:ResSumKpi} and comparing to~\Eq{eq:KpivectFF}, we get for the scalar form factor:
\eq{
f_0(s) =  \frac{1}{m_K^2-m_\pi^2} \sum_R \frac{m_{R}^2 \,f_R\, g_{RK\pi} \ e^{i\phi_R(s)}}{m_{R}^2 - s -i \sqrt{s}\, \Gamma_{R}(s)}\ .
\label{eq:f0model}
}
Even though we do not attempt at using this model for phenomenological purposes, it may be interesting to estimate some of its parameters.
From \Reff{ParticleDataGroup:2022pth} we have $\B(K^*_0(700) \to K^- \pi^+)= 2/3$ in the isospin-limit prediction, and $\B(K^*_0(1430) \to K^- \pi^+)\simeq 2/3 \times 0.93 = 0.62$. We also take 
$M_{K^*_0(700)}=0.68\pm 0.05$ GeV, $\Gamma^{\rm tot}_{K^*_0(700)}=0.30 \pm 0.04$ GeV, $M_{K^*_0(1430)}=1.425\pm 0.050$ GeV,
$\Gamma_{K^*_0(1430)}=0.270\pm 0.080$ GeV, so that we obtain for the central values of the couplings
$g_{K^*_0(700)K\pi}=4.75\ {\rm GeV}$ and $g_{K^*_0(1430)K\pi}=3.72\ {\rm GeV}$.

A fit to such Breit-Wigner parametrisations (for both $S$ and $P$-waves) was performed by the Belle collaboration using the $\tau\to K_S\pi\nu_\tau$ data~\cite{Belle:2007goc}. The resonances included in the fits were  either $K^*_0(700)$, $K^*(892)$ and $K^*(1410)$, or $K^*_0(700)$, $K^*(892)$ and $K^*_0(1430)$. Each of the two fits were limited to three resonances as it was not possible to obtain a satisfactory fit with all four of them. 
As pointed out in \Reff{VonDetten:2021rax}, these descriptions may be qualitatively useful, but do not meet some model-independent constraints such as the value of the phase imposed by unitarity in the elastic regime and the Callan-Treiman theorem.We will thus not use this description for phenomenological studies, but we find it illustrative to describe 
how this parametrisation could be extended in the case of $B\to K\pi$ form factors.

\subsubsection{$B\to K\pi$ form factors for the $K\pi$ S-wave}

In the case of the $B\to K\pi$ form factors and along the same lines we have:
\eq{
\langle K^-(k_1) \pi^+(k_2) | \bar s \Gamma b |\bar{B}^0(q+k) \rangle = 
\sum_R  BW_R(k^2) \langle K^-(k_1) \pi^+(k_2) | R(k) \rangle \langle R(k) | \bar s \Gamma b |\bar{B}^0(q+k) \rangle\ ,
\label{eq:ResSum}
}
for a generic Dirac structure $\Gamma$. Once again we will focus on the contributions to the $K\pi$ S-wave component of this matrix element, and thus the resonances $R$ considered are scalar.
The third factor in the right-hand side is related to $B\to R$ form factors, defined as:
\eqa{
  \langle R(k) | \bar s \gamma_\mu\gamma_5 b |B(q+k) \rangle
  &=&i[F_{R,+}\,(q^2)(q+2k)_\mu + F_{R,-}(q^2)\, q_\mu]\\
 \langle R(k) | \bar s \sigma_{\mu\nu}\gamma_5 q^\nu b |B(q+k) \rangle
  &=&F_{R,+}^{T}\,(q^2)\times \left[q^2 k_\mu -(k \cdot q) q_\mu\right]
}
Plugging~\Eq{eq:gRKpi} into~\Eq{eq:ResSum} and defining
\eq{
\G_{R,0}(q^2)=F_{R,+}(q^2)\,, \quad \G_{R,t}(q^2)=
\frac{1}{2}[(m_B^2-m_R^2)F_{R,+}(q^2)+q^2 F_{R,-}(q^2)]\,,\quad
\G_{R,T}(q^2)=\frac{1}{2} F_{R,+}^{T}(q^2)\,.
}
we obtain the following expression for the $S$-wave $B\to K\pi$ form factors:
\eq{
F_i^{(\ell=0)}(s,q^2) =
\sum_R  \frac{X_{R,i}(s,q^2)\, g_{RK\pi} \, \G_{R,i}(q^2) \,e^{i\phi_R(s)}}{m_R^2 - s - i\sqrt{s}\, \Gamma_R(s)}
\label{eq:FFmodels1}
}
with $i=\{ 0,t,T \}$, and the weights
\begin{equation}
X_{R,0} =  \frac{\sqrt{\lambda}}{\sqrt{q^2}}  \ , \qquad X_{R,t} = \frac{2}{\sqrt{q^2}}\ ,\qquad
X_{R,T} =  \sqrt{\lambda}\sqrt{q^2} 
\label{eq:Ys}
\end{equation}
depending on $s$ and $q^2$ also implicitly through the function $\lambda\equiv \lambda(m_B^2,q^2,s)$.
As in \Reff{Descotes-Genon:2019bud}, we assume ansatz that the phase cancellation between $f_0$ and the form factors $\F_{R,i}$ that follows from unitarity happens at the level of the individual resonances~\cite{Cheng:2017smj}, so that:
\eq{
\tan \big[ \delta_{K\pi}^0(s) - \phi_R(s) \big] = \frac{\sqrt{s}\, \Gamma_R(s)}{m_R^2 - s}
\ ,
\label{eq:phasecondition}
}
where $\delta_{K\pi}^0(s)$ as the phase of the $K\pi$ form factor:
\eq{
f_0(s) = |f_0(s)| e^{i\delta_{K\pi}^0(s)}\ .
}
Note that this assumption also implies that the phases $\phi_R(s)$ are
$q^2$-independent.

The sum rules in~\Eq{eq:AllLCSRs} can then be reexpressed as:
\eq{
\sum_R \G_{R,i}(q^2) H_R(s_0,M^2) = {\cal S}_i^{\rm OPE}(q^2,\sigma_0,M^2)\ ,
\label{eq:srBW}
}
with
\eq{
H_R(s_0,M^2) = \frac{3}{16\pi^2} \int_{s_{\rm th}}^{s_0} ds \ e^{-s/M^2}\  \frac{(m_K^2-m_\pi^2) g_{RK\pi}\,\lambda_{K\pi}^{1/2}(s) \lambda(s) \, |f_0(s)|}{s \sqrt{(m_R^2-s)^2 + s\,\Gamma_R^2(s)}}\ .
\label{eq:IR}
}

Following \Reff{Cheng:2017smj,Descotes-Genon:2019bud}, we could parametrize the $q^2$-dependence of the $B\to R$ form factors $\G_{R,i}^{(T)}(q^2)$ entering~\Eq{eq:FFmodels1} with a standard $z$-series expansion and work out the consequences of the sum rules of~\Eq{eq:srBW}. We  refrain from following this path as we adopt a different model, better suited for the description of the complicated dynamics of the $K\pi$ $S$-wave.

%%%%%%%%%%%%%%%%%%%%%%%%%%%%%%%%%%%%%%%%%%%%%%%%%%%%%%%%%%
\subsection{Two-channel dispersive model for the $K\pi$ scalar form factor}\label{app:dispmodel}

For completeness, we briefly recall the formalism developed in \Reff{VonDetten:2021rax} and used to obtain the scalar $K\pi$ form factor in~\Sec{sec:ffmodel}. Due to the small impact of the $K\eta$ channel, only two channels, $K\pi$ and $K\eta'$ ($a=1,2$), are considered. The scalar form factors for both channels gathered in a two-component vector ${\mathbf f_0}$ are obtained as
\begin{equation}
{\mathbf f_0}(s)=\Omega(s)[{\bf 1}-V_R(s)\Sigma(s)]^{-1}M(s) \equiv B(s)M(s)\ .
\end{equation}
In this equation, the Omn\`es function is given by
\begin{equation}    \Omega(s)=\left(\begin{array}{cc}
        \Omega_{11}(s) & 0  \\
        0 & 1
    \end{array}\right)\ ,
    \qquad \Omega_{11}=\exp\left(\frac{s}{\pi}\int_{s_{\rm th}}^\infty dz\frac{\delta_0(s)}{z(z-s)}\right)\ ,
    \end{equation}
with $s_{\rm th}=(m_K+m_\pi)^2$. The phase $\delta_0$ is obtained from the low-energy $K\pi$ scattering data constrained with dispersion relations~\cite{Pelaez:2016tgi}.

The dressed loop operator $\Sigma$ is obtained from another dispersion integral
\begin{equation}
    \Sigma_{ab}=\frac{s}{2i\pi}\int_{s_{\rm th}}^\infty dz\frac{{\rm disc}\ \Sigma_{ab}(s)}{z(z-s)}\ ,
\qquad {\rm disc}\ \Sigma_{ab}=\Omega_{ac}^\dagger {\rm disc}\ G_{cc}\Omega_{cb}\ ,
\end{equation}
with the discontinuity of the loop operator in the case of two-particle states:
\begin{equation}
{\rm disc}\ G_{cc}=2i\rho_c(s)\ , \qquad \rho_c(s)=\frac{\sqrt{\lambda_{ij}(s)}}{16\pi s} \ ,
\end{equation}
where $\lambda_{ij}(s)$ is the K\"all\'en function corresponding to the two particles of masses $m_i$ and $m_j$ involved in the channel $c$.
The interaction potential reads
\begin{equation}
    V_{R,ab}(s)=\sum_r g_a^{(r)} \frac{s-s_{K\eta}}{(s-\tilde{M}^2_{(r)})(s_{K\eta}-\tilde{M}^2_{(r)})}
    g_b^{(r)}\ ,
\end{equation}
where $s_{K\eta}$ is chosen at $(m_K+m_\eta)^2$ and the masses $\tilde{M}^2_{(r)}$ of the resonances and their couplings $g_i^{(r)}$ to the $\pi K$ and $\eta' K$ channels are obtained from a fit to scattering data.

Finally, the source term for the scalar form factor is given by
\begin{equation}
\label{eq:Msource}
    M_a=\sum_{k=0}^{k_{\rm max}} c_a^{(k)}s^k
    -\sum_r g_a^{(r)} \frac{s-s_{K\eta}}{(s-\tilde{M}^2_{(r)})(s_{K\eta}-\tilde{M}^2_{(r)})}
    \alpha^{(r)}\ .
\end{equation}
The coefficients $c^{(k)}$ and the resonance couplings $\alpha^{(r)}$ depend on the process considered. The order of the polynomial $k_{\rm max}$ is also part of the model, potentially improving the description at intermediate energies at the expense of changing the high-energy behaviour.

In~\Reff{VonDetten:2021rax}, the authors
consider the scalar $K\pi$ form factor\footnote{Note that \Reff{VonDetten:2021rax} defines both $f_0$ and $f_+$ from
$\bar{K}^0 \pi^-(p_\pi)$ whereas we use $K^-\pi^+$. Due to the isospin relations in~\Eq{eq:isoKpi}, the two sets of definition are equivalent up to an overall (-1) factor.}
normalised at zero:
$$
\bar{f}_0 (s) \equiv {f_0(s)}/{f_0(0)}\,,
$$ where the normalisation is $f_0(0)=f_+(0)$. They determine the parameters of the model in the following way.
First, the masses $\tilde{M}_{(r)}$ of the resonances and their couplings $g_a^{(r)}$ to the $\pi K$ and $\eta' K$ channels are determined from a fit to scattering data~\cite{Pelaez:2016tgi}. Afterwards,  $\tau^-\to K_S\pi^-\nu_\tau$ data from the Belle experiment~\cite{Belle:2007goc} is exploited in a joint fit of their parametrisation of the normalised scalar $K\pi$ form factor $\bar{f}_0 (s)$ together with a parametrisation of the vector $K\pi$ form factor. The latter is based on Resonance Chiral Theory~\cite{Ecker:1988te} and it has a similar structure as the Model II considered in \Reff{Descotes-Genon:2019bud}, although with a slower decrease at large energies. Four different assumptions are considered for the polynomial term in~\Eq{eq:Msource}, leading to four different descriptions of the scalar form factor.

%%%%%%%%%%%%%%%%%%%%%%%%%%%%%%%%%%%%%%%%%%%%%%%%%%%%%%%%%%%%%%%%%%%%%%%%%%%%%%%%%%%%
\section{Two-point sum rule in the scalar channel}
\label{app:2ptSR}

Here we estimate the duality threshold $s_0$ for the $S$-wave $K\pi$ state in the LCSRs of~\Eq{eq:AllLCSRs}. Following the procedure adopted in \cite{Cheng:2017smj,Descotes-Genon:2019bud}, we use the QCD sum rule for the two-point correlation function of the interpolating currents:
\eq{
\Pi_S(q^2)=i\int d^4x e^{iqx}\langle 0|T\{j_S(x),j^\dagger_S(0) \}|0\rangle\,,
\label{eq:2pointcorr}
}
where $j_S$ is the scalar current with strangeness defined in (\ref{eq:scalcurr}).
Note that the above correlation function contains no Lorentz indices and 
hence directly depends on $q^2$.

A QCD sum rule for this correlation function is usually derived 
(see e.g. \cite{Jamin:1994vr,Chetyrkin:1996xa}),
from the doubly differentiated  dispersion relation in the variable $q^2$:
\eq{
\frac{d^2}{d(q^2)^2}\Pi_S(q^2)=
\frac{2}{\pi}\int\limits_0^\infty\,ds \frac{\rm{Im} \Pi_S(s)}{(s-q^2)^3}\,.
\label{eq:2deriv}
}
After Borel transformation $q^2\to M^2$ we have:
\eq{
\Pi_S^{''}(M^2)\equiv {\cal B}_M\bigg[\frac{d^2}{d(q^2)^2}\Pi_S(q^2)\bigg]=
\frac{1}{\pi M^4}\int\limits_0^\infty\!ds \,e^{-s/M^2}\rm{Im} \,\Pi_S(s)\,.
\label{eq:Borel2pt}
}
At sufficiently large $M^2$, the l.h.s. of this relation is calculated 
from the OPE in terms of perturbative part and vacuum condensate contributions.
The integral on r.h.s. is taken  
over the spectral density 
$$\rho_S(s)=\frac{1}{\pi}\rm{Im}\, \Pi_S(s)\,$$ of the hadronic states, starting 
 with the contribution of the  $S$-wave $K\pi$ state. 
\eq{
\rho_S^{(K\pi)}(s)=\frac{3}{32\pi^2}|f_0(s)|^2(m_K^2-m_\pi^2)^2
\frac{\sqrt{\lambda_{K\pi}(s)}}{s}\,\theta(s-(m_K+m_\pi)^2)\,,
\label{eq:hadrdens}
}
where  the definition  (\ref{eq:KpivectFF}) of the scalar $K\pi$ form factor is used and the factor 3/2 accounting for the two isospin related states $K^-\pi^+$ and $\bar{K}^0\pi^0$ is included. We then assume that the sum over all other contributions to the hadronic density is approximated with the spectral density calculated from OPE and integrated above an effective threshold $s_0$, so that~\Eq{eq:Borel2pt} turns into:
\eq{
M^4\Pi_S^{''(OPE)}(M^2)=\int\limits_0^{s_0}\!ds \,e^{-s/M^2}\rho_S^{(K\pi)}(s)
+\int\limits_{s_0}^\infty\!ds \,e^{-s/M^2}\rho_S^{(OPE)}(s)\,.
\label{eq:2ptdual}
}
Using~\Eq{eq:hadrdens}, we obtain the desired sum rule in the form\footnote{Note that it is more convenient 
to represent the r.h.s.  in terms of two parts, rather than
as an integral over $\rho_S^{(OPE)}(s)$ taken from $m_s^2$ to $s_0$. The reason is 
a rather complicated expression of the OPE spectral density in the vicinity of the threshold.}:
\begin{eqnarray}
\frac{3}{32\pi^2}(m_K^2\!-\!  m_\pi^2)^2\!\!\!\!\!\!\!\!\int\limits_{(m_K+m_\pi)^2}^{s_0}
\!\!\!\!\!\!\!\!\!ds \,e^{-s/M^2}
|f_0(s)|^2
\frac{\sqrt{\lambda_{K\pi}(s)}}{s}
%\nonumber\\
=M^4\Pi_S^{''(OPE)}(M^2)\!-\!\!
\int\limits_{s_0}^\infty\!ds \,e^{-s/M^2}\!\rho_S^{(OPE)}(s)\,.
\nonumber\\
\label{eq:fitSR}
\end{eqnarray}
The expressions for $\Pi_S^{''(OPE)}(M^2)$ and $\rho_S^{(OPE)}(s)$ in this sum rule are taken from 
the literature \cite{Jamin:1994vr,Chetyrkin:1996xa} (see also \cite{Chetyrkin:2005kn}). They include the 
perturbative part (the simple loop and gluon radiative corrections) up to $O(\alpha_s^3)$ 
and the condensate contributions up to dimension $d= 6$. 
For simplicity, we omit the known but numerucally very small 
$O(\alpha_s^4)$ terms in the perturbative part. Note that, apart from the overall factor $(m_s-m_d)^2$, the $d$-quark
mass is neglected and the expansion in the numerically small ratio $m_s^2/M^2$
is applied. We use: 
\eq{
\Pi_S^{''(OPE)}(M^2)=\Pi_S^{''(d=0,2)}(M^2)+\Pi_S^{''(d=4,6)}(M^2)\,.
\label{eq:subdiv}
}
The part with $d=0,2$ terms originating from the perturbative contributions
and $O(m_s^2/M^2)$ corrections is 
\eq{
\Pi_S^{''(d=0,2)}(M^2)=
\frac{3(m_s-m_d)^2}{8\pi^2}\Bigg\{1+
\sum\limits_{n=1}^3 b_{0,n}\, \left(\frac{\alpha_s}{\pi}\right)^n
- 2 \frac{m_s^2}{M^2}
\left(
1+\sum\limits_{n=1}^2 b_{2,n} \, \left(\frac{\alpha_s}{\pi}\right)^n
\right)
\Bigg \} \,,
\label{eq:BpiOPE} 
}
where the coefficients of $\alpha_s$-expansion are 
\begin{eqnarray}
b_{0,1} =  \frac{11}{3} + 2\,\gamma_E-2  \,l_{M}
{},
\nonumber
\end{eqnarray}
\begin{eqnarray}
b_{0,2} =   
\frac{5071}{144} 
+\frac{139}{6}  \, \gamma_E
+\frac{17}{4}  \,\gamma_E^2
-\frac{17}{24}  \pi^2
-\frac{35}{2}  \,\zeta_{3}
-\frac{139}{6}  \,l_{M}
-\frac{17}{2}  \, \gamma_E \,l_{M}
+\frac{17}{4}  \,l_M^2\,,
\label{b02}
\end{eqnarray}
\begin{eqnarray}
b_{0,3} &=&   
\frac{1995097}{5184} 
+\frac{2720}{9}  \, \gamma_E
+\frac{695}{8}  \,\gamma_E^2
+\frac{221}{24}  \,\gamma_E^3
-\frac{695}{48}  \pi^2
-\frac{221}{48}  \, \gamma_E \pi^2
\nonumber
\\
&{}&
\phantom{+}
-\frac{1}{36}  \pi^4
-\frac{61891}{216}  \,\zeta_{3}
-\frac{475}{4}  \, \gamma_E \,\zeta_{3}
+\frac{715}{12}  \,\zeta_{5}
\nonumber\\
&{+}& \,l_{M}
\left[
-\frac{2720}{9} 
-\frac{695}{4}  \, \gamma_E
-\frac{221}{8}  \,\gamma_E^2
+\frac{221}{48}  \pi^2
+\frac{475}{4}  \,\zeta_{3}
%zero == 0
\right]
\nonumber\\
&{+}& \,l_M^2
\left[
\frac{695}{8} 
+\frac{221}{8}  \, \gamma_E
\right]
-\frac{221}{24} \,l_M^3
{},
\label{b03}
\end{eqnarray}
%

%%%%%%%%%%%%%%%%%%%%%%%%%%%%%%%%%%%%
\begin{eqnarray}
b_{2,1} =   
\frac{16}{3} 
+4  \, \gamma_E
-4  \,l_{M}
{},
\label{b21}
\end{eqnarray}
%zero == 0
\begin{eqnarray}
b_{2,2} =  
\frac{5065}{72} 
+\frac{97}{2}  \, \gamma_E
+\frac{25}{2}  \,\gamma_E^2
-\frac{25}{12}  \pi^2
-\frac{77}{3}  \,\zeta_{3}
-\frac{97}{2}  \,l_{M}
-25  \, \gamma_E \,l_{M}
+\frac{25}{2}  \,l_M^2\,.
%zero == 0
\label{b22}
\end{eqnarray}
In the above, $l_M = \log\frac{M^2}{\mu^2}$, $\zeta_n\equiv \zeta(n)$ is the
Riemann's zeta-function, $\gamma_E$ is the Euler constant and
the $\overline{\rm MS}$ quark masses are used. 

The  part containing power corrections $\sim 1/M^d$ with $d=4,6$ is:
\begin{eqnarray}
\Pi_S^{''(d=4,6)}(M^2) =
\frac{(m_s-m_d)^2}{2M^4}\Bigg\{2m_s\langle \bar q q\rangle
\left[1 + \frac{\alpha_s}{\pi}\left(\frac{14}{3}+2\gamma_E-2l_M\right)
\right]
\nonumber
\\
-\frac19 I_G\left[1 + \frac{\alpha_s}{\pi}\left(
\frac{67}{18} + 2\gamma_E-2l_M\right)\right]
+ I_s\left[1+ \frac{\alpha_s}{\pi}\left(\frac{37}{9}+2\gamma_E-
2l_M\right)\right] 
\nonumber\\
-\frac{3}{7\pi^2}m_s^4\left(\frac{\pi}{\alpha_s}+
\frac{5}{6} + \frac{15}{4}\gamma_E-\frac{15}{4}l_M
\right)+\frac{I_6}{3M^2}\Bigg\}
\,,\label{BpiOPEnp} 
\end{eqnarray}
where  the 
combinations  of condensate densities and  $m_s$-power corrections with $d=4$ are:
\eq{
I_s=m_s\langle \bar{s}s\rangle+\frac{3}{7\pi^2}\,
m_s^4\left(\frac{\pi}{\alpha_s}-\frac{53}{24}\right )\,,
}

\eq{
I_G=-\frac{9}{4}\langle \frac{\alpha_s}{\pi}
G^2\rangle\left (1+\frac{16}{9}\frac{\alpha_s}{\pi}\right)+
4\frac{\alpha_s}{\pi}\left(1+\frac{91}{24}\frac{\alpha_s}{\pi}\right)m_s\langle \bar{s}s\rangle+\frac{3}{4\pi^2}\left(1+\frac{4}{3}\frac{\alpha_s}{\pi}\right)
m_s^4\,,
}
and the one with $d=6$  is 
\eq{
I_6 (\mu)= 3m_s(\mu)\langle\bar{q}qG\rangle-
\frac{32}9\pi^2\frac{\alpha_s(\mu)}{\pi}r_v\Big (\langle\bar{q}q\rangle^2
+\langle\bar{s}s\rangle^2+9\langle\bar{q}q\rangle
\langle\bar{s}s\rangle\Big) (\mu)\,.
\label{eq:I6}
}
Here we use the following shorthand notation
for the quark and gluon condensate densities:
\begin{eqnarray}
&&\langle \bar{q}q\rangle \equiv 
\langle 0|\bar{d}d |0\rangle\simeq \langle 0|\bar{u}u |0\rangle, \langle \bar{s}s\rangle \equiv 
\langle 0|\bar{s}s |0\rangle,
\nonumber\\
&&\langle \frac{\alpha_s}{\pi}
G^2\rangle\equiv\frac{\alpha_s}{\pi}\langle 0| G^a_{\mu\nu} G^{a\,\mu\nu} |0\rangle\,,
\nonumber
\end{eqnarray}
and the standard parametrization for the quark-gluon condensate density: 
\begin{eqnarray}
\langle\bar{q}qG\rangle \equiv \langle 0 |g_s\bar{q} G^a_{\mu\nu}t^a \sigma^{\mu\nu}q|0 \rangle = m_0^2 \langle \bar{q}q\rangle\,.
\nonumber
\end{eqnarray}
Finally, the four-quark condensate contribution in~\Eq{eq:I6} is 
factorized according to the vacuum dominance
ansatz \cite{Shifman:1978bx} and the parameter $r_v$ reflects
the uncertainty of this approximation. We use $r_v =[0.1,1]$, with a default value at $r_v=1$. Note, that apart from $\alpha_s$, the $s$-quark mass and condensate density are the only scale-dependent parameters,
since we neglect the inessential scale-dependence of the quark-gluon and 
four-quark condensate terms \cite{Ioffe:2005ym}. Hence, the condensates and $\alpha_s$ 
in~\Eq{eq:I6} are taken at the fixed scale $\mu=1$ GeV.

In addition, we need the spectral function 
calculated from OPE
with the same $O(\alpha_s^3)$ accuracy:
\begin{eqnarray}
&&\rho_S^{\rm OPE}(s) =\frac{1}{\pi}\mbox{Im}\Pi_S(s)=
\frac{3(m_s-m_d)^2}{8\pi^2}\,s\,\Bigg\{1+
\sum\limits_{n=1}^3 r_{0,n}\, \left(\frac{\alpha_s}{\pi}\right)^n
- 2 \frac{m_s^2}{s}
\left(1+\sum\limits_{n=1}^2 r_{2,n} \, \left(\frac{\alpha_s}{\pi}\right)^n 
\right)
\Bigg \} 
\nonumber\\
&&+\frac{m_s^2}{s}\left\{ \frac{45}{56\pi^2}m_s^4
-
2\frac{\alpha_s}{\pi}m_s\langle \bar qq\rangle +
\frac{\alpha_s}{9\pi}I_G - \frac{\alpha_s}{\pi}I_s\right\}
%\nonumber
\,,
\label{ImpiOPE}
\end{eqnarray}
where $l_s = \log\frac{s}{\mu^2}$ and the coefficients are:
\begin{eqnarray}
r_{0,1} =  
\frac{17}{3} 
-2  \, l_s,
\ \ 
{r_{0,2} =  } 
\frac{9631}{144} 
-\frac{17}{12}  \pi^2
-\frac{35}{2}  \,\zeta_{3}
-\frac{95}{3}  \, l_s
+\frac{17}{4}  \,l_s^2
{},
\label{r02}
\end{eqnarray}

\begin{eqnarray}
r_{0,3} =  
&{}&
\frac{4748953}{5184} 
-\frac{229}{6}  \pi^2
-\frac{1}{36}  \pi^4
-\frac{91519}{216}  \,\zeta_{3}
+\frac{715}{12}  \,\zeta_{5}
-\frac{4781}{9}  \, l_s
+\frac{221}{24}  \pi^2 \, l_s
\nonumber
\\
&{}&
\phantom{+}
+\frac{475}{4}  \,\zeta_{3} \, l_s
+\frac{229}{2}  \,l_s^2
-\frac{221}{24}  \,l_s^3
{},
\label{r03}
\end{eqnarray}
\begin{eqnarray}
r_{2,1} =  
\frac{16}{3} 
-4  \, l_s
{},
\ \ 
r_{2,2} =  
\frac{5065}{72} 
-\frac{25}{6}  \pi^2
-\frac{77}{3}  \,\zeta_{3}
-\frac{97}{2}  \, l_s
+\frac{25}{2}  \,l_s^2
%zero == 0
{}.
\label{r21_22}
\end{eqnarray}
Note that this form of the spectral density is adjusted to the 
integration above $s_0\gg m_s^2$.

In principle, we could now determine $s_0$ for fixed $M^2$ by equating both sides of the sum rule in~\Eq{eq:fitSR}. A similar procedure was followed in \cite{Descotes-Genon:2019bud}. For our sum rule, this entails using the four models for $f_0$ introduced in~\Sec{sec:ffmodel} on the left-hand side of \eqref{eq:fitSR}. For the OPE contribution on the right-hand side, we use the input parameters within their ranges indicated in~\Tab{tab:OPEinputs}. 
We use the four-loop renormalization of the strong
coupling and of the quark mass from \cite{Chetyrkin:2000yt}. Furthermore, we adopt the interval of Borel parameter 
squared $1.0<M^2<1.5$ GeV$^2$,
close to the one used in the case of the $P$-wave
$K\pi$ state \cite{Descotes-Genon:2019bud}. We have checked that
for $M^2>1.0$ GeV$^2$ the contributions of 
power corrections in the OPE are very small, so that  
\begin{eqnarray}
&&\frac{\Pi_S^{''(d=4,6)}(M^2)}{\Pi_S^{''(d=0,2)}(M^2)} <6.5\%\,. 
\label{eq:ratio2}
\end{eqnarray}
In~\Eq{eq:fitSR}
we adopt $\mu=M$ and allow a variation:
$$M^2/2<\mu^2<2M^2.$$ 
 Within this range, the convergence of the perturbative expansion in $\alpha_s$ is quite satisfactory, manifested by the tiny $O(\alpha_s^4)$ terms not included in our analysis.

For a fixed value of $M^2$, we can then determine the value of $s_0$ for each $f_0$ model. Doing so, we find broad intervals for $s_0$ that all satisfy the two point sum rule within uncertainty of the latter determined by 
varying the input parameters. This is mainly caused by the still comparatively large uncertainty of $m_s$.  Therefore, in our numerical analysis, we fix $M^2$ 
at the central value of the adopted interval and 
take a single corresponding value of $s_0$ 
for which the two-point sum rule is satisfied for all four models. Our resulting choice is 
\begin{equation}
M^2 = 1.25 \;{\rm GeV}^2 \ , \quad\quad 
s_0 = 1.8\; {\rm GeV}^2, \end{equation}
and we not vary these parameters. Additionally, for this choice the contribution of 
higher states in the sum rule \eqref{eq:fitSR}
estimated via duality remains moderate:
\begin{eqnarray}
&&\frac{\int\limits_{s_0}^\infty\!ds \,e^{-s/M^2}\rho_S^{(OPE)}(s)}{M^4\Pi_S^{''(OPE)}(M^2)}<40 \%\ ,
\label{eq:ratios0}
\end{eqnarray}
similar to what is found in the case of the LCSRs.

\begin{table}
\centering
\def\arraystretch{1.5}
\setlength{\tabcolsep}{7pt}
\begin{tabular}{@{}ccc@{}}
\toprule[0.7mm]
$\alpha_s(m_Z)$ & $0.1179\pm 0.00105$ & 
\cite{ParticleDataGroup:2022pth}\\
\hline
$\langle \bar{q}q \rangle(\mu=2\, \mbox{GeV} )$ &
$-\left(286\pm 23~\mbox{MeV} \right)^3 $ &\cite{Aoki:2019cca}\\
\hline
$\langle \bar{s}s\rangle /\langle \bar{q}q \rangle$ & $0.8\pm 0.3$ &\\
$\langle GG\rangle$ & $0.012^{+0.006}_{-0.012} ~\mbox{GeV}^4$ & \cite{Ioffe:2005ym}\\
$m_0^2$ & $0.8 \pm 0.2 ~\mbox{GeV}^2\,,$ &\\
%$r_v$ & $0.55 \pm 0.45$ &\\
\bottomrule[0.7mm]
\end{tabular}
\caption{\it Inputs used in the two-point sum rule in addition to the ones presented in~\Tab{tab:inputs}.}
\label{tab:OPEinputs}
\end{table}

\end{document}